\documentclass[]{IEEEtran}

\usepackage{cite,graphicx,epsfig,epstopdf,subfigure,amsmath,amssymb,eufrak,mathrsfs,epsf}

\newcommand{\modLambda}{\textrm{ mod }\Lambda}
\newcommand{\halflog}{\frac{{1}}{{2}}\log}
\newcommand{\snr}{\textrm{SNR}}
\newcommand{\inr}{\textrm{INR}}

\newtheorem{theorem}{Theorem}

\newtheorem{lemma}{Lemma}

\newtheorem{definition}{Definition}

\begin{document}
\title{Approximate Capacity of a Class of\\ Partially Connected Interference Channels}
\author{Muryong~Kim, Yitao~Chen, and Sriram~Vishwanath,~\IEEEmembership{Senior Member,~IEEE}\\% <-this % stops a space
%\normalsize{E-mail: muryong@utexas.edu}
%\thanks{This work was supported in part by the Office of Naval Research (ONR).}
\thanks{The authors are with the University of Texas at Austin, Austin, TX 78701 USA (e-mail: muryong@utexas.edu, yitaochen@utexas.edu, sriram@ece.utexas.edu).}}

%\markboth{IEEE TRANSACTIONS ON INFORMATION
%THEORY}{Shell \MakeLowercase{\textit{et al.}}: }

\maketitle

\begin{abstract}
We derive inner and outer bounds on the capacity region for a class of three-user partially connected interference channels. We focus on the impact of topology, interference alignment, and interplay between interference and noise. The representative channels we consider are the ones that have clear interference alignment gain. For these channels, Z-channel type outer bounds are tight to within a constant gap from capacity. We present near-optimal achievable schemes based on rate-splitting and lattice alignment.
\end{abstract}


\IEEEpeerreviewmaketitle

%\newpage

\section{Introduction}
\subsection{Motivation}
The capacity of the Interference channel remains one of the most challenging open problems in the domain of network information theory. The capacity region is not known in general, except for a specific range of channel parameters. For the two-user scalar Gaussian interference channel, where the interference alignment is not required, the approximate capacity region to within one bit is known \cite{EtkinTseWang08}. For the channels where interference alignment is required such as the $K$-user Gaussian interference channel \cite{MotahariGharanMaddahAliKhandani14,SridharanVishwanathJafar08,SridharanJafarianVishwanathJafarShamai08,BreslerParekhTse10,JafarVishwanath10,OrdentlichErezNazer14} and the Gaussian X-channel \cite{MotahariGharanMaddahAliKhandani14,HuangCadambeJafar12,NiesenMaddahAli13}, a tight characterization of the capacity region is not known, even for symmetric channel cases.

A tractable approach to the capacity of interference channels is to consider partial connectivity of interference links and analyze the impact of topology on the capacity. Topological interference management \cite{Jafar2014} approach gives important insights on the degrees-of-freedom (DoF) of partially connected interference channels and their connection to index coding problems \cite{BirkKol98,Bar-YossefBirkJayramKol06,Bar-YossefBirkJayramKol11,AlonLubetzkyStavWeinsteinHassidim08,MalekiCadambeJafar2012,ArbabjolfaeiBandemerKimSasogluWang13,Ong14,EffrosElRouayhebLangberg15}. It is shown that the symmetric DoF of a partially connected interference channel can be found by solving the corresponding index coding problem.

In this paper, we consider a class of three-user partially connected interference channels and characterize approximate capacity regions at finite SNR. We focus on the impact of interference topology, interference alignment, and interplay between interference and noise. We choose a few representative topologies where we can achieve clear interference alignment gain. For these topologies, Z-channel type outer bounds are tight to within a constant gap from the corresponding inner bound. For each topology, we present an achievable scheme based on rate-splitting, lattice alignment, and successive decoding.

\subsection{Related Work}
Lattice coding based on nested lattices is shown to achieve the capacity of the single user Gaussian channel in \cite{ErezZamir04,Zamir14}. The idea of lattice-based interference alignment by decoding the sum of lattice codewords appeared in the conference version of \cite{BreslerParekhTse10}. This lattice alignment technique is used to derive capacity bounds for three-user interference channel in \cite{SridharanVishwanathJafar08,SridharanJafarianVishwanathJafarShamai08}. The idea of decoding the sum of lattice codewords is also used in \cite{WilsonNarayananPfisterSprintson10,NamChungLee10,NamChungLee11} to derive the approximate capacity of the two-way relay channel. An extended approach, compute-and-forward \cite{NazerGastpar11,GastparNazer11} enables to first decode some linear combinations of lattice codewords and then solve the lattice equation to recover the desired messages. This approach is also used in \cite{OrdentlichErezNazer14} to characterize approximate sum-rate capacity of the fully connected $K$-user interference channel.

The idea of sending multiple copies of the same sub-message at different signal levels, so-called Zigzag decoding, appeared in \cite{JafarVishwanath10} where receivers collect side information and use them for interference cancellation.

The $K$-user cyclic Gaussian interference channel is considered in \cite{ZhouYu13} where an approximate capacity for the weak interference regime ($\snr_k\geq \inr_k$ for all $k$) and the exact capacity for the strong interference regime ($\snr_k\leq \inr_k$ for all $k$) are derived. Our type 4 and 5 channels are $K=3$ cases in \emph{mixed} interference regimes, which were not considered in \cite{ZhouYu13}.

\subsection{Main Results}
\begin{figure*}[tp]
  \begin{center}
    \mbox{
      \subfigure[Type 1]{\includegraphics[width=0.18\textwidth]{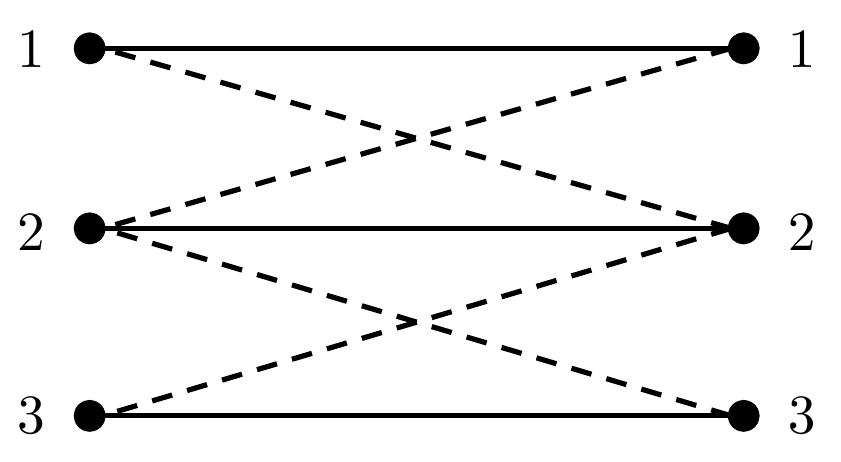}}
      }
    \mbox{
      \subfigure[Type 2]{\includegraphics[width=0.18\textwidth]{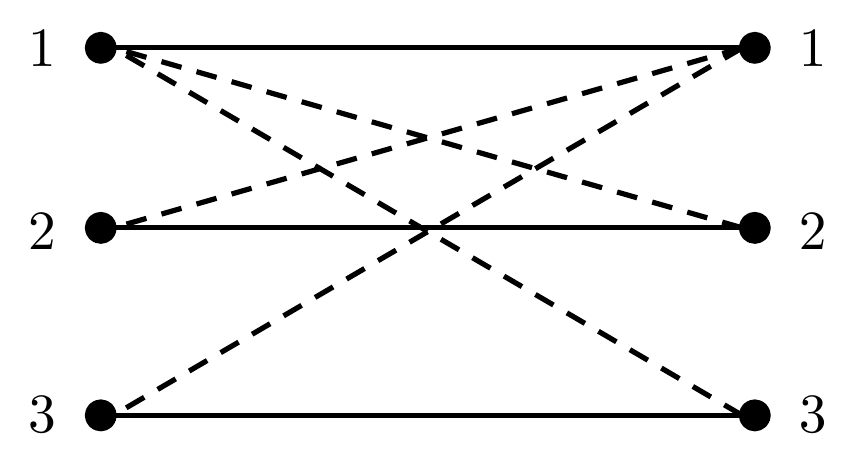}}
      }
    \mbox{
      \subfigure[Type 3]{\includegraphics[width=0.18\textwidth]{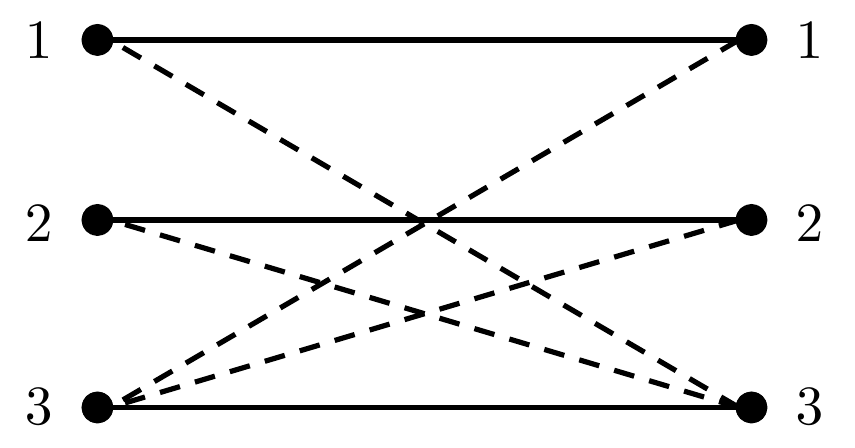}}
      }
    \mbox{
      \subfigure[Type 4]{\includegraphics[width=0.18\textwidth]{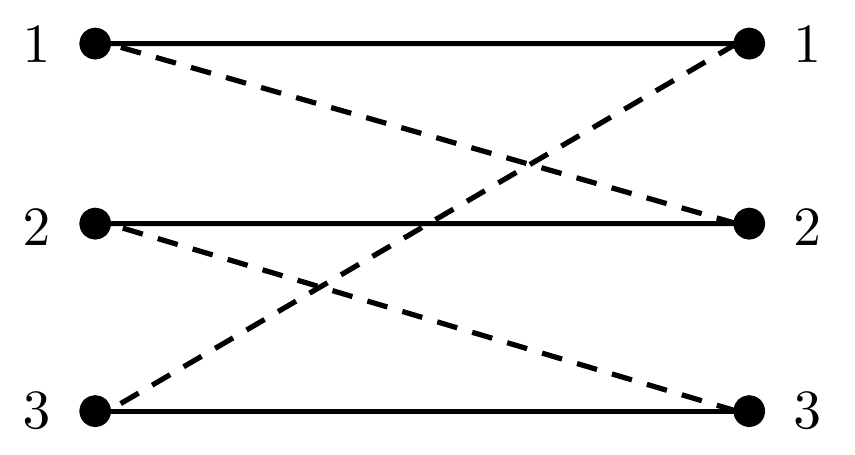}}
      }
    \mbox{
      \subfigure[Type 5]{\includegraphics[width=0.18\textwidth]{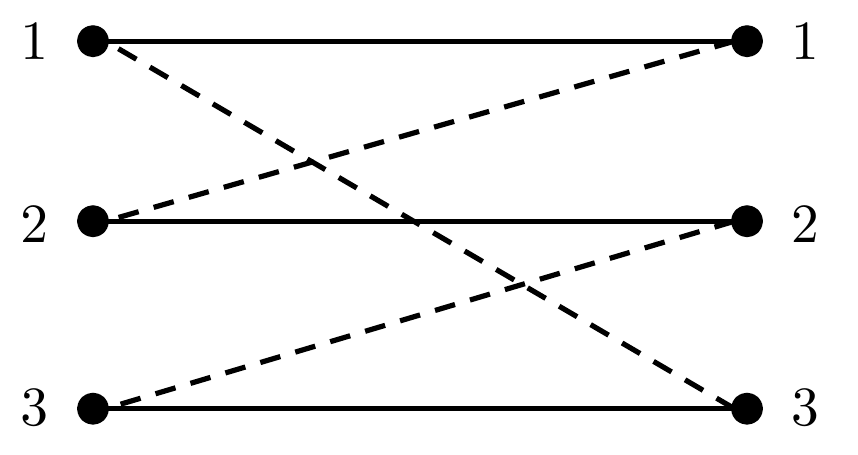}}
      }
    \mbox{
      \subfigure[$\mathcal{G}_1$]{\includegraphics[width=0.18\textwidth]{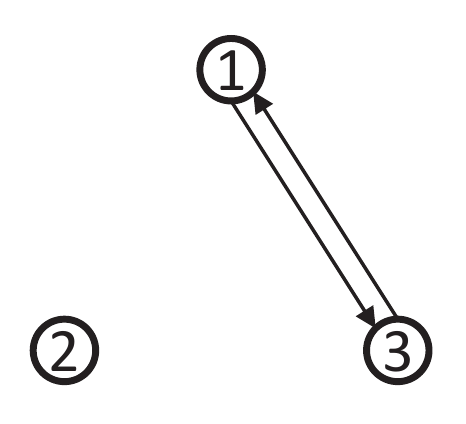}}
      }
    \mbox{
      \subfigure[$\mathcal{G}_2$]{\includegraphics[width=0.18\textwidth]{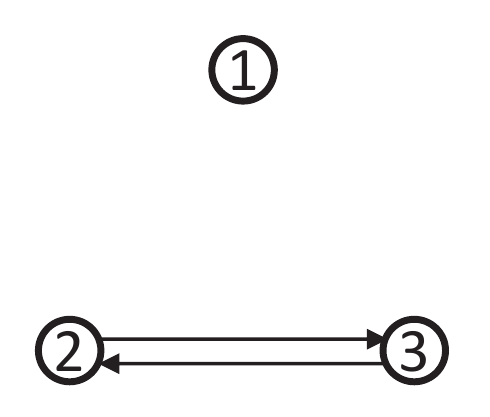}}
      }
    \mbox{
      \subfigure[$\mathcal{G}_3$]{\includegraphics[width=0.18\textwidth]{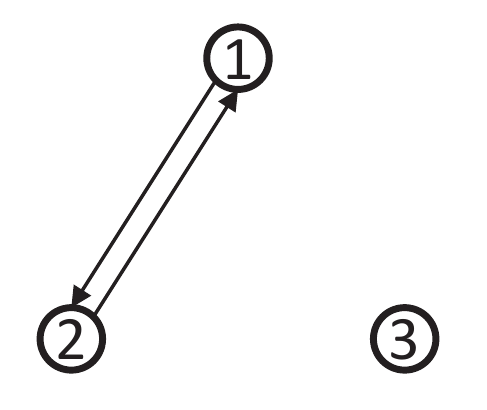}}
      }
    \mbox{
      \subfigure[$\mathcal{G}_4$]{\includegraphics[width=0.18\textwidth]{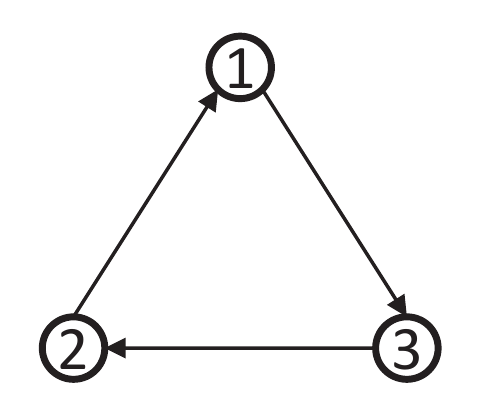}}
      }
    \mbox{
      \subfigure[$\mathcal{G}_5$]{\includegraphics[width=0.18\textwidth]{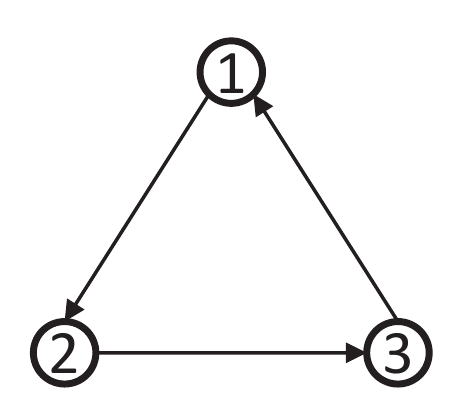}}
      }
\caption{Five channel types and their side information graphs: $\mathcal{G}_1=\{(1|3),(2),(3|1)\}$, $\mathcal{G}_2=\{(1),(2|3),(3|2)\}$, $\mathcal{G}_3=\{(1|2),(2|1),(3)\}$, $\mathcal{G}_4=\{(1|2),(2|3),(3|1)\}$, and $\mathcal{G}_5=\{(1|3),(2|1),(3|2)\}$.}
\label{fig:channelType}
  \end{center}
\end{figure*}

We consider five channel types defined in Table \ref{tab:TxRxSignals} and described in Fig. \ref{fig:channelType} (a)--(e). Each channel type is a partially connected three-user Gaussian interference channel. Each transmitter is subject to power constraint $\mathbb{E}[X_k^2]\leq P_k=P$. Let us denote the noise variance by $N_k=\mathbb{E}[Z_k^2]$. Without loss of generality, we assume that $N_1\leq N_2\leq N_3$.

\begin{definition}[side information graph]
The side information graph representation of an interference channel satisfies the following.
\begin{itemize}
\item A node represents a transmitter-receiver pair, or equivalently, the message.
\item There is a directed edge from node $i$ to node $j$ if transmitter $i$ does not interfere at receiver $j$.
\end{itemize}
\end{definition}
The side information graphs for five channel types are described in Fig. \ref{fig:channelType} (f)--(j). We state the main results in the following two theorems, of which the proofs will be given in the main body of the paper.

\begin{theorem}[Capacity region outer bound]
For the five channel types, if $(R_1,R_2,R_3)$ is achievable, it must satisfy
\begin{eqnarray}
\sum_{j\in\mathcal{K}} R_j \leq \halflog\left(1+\frac{|\mathcal{K}|P}{\min_{j\in\mathcal{K}}\{N_j\}}\right)
\end{eqnarray}
for every subset $\mathcal{K}$ of the nodes $\{1,2,3\}$ that does not include a directed cycle in the side information graph over the subset.
\end{theorem}

\begin{theorem}[Capacity region to within one bit]\ \\
For any rate triple $(R_1,R_2,R_3)$ on the boundary of the outer bound region, the point $(R_1-1,R_2-1,R_3-1)$ is achievable.
%For any rate triple $(R_1,R_2,R_3)$ on the boundary of the achievable rate region, the rate point $(R_1+1,R_2+1,R_3+1)$ is not achievable.
\end{theorem}

\begin{table}
\begin{center}
\begin{tabular}{|c|l|}
\hline
Type & \ \ \ \ \ \ \ \ Channel model \\
\hline
$1$ 
& $\left. {\begin{array}{*{20}l}
  Y_1=X_1+X_2+Z_1\\
  Y_2=X_1+X_2+X_3+Z_2\\
  Y_3=X_2+X_3+Z_3\\
\end{array} } \right.$
\\
\hline
$2$ 
& $\left. {\begin{array}{*{20}l}
  Y_1=X_1+X_2+X_3+Z_1\\
  Y_2=X_1+X_2+Z_2\\
  Y_3=X_1+X_3+Z_3\\\end{array} } \right.$
\\
\hline
$3$ 
& $\left. {\begin{array}{*{20}l}
  Y_1=X_1+X_3+Z_1\\
  Y_2=X_2+X_3+Z_2\\
  Y_3=X_1+X_2+X_3+Z_3\\
\end{array} } \right.$
\\
\hline
$4$ 
& $\left. {\begin{array}{*{20}l}
  Y_1=X_1+X_3+Z_1\\
  Y_2=X_1+X_2+Z_2\\
  Y_3=X_2+X_3+Z_3\\
\end{array} } \right.$
\\
\hline
$5$ 
& $\left. {\begin{array}{*{20}l}
  Y_1=X_1+X_2+Z_1\\
  Y_2=X_2+X_3+Z_2\\
  Y_3=X_1+X_3+Z_3\\
\end{array} } \right.$
\\
\hline
\end{tabular}
\caption{Five channel types}
  \label{tab:TxRxSignals}
\end{center}
\end{table}

\subsection{Paper Organization and Notation}
The capacity outer bounds are derived in Section II. The inner bounds for each channel type and the corresponding gap analysis are given in Section III, IV, V, VI, VII, respectively. Section VIII concludes the paper. While lattice coding-based achievable rate regions for channel types 4 and 5 are presented in Section VI and VII, random coding achievability is given in Appendix.

Signal $\mathbf{x}_{ij}$ is a coded version of message $M_{ij}$ with code rate $R_{ij}$ unless otherwise stated. The single user capacity at receiver $k$ is denoted by $C_k=\halflog\left(1+\frac{P}{N_k}\right)$. Let $\mathcal{C}$ denote the capacity region of an interference channel. Also, let $\mathcal{R}_i$ and $\mathcal{R}_o$ denote the capacity inner bound and the capacity outer bound, respectively. Thus, $\mathcal{R}_i\subset\mathcal{C}\subset\mathcal{R}_o$. Let $\delta_k$ denote the gap on the rate $R_k$ between $\mathcal{R}_i$ and $\mathcal{R}_o$. Let $\delta_{jk}$ denote the gap on the sum-rate $R_j+R_k$ between $\mathcal{R}_i$ and $\mathcal{R}_o$. For example, if
\begin{eqnarray}
&&\mathcal{R}_i=\{(R_j,R_k): R_k\leq L_k, R_j+R_k\leq L_{jk}\}\\
&&\mathcal{R}_o=\{(R_j,R_k): R_k\leq U_k, R_j+R_k\leq U_{jk}\},
\end{eqnarray}
then $\delta_k=U_k-L_k$ and $\delta_{jk}=U_{jk}-L_{jk}$. For side information graph, we use graph notation of \cite{ArbabjolfaeiBandemerKimSasogluWang13}. For example, $\mathcal{G}_1=\{(1|3),(2),(3|1)\}$ means that node 1 has an incoming edge from node 3, that node 2 has no incoming edge, and that node 3 has an incoming edge from node 1.

\section{Capacity outer bounds}
\begin{figure}[tp]
  \begin{center}
    \mbox{
      \subfigure[Channel type 1]{\includegraphics[width=0.22\textwidth]{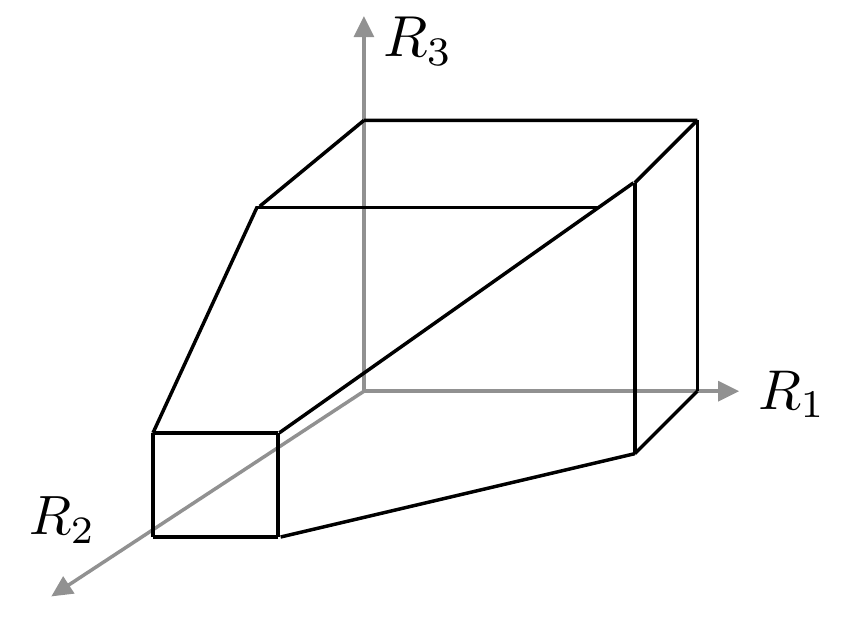}}
      }
    \mbox{
      \subfigure[Channel types 4 and 5]{\includegraphics[width=0.22\textwidth]{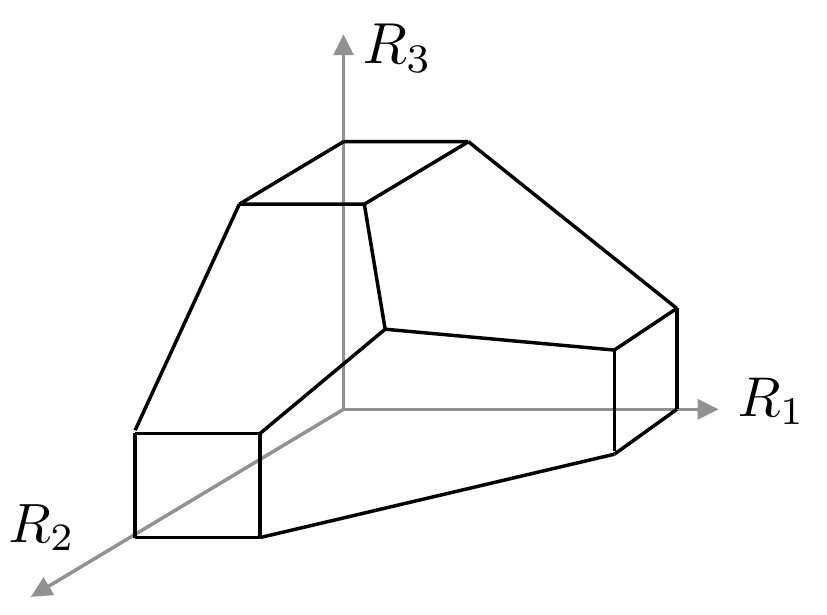}}
      }
\caption{The shape of the outer bound region. The regions for channel types 2 and 3 look similar to the one for channel type 1 (with change of axis).}
\label{fig:outerBoundShape}
  \end{center}
\end{figure}

We prove the capacity outer bound in \emph{Theorem 1} for each channel type. The result is summarized in Table \ref{tab:outerBounds}. The shape of the outer bound region is illustrated in Fig. \ref{fig:outerBoundShape}. For all channel types, we assume $P_1=P_2=P_3=P$ and $N_1\leq N_2\leq N_3$.

\subsection{Channel Type 1}
In this section, we present an outer bound on the capacity region of Type 1 channel defined by
\[
\left[ {\begin{array}{*{20}c}
   Y_1\\
   Y_2\\
   Y_3\\
\end{array} } \right]
=\left[ {\begin{array}{*{20}c}
   1 & 1 & 0\\
   1 & 1 & 1\\
   0 & 1 & 1\\
\end{array} } \right]
\left[ {\begin{array}{*{20}c}
   X_1\\
   X_2\\
   X_3\\
\end{array} } \right]
+\left[ {\begin{array}{*{20}c}
   Z_1\\
   Z_2\\
   Z_3\\
\end{array} } \right].
\]
We state the outer bound in the following theorem.
\begin{theorem}
The capacity region of Type 1 channel is contained in the following outer bound region:
\begin{eqnarray*}
&&\ \ \ \ \ \ \ R_k\leq C_k,\ k=1,2,3\\
&&R_1+R_2\leq \halflog\left(1+\frac{P}{N_1}\right)+\halflog\left(\frac{2P+N_2}{P+N_2}\right)\\
&&R_2+R_3\leq \halflog\left(1+\frac{P}{N_2}\right)+\halflog\left(\frac{2P+N_3}{P+N_3}\right).
\end{eqnarray*}
\begin{IEEEproof}
The individual rate bounds are obvious. We proceed to sum-rate bounds.
\begin{eqnarray*}
&&n(R_1+R_2-\epsilon)\\
&&\ \ \leq I(X_1^n;Y_1^n)+I(X_2^n;Y_2^n)\\
&&\ \ \leq I(X_1^n;Y_1^n|X_2^n)+I(X_2^n;Y_2^n|X_3^n)\\
&&\ \ = h(Y_1^n|X_2^n)-h(Y_1^n|X_1^n,X_2^n)\\
&&\ \ \ \ \ \ \ +h(Y_2^n|X_3^n)-h(Y_2^n|X_2^n,X_3^n)\\
&&\ \ = h(X_1^n+Z_1^n)-h(Z_1^n)\\
&&\ \ \ \ \ \ \ +h(X_1^n+X_2^n+Z_2^n)-h(X_1^n+Z_2^n)\\
&&\ \ \leq \frac{n}{2}\log\left(\frac{P+N_1}{N_1}\right)+\frac{n}{2}\log\left(\frac{2P+N_2}{P+N_2}\right)
\end{eqnarray*}
where the first inequality is by Fano's inequality, the second inequality due to the independence of $X_1,X_2,X_3$. The third inequality holds from the fact that Gaussian distribution maximizes differential entropy and that $h(X_1^n+Z_1^n)-h(X_1^n+Z_2^n)$ is also maximized by Gaussian distribution. Similarly,
\begin{eqnarray*}
&&n(R_2+R_3-\epsilon)\\
&&\ \ \leq I(X_2^n;Y_2^n)+I(X_3^n;Y_3^n)\\
&&\ \ \leq I(X_2^n;Y_2^n|X_1^n,X_3^n)+I(X_3^n;Y_3^n)\\
&&\ \ = h(Y_2^n|X_1^n,X_3^n)-h(Y_2^n|X_1^n,X_2^n,X_3^n)\\
&&\ \ \ \ \ \ \ +h(Y_3^n)-h(Y_3^n|X_3^n)\\
&&\ \ = h(X_2^n+Z_2^n)-h(Z_2^n)\\
&&\ \ \ \ \ \ \ +h(X_2^n+X_3^n+Z_3^n)-h(X_2^n+Z_3^n)\\
&&\ \ \leq \frac{n}{2}\log\left(\frac{P+N_2}{N_2}\right)+\frac{n}{2}\log\left(\frac{2P+N_3}{P+N_3}\right).
\end{eqnarray*}
\end{IEEEproof}
\end{theorem}

\subsection{Channel Type 2}
In this section, we present an outer bound on the capacity region of Type 2 channel defined by
\[
\left[ {\begin{array}{*{20}c}
   Y_1\\
   Y_2\\
   Y_3\\
\end{array} } \right]
=\left[ {\begin{array}{*{20}c}
   1 & 1 & 1\\
   1 & 1 & 0\\
   1 & 0 & 1\\
\end{array} } \right]
\left[ {\begin{array}{*{20}c}
   X_1\\
   X_2\\
   X_3\\
\end{array} } \right]
+\left[ {\begin{array}{*{20}c}
   Z_1\\
   Z_2\\
   Z_3\\
\end{array} } \right].
\]
We state the outer bound in the following theorem.
\begin{theorem}
The capacity region of Type 2 channel is contained in the following outer bound region:
\begin{eqnarray*}
&&\ \ \ \ \ \ \ R_k\leq C_k,\ k=1,2,3\\
&&R_1+R_2\leq \halflog\left(1+\frac{P}{N_1}\right)+\halflog\left(\frac{2P+N_2}{P+N_2}\right)\\
&&R_1+R_3\leq \halflog\left(1+\frac{P}{N_1}\right)+\halflog\left(\frac{2P+N_3}{P+N_3}\right).
\end{eqnarray*}
\begin{IEEEproof}
\begin{eqnarray*}
&&n(R_1+R_2-\epsilon)\\
&&\ \ \leq I(X_1^n;Y_1^n)+I(X_2^n;Y_2^n)\\
&&\ \ \leq I(X_1^n;Y_1^n|X_2^n,X_3^n)+I(X_2^n;Y_2^n)\\
&&\ \ = h(Y_1^n|X_2^n,X_3^n)-h(Y_1^n|X_1^n,X_2^n,X_3^n)\\
&&\ \ \ \ \ \ \ +h(Y_2^n)-h(Y_2^n|X_2^n)\\
&&\ \ = h(X_1^n+Z_1^n)-h(Z_1^n)\\
&&\ \ \ \ \ \ \ +h(X_1^n+X_2^n+Z_2^n)-h(X_1^n+Z_2^n)\\
&&\ \ \leq \frac{n}{2}\log\left(\frac{P+N_1}{N_1}\right)+\frac{n}{2}\log\left(\frac{2P+N_2}{P+N_2}\right).
\end{eqnarray*}
\begin{eqnarray*}
&&n(R_1+R_3-\epsilon)\\
&&\ \ \leq I(X_1^n;Y_1^n)+I(X_3^n;Y_3^n)\\
&&\ \ \leq I(X_1^n;Y_1^n|X_2^n,X_3^n)+I(X_3^n;Y_3^n)\\
&&\ \ = h(Y_1^n|X_2^n,X_3^n)-h(Y_1^n|X_1^n,X_2^n,X_3^n)\\
&&\ \ \ \ \ \ \ +h(Y_3^n)-h(Y_3^n|X_3^n)\\
&&\ \ = h(X_1^n+Z_1^n)-h(Z_1^n)\\
&&\ \ \ \ \ \ \ +h(X_1^n+X_3^n+Z_3^n)-h(X_1^n+Z_3^n)\\
&&\ \ \leq \frac{n}{2}\log\left(\frac{P+N_1}{N_1}\right)+\frac{n}{2}\log\left(\frac{2P+N_3}{P+N_3}\right).
\end{eqnarray*}
\end{IEEEproof}
\end{theorem}

\subsection{Channel Type 3}
In this section, we present an outer bound on the capacity region of Type 3 channel defined by
\[
\left[ {\begin{array}{*{20}c}
   Y_1\\
   Y_2\\
   Y_3\\
\end{array} } \right]
=\left[ {\begin{array}{*{20}c}
   1 & 0 & 1\\
   0 & 1 & 1\\
   1 & 1 & 1\\
\end{array} } \right]
\left[ {\begin{array}{*{20}c}
   X_1\\
   X_2\\
   X_3\\
\end{array} } \right]
+\left[ {\begin{array}{*{20}c}
   Z_1\\
   Z_2\\
   Z_3\\
\end{array} } \right].
\]
We state the outer bound in the following theorem.
\begin{theorem}
The capacity region of Type 3 channel is contained in the following outer bound region:
\begin{eqnarray*}
&&\ \ \ \ \ \ \ R_k\leq C_k,\ k=1,2,3\\
&&R_1+R_3\leq \halflog\left(1+\frac{P}{N_1}\right)+\halflog\left(\frac{2P+N_3}{P+N_3}\right)\\
&&R_2+R_3\leq \halflog\left(1+\frac{P}{N_2}\right)+\halflog\left(\frac{2P+N_3}{P+N_3}\right).
\end{eqnarray*}
\begin{IEEEproof}
\begin{eqnarray*}
&&n(R_1+R_3-\epsilon)\\
&&\ \ \leq I(X_1^n;Y_1^n)+I(X_3^n;Y_3^n)\\
&&\ \ \leq I(X_1^n;Y_1^n|X_3^n)+I(X_3^n;Y_3^n|X_2^n)\\
&&\ \ = h(Y_1^n|X_3^n)-h(Y_1^n|X_1^n,X_3^n)\\
&&\ \ \ \ \ \ \ +h(Y_3^n|X_2^n)-h(Y_3^n|X_2^n,X_3^n)\\
&&\ \ = h(X_1^n+Z_1^n)-h(Z_1^n)\\
&&\ \ \ \ \ \ \ +h(X_1^n+X_3^n+Z_3^n)-h(X_1^n+Z_3^n)\\
&&\ \ \leq \frac{n}{2}\log\left(\frac{P+N_1}{N_1}\right)+\frac{n}{2}\log\left(\frac{2P+N_3}{P+N_3}\right).
\end{eqnarray*}
\begin{eqnarray*}
&&n(R_2+R_3-\epsilon)\\
&&\ \ \leq I(X_2^n;Y_2^n)+I(X_3^n;Y_3^n)\\
&&\ \ \leq I(X_2^n;Y_2^n|X_3^n)+I(X_3^n;Y_3^n|X_1^n)\\
&&\ \ = h(Y_2^n|X_3^n)-h(Y_2^n|X_2^n,X_3^n)\\
&&\ \ \ \ \ \ \ +h(Y_3^n|X_1^n)-h(Y_3^n|X_1^n,X_3^n)\\
&&\ \ = h(X_2^n+Z_2^n)-h(Z_2^n)\\
&&\ \ \ \ \ \ \ +h(X_2^n+X_3^n+Z_3^n)-h(X_2^n+Z_3^n)\\
&&\ \ \leq \frac{n}{2}\log\left(\frac{P+N_2}{N_2}\right)+\frac{n}{2}\log\left(\frac{2P+N_3}{P+N_3}\right).
\end{eqnarray*}
\end{IEEEproof}
\end{theorem}

\subsection{Channel Type 4}
In this section, we present an outer bound on the capacity region of Type 4 channel defined by
\[
\left[ {\begin{array}{*{20}c}
   Y_1\\
   Y_2\\
   Y_3\\
\end{array} } \right]
=\left[ {\begin{array}{*{20}c}
   1 & 0 & 1\\
   1 & 1 & 0\\
   0 & 1 & 1\\
\end{array} } \right]
\left[ {\begin{array}{*{20}c}
   X_1\\
   X_2\\
   X_3\\
\end{array} } \right]
+\left[ {\begin{array}{*{20}c}
   Z_1\\
   Z_2\\
   Z_3\\
\end{array} } \right].
\]
This is a cyclic Gaussian interference channel \cite{ZhouYu13}. We first show that channel type 4 is in the mixed interference regime. By normalizing the noise variances, we get the equivalent channel given by
\[
\left[ {\begin{array}{*{20}c}
   Y_1'\\
   Y_2'\\
   Y_3'\\
\end{array} } \right]
=\left[ {\begin{array}{*{20}c}
   h_{11} & h_{12} & h_{13}\\
   h_{21} & h_{22} & h_{23}\\
   h_{31} & h_{32} & h_{33}\\
\end{array} } \right]
\left[ {\begin{array}{*{20}c}
   X_1\\
   X_2\\
   X_3\\
\end{array} } \right]
+\left[ {\begin{array}{*{20}c}
   Z_1'\\
   Z_2'\\
   Z_3'\\
\end{array} } \right]
\]
where $Y_k'=\frac{1}{\sqrt{N_k}} Y_k$, $Z_k'=\frac{1}{\sqrt{N_k}}Z_k$, $N_0=\mathbb{E}[Z_k'^2]=1$, $\mathbb{E}[X_k^2]\leq P_k=P$ and
\[
\left[ {\begin{array}{*{20}c}
   h_{11} & h_{12} & h_{13}\\
   h_{21} & h_{22} & h_{23}\\
   h_{31} & h_{32} & h_{33}\\
\end{array} } \right]
=\left[ {\begin{array}{*{20}c}
   \frac{1}{\sqrt{N_1}} & 0 & \frac{1}{\sqrt{N_1}}\\
   \frac{1}{\sqrt{N_2}} & \frac{1}{\sqrt{N_2}} & 0\\
   0 & \frac{1}{\sqrt{N_3}} & \frac{1}{\sqrt{N_3}}\\
\end{array} } \right].
\]
With the usual definitions of $\snr_k=\frac{h_{kk}^2 P_k}{N_0}$ and\\ $\inr_k=\frac{h_{jk}^2 P_k}{N_0}$ for $j\neq k$ as in \cite{EtkinTseWang08,ZhouYu13},
\begin{eqnarray}
&&\snr_1=\frac{P}{N_1} \geq \inr_1=\frac{P}{N_2}\\
&&\snr_2=\frac{P}{N_2} \geq \inr_2=\frac{P}{N_3}\\
&&\snr_3=\frac{P}{N_3} \leq \inr_3=\frac{P}{N_1}.
\end{eqnarray}
We state the outer bound in the following theorem.
\begin{theorem}
The capacity region of Type 4 channel is contained in the following outer bound region:
\begin{eqnarray*}
&&\ \ \ \ \ \ \ R_k\leq C_k,\ k=1,2,3\\
&&R_1+R_2\leq \halflog\left(1+\frac{P}{N_1}\right)+\halflog\left(\frac{2P+N_2}{P+N_2}\right)\\
&&R_1+R_3\leq \halflog\left(1+\frac{2P}{N_1}\right)\\
&&R_2+R_3\leq \halflog\left(1+\frac{P}{N_2}\right)+\halflog\left(\frac{2P+N_3}{P+N_3}\right).
\end{eqnarray*}
\begin{IEEEproof}
\begin{eqnarray*}
&&n(R_1+R_2-\epsilon)\\
&&\ \ \leq I(X_1^n;Y_1^n)+I(X_2^n;Y_2^n)\\
&&\ \ \leq I(X_1^n;Y_1^n|X_3^n)+I(X_2^n;Y_2^n)\\
&&\ \ = h(Y_1^n|X_3^n)-h(Y_1^n|X_1^n,X_3^n)\\
&&\ \ \ \ \ \ \ +h(Y_2^n)-h(Y_2^n|X_2^n)\\
&&\ \ = h(X_1^n+Z_1^n)-h(Z_1^n)\\
&&\ \ \ \ \ \ \ +h(X_1^n+X_2^n+Z_2^n)-h(X_1^n+Z_2^n)\\
&&\ \ \leq \frac{n}{2}\log\left(\frac{P+N_1}{N_1}\right)+\frac{n}{2}\log\left(\frac{2P+N_2}{P+N_2}\right).
\end{eqnarray*}
\begin{eqnarray*}
&&n(R_2+R_3-\epsilon)\\
&&\ \ \leq I(X_2^n;Y_2^n)+I(X_3^n;Y_3^n)\\
&&\ \ \leq I(X_2^n;Y_2^n|X_1^n)+I(X_3^n;Y_3^n)\\
&&\ \ = h(Y_2^n|X_1^n)-h(Y_2^n|X_1^n,X_2^n)\\
&&\ \ \ \ \ \ \ +h(Y_3^n)-h(Y_3^n|X_3^n)\\
&&\ \ = h(X_2^n+Z_2^n)-h(Z_2^n)\\
&&\ \ \ \ \ \ \ +h(X_2^n+X_3^n+Z_3^n)-h(X_2^n+Z_3^n)\\
&&\ \ \leq \frac{n}{2}\log\left(\frac{P+N_2}{N_2}\right)+\frac{n}{2}\log\left(\frac{2P+N_3}{P+N_3}\right).
\end{eqnarray*}
\begin{eqnarray*}
&&n(R_1+R_3-\epsilon)\\
&&\ \ \leq I(X_1^n;Y_1^n)+I(X_3^n;Y_3^n)\\
&&\ \ \leq I(X_1^n;Y_1^n)+I(X_3^n;Y_3^n|X_2^n)\\
&&\ \ \leq I(X_1^n;Y_1^n)+I(X_3^n;Y_1^n|X_1^n)\\
&&\ \ \leq I(X_1^n,X_3^n;Y_1^n)\\
&&\ \ = h(Y_1^n)-h(Y_1^n|X_1^n,X_3^n)\\
&&\ \ = h(X_1^n+X_3^n+Z_1^n)-h(Z_1^n)\\
&&\ \ \leq \frac{n}{2}\log\left(\frac{2P+N_1}{N_1}\right)
\end{eqnarray*}
where we used the fact that $I(X_3^n;Y_3^n|X_2^n)=I(X_3^n;X_3^n+Z_3^n)\leq I(X_3^n;X_3^n+Z_1^n)=I(X_3^n;Y_1^n|X_1^n)$.
%\begin{eqnarray}
%&&n(R_1+R_3-\epsilon)\\
%&&\ \ \leq I(X_1^n;Y_1^n)+I(X_3^n;Y_3^n)\\
%&&\ \ \leq I(X_1^n;Y_1^n)+I(X_3^n;Y_3^n|X_2^n)\\
%&&\ \ = h(Y_1^n)-h(Y_1^n|X_1^n)\\
%&&\ \ \ \ \ \ \ +h(Y_3^n|X_2^n)-h(Y_3^n|X_2^n,X_3^n)\\
%&&\ \ = h(X_1^n+X_3^n+Z_1^n)-h(X_3^n+Z_1^n)\\
%&&\ \ \ \ \ \ \ +h(X_3^n+Z_3^n)-h(Z_3^n)\\
%&&\mbox{EPI fail}
%\end{eqnarray}
\end{IEEEproof}
\end{theorem}

\begin{table*}
\begin{center}
\begin{tabular}{|c|c|c|c|}
\hline
Type & Outer bound region $\mathcal{R}_o$ &  Relaxed outer bound region $\mathcal{R}_o'$ &  Two-dimensional cross-section of $\mathcal{R}_o'$ \\
\hline
$1$ 
& $\left. {\begin{array}{*{20}l}
\ \ \ \ \ \ \ R_k\leq C_k,\ k=1,2,3\\
 R_1+R_2\leq \halflog\left(\frac{P+N_1}{N_1}\cdot\frac{2P+N_2}{P+N_2}\right)\\
 R_2+R_3\leq \halflog\left(\frac{P+N_2}{N_2}\cdot\frac{2P+N_3}{P+N_3}\right)\\
% R_1+R_2\leq \halflog\left(\frac{P+N_1}{N_1}\right)+\halflog\left(\frac{2P+N_2}{P+N_2}\right)\\
% R_2+R_3\leq \halflog\left(\frac{P+N_2}{N_2}\right)+\halflog\left(\frac{2P+N_3}{P+N_3}\right)
\end{array} } \right.$
& $\left. {\begin{array}{*{20}l}
\ \ \ \ \ \ \ R_k\leq \halflog\left(\frac{P}{N_k}\cdot\frac{4}{3}\right)\\
 R_1+R_2\leq \halflog\left(\frac{P}{N_1}\cdot\frac{7}{3}\right)\\
 R_2+R_3\leq \halflog\left(\frac{P}{N_2}\cdot\frac{7}{3}\right)\\
\end{array} } \right.$
& $\left. {\begin{array}{*{20}l}
 \textrm{At some } R_2 \in [0,C_2],\\
 R_1\leq \min\left\{\halflog\left(\frac{P}{N_1}\cdot\frac{7}{3}\right)-R_2, \halflog\left(\frac{P}{N_1}\cdot\frac{4}{3}\right)\right\}\\
 R_3\leq \min\left\{\halflog\left(\frac{P}{N_2}\cdot\frac{7}{3}\right)-R_2, \halflog\left(\frac{P}{N_3}\cdot\frac{4}{3}\right)\right\}\\
\end{array} } \right.$
\\
\hline
$2$ 
& $\left. {\begin{array}{*{20}l}
\ \ \ \ \ \ \ R_k\leq C_k,\ k=1,2,3\\
 R_1+R_2\leq \halflog\left(\frac{P+N_1}{N_1}\cdot\frac{2P+N_2}{P+N_2}\right)\\
 R_1+R_3\leq \halflog\left(\frac{P+N_1}{N_1}\cdot\frac{2P+N_3}{P+N_3}\right)\\
% R_1+R_2\leq \halflog\left(\frac{P+N_1}{N_1}\right)+\halflog\left(\frac{2P+N_2}{P+N_2}\right)\\
% R_1+R_3\leq \halflog\left(\frac{P+N_1}{N_1}\right)+\halflog\left(\frac{2P+N_3}{P+N_3}\right)
\end{array} } \right.$
& $\left. {\begin{array}{*{20}l}
\ \ \ \ \ \ \ R_k\leq \halflog\left(\frac{P}{N_k}\cdot\frac{4}{3}\right)\\
 R_1+R_2\leq \halflog\left(\frac{P}{N_1}\cdot\frac{7}{3}\right)\\
 R_1+R_3\leq \halflog\left(\frac{P}{N_1}\cdot\frac{7}{3}\right)\\
\end{array} } \right.$
& $\left. {\begin{array}{*{20}l}
 \textrm{At some } R_1 \in [0,C_1],\\
 R_2\leq \min\left\{\halflog\left(\frac{P}{N_1}\cdot\frac{7}{3}\right)-R_1, \halflog\left(\frac{P}{N_2}\cdot\frac{4}{3}\right)\right\}\\
 R_3\leq \min\left\{\halflog\left(\frac{P}{N_1}\cdot\frac{7}{3}\right)-R_1, \halflog\left(\frac{P}{N_3}\cdot\frac{4}{3}\right)\right\}\\
\end{array} } \right.$
\\
\hline
$3$ 
& $\left. {\begin{array}{*{20}l}
\ \ \ \ \ \ \ R_k\leq C_k,\ k=1,2,3\\
 R_1+R_3\leq \halflog\left(\frac{P+N_1}{N_1}\cdot\frac{2P+N_3}{P+N_3}\right)\\
 R_2+R_3\leq \halflog\left(\frac{P+N_2}{N_2}\cdot\frac{2P+N_3}{P+N_3}\right)\\
% R_1+R_3\leq \halflog\left(\frac{P+N_1}{N_1}\right)+\halflog\left(\frac{2P+N_3}{P+N_3}\right)\\
% R_2+R_3\leq \halflog\left(\frac{P+N_2}{N_2}\right)+\halflog\left(\frac{2P+N_3}{P+N_3}\right)
\end{array} } \right.$
& $\left. {\begin{array}{*{20}l}
\ \ \ \ \ \ \ R_k\leq \halflog\left(\frac{P}{N_k}\cdot\frac{4}{3}\right)\\
 R_1+R_3\leq \halflog\left(\frac{P}{N_1}\cdot\frac{7}{3}\right)\\
 R_2+R_3\leq \halflog\left(\frac{P}{N_2}\cdot\frac{7}{3}\right)\\
\end{array} } \right.$
& $\left. {\begin{array}{*{20}l}
 \textrm{At some } R_3 \in [0,C_3],\\
 R_1\leq \min\left\{\halflog\left(\frac{P}{N_1}\cdot\frac{7}{3}\right)-R_3, \halflog\left(\frac{P}{N_1}\cdot\frac{4}{3}\right)\right\}\\
 R_2\leq \min\left\{\halflog\left(\frac{P}{N_2}\cdot\frac{7}{3}\right)-R_3, \halflog\left(\frac{P}{N_2}\cdot\frac{4}{3}\right)\right\}\\
\end{array} } \right.$
\\
\hline
$4$ 
& $\left. {\begin{array}{*{20}l}
\ \ \ \ \ \ \ R_k\leq C_k,\ k=1,2,3\\
 R_1+R_2\leq \halflog\left(\frac{P+N_1}{N_1}\cdot\frac{2P+N_2}{P+N_2}\right)\\
 R_1+R_3\leq \halflog\left(\frac{2P+N_1}{N_1}\right)\\
 R_2+R_3\leq \halflog\left(\frac{P+N_2}{N_2}\cdot\frac{2P+N_3}{P+N_3}\right)\\
% R_1+R_2\leq \halflog\left(\frac{P+N_1}{N_1}\right)+\halflog\left(\frac{2P+N_2}{P+N_2}\right)\\
% R_1+R_3\leq \halflog\left(\frac{2P+N_1}{N_1}\right)\\
% R_2+R_3\leq \halflog\left(\frac{P+N_2}{N_2}\right)+\halflog\left(\frac{2P+N_3}{P+N_3}\right)\\
\end{array} } \right.$
& $\left. {\begin{array}{*{20}l}
\ \ \ \ \ \ \ R_k\leq \halflog\left(\frac{P}{N_k}\cdot\frac{4}{3}\right)\\
 R_1+R_2\leq \halflog\left(\frac{P}{N_1}\cdot\frac{7}{3}\right)\\
 R_1+R_3\leq \halflog\left(\frac{P}{N_1}\cdot\frac{7}{3}\right)\\
 R_2+R_3\leq \halflog\left(\frac{P}{N_2}\cdot\frac{7}{3}\right)\\
\end{array} } \right.$
& $\left. {\begin{array}{*{20}l}
 \textrm{At some } R_1 \in [0,C_1],\\
 R_2\leq \min\left\{\halflog\left(\frac{P}{N_1}\cdot\frac{7}{3}\right)-R_1, \halflog\left(\frac{P}{N_2}\cdot\frac{4}{3}\right)\right\}\\
 R_3\leq \min\left\{\halflog\left(\frac{P}{N_1}\cdot\frac{7}{3}\right)-R_1, \halflog\left(\frac{P}{N_3}\cdot\frac{4}{3}\right)\right\}\\
 R_2+R_3\leq \halflog\left(\frac{P}{N_2}\cdot\frac{7}{3}\right)\\
\end{array} } \right.$
\\
\hline
$5$ 
& $\left. {\begin{array}{*{20}l}
\ \ \ \ \ \ \ R_k\leq C_k,\ k=1,2,3\\
 R_1+R_2\leq \halflog\left(\frac{2P+N_1}{N_1}\right)\\
 R_2+R_3\leq \halflog\left(\frac{2P+N_2}{N_2}\right)\\
 R_1+R_3\leq \halflog\left(\frac{P+N_1}{N_1}\cdot\frac{2P+N_3}{P+N_3}\right)\\
% R_1+R_2\leq \halflog\left(\frac{2P+N_1}{N_1}\right)\\
% R_1+R_3\leq \halflog\left(\frac{P+N_1}{N_1}\right)+\halflog\left(\frac{2P+N_3}{P+N_3}\right)\\
% R_2+R_3\leq \halflog\left(\frac{2P+N_2}{N_2}\right)\\
\end{array} } \right.$
& $\left. {\begin{array}{*{20}l}
\ \ \ \ \ \ \ R_k\leq \halflog\left(\frac{P}{N_k}\cdot\frac{4}{3}\right)\\
 R_1+R_2\leq \halflog\left(\frac{P}{N_1}\cdot\frac{7}{3}\right)\\
 R_2+R_3\leq \halflog\left(\frac{P}{N_2}\cdot\frac{7}{3}\right)\\
 R_1+R_3\leq \halflog\left(\frac{P}{N_1}\cdot\frac{7}{3}\right)\\
\end{array} } \right.$
& $\left. {\begin{array}{*{20}l}
\textrm{At some } R_2 \in [0,C_2],\\
 R_1\leq \min\left\{\halflog\left(\frac{P}{N_1}\cdot\frac{7}{3}\right)-R_2, \halflog\left(\frac{P}{N_1}\cdot\frac{4}{3}\right)\right\}\\
 R_3\leq \min\left\{\halflog\left(\frac{P}{N_2}\cdot\frac{7}{3}\right)-R_2, \halflog\left(\frac{P}{N_3}\cdot\frac{4}{3}\right)\right\}\\
 R_1+R_3\leq \halflog\left(\frac{P}{N_1}\cdot\frac{7}{3}\right)\\
\end{array} } \right.$
\\
\hline
\end{tabular}
\caption{Capacity outer bounds}
  \label{tab:outerBounds}
\end{center}
\end{table*}

\subsection{Channel Type 5}
In this section, we present an outer bound on the capacity region of Type 5 channel defined by
\[
\left[ {\begin{array}{*{20}c}
   Y_1\\
   Y_2\\
   Y_3\\
\end{array} } \right]
=\left[ {\begin{array}{*{20}c}
   1 & 1 & 0\\
   0 & 1 & 1\\
   1 & 0 & 1\\
\end{array} } \right]
\left[ {\begin{array}{*{20}c}
   X_1\\
   X_2\\
   X_3\\
\end{array} } \right]
+\left[ {\begin{array}{*{20}c}
   Z_1\\
   Z_2\\
   Z_3\\
\end{array} } \right].
\]
This is a cyclic Gaussian interference channel \cite{ZhouYu13}. We first show that channel type 5 is in the mixed interference regime. By normalizing the noise variances, we get the equivalent channel given by
\[
\left[ {\begin{array}{*{20}c}
   Y_1'\\
   Y_2'\\
   Y_3'\\
\end{array} } \right]
=\left[ {\begin{array}{*{20}c}
   \frac{1}{\sqrt{N_1}} & \frac{1}{\sqrt{N_1}} & 0\\
   0 & \frac{1}{\sqrt{N_2}} & \frac{1}{\sqrt{N_2}}\\
   \frac{1}{\sqrt{N_3}} & 0 & \frac{1}{\sqrt{N_3}}\\
\end{array} } \right]
\left[ {\begin{array}{*{20}c}
   X_1\\
   X_2\\
   X_3\\
\end{array} } \right]
+\left[ {\begin{array}{*{20}c}
   Z_1'\\
   Z_2'\\
   Z_3'\\
\end{array} } \right].
\]
We can see that
\begin{eqnarray}
&&\snr_1=\frac{P}{N_1} \geq \inr_1=\frac{P}{N_3}\\
&&\snr_2=\frac{P}{N_2} \leq \inr_2=\frac{P}{N_1}\\
&&\snr_3=\frac{P}{N_3} \leq \inr_3=\frac{P}{N_2}.
\end{eqnarray}
We state the outer bound in the following theorem.
\begin{theorem}
The capacity region of Type 5 channel is contained in the following outer bound region:
\begin{eqnarray*}
&&\ \ \ \ \ \ \ R_k\leq C_k,\ k=1,2,3\\
&&R_1+R_2\leq \halflog\left(1+\frac{2P}{N_1}\right)\\
&&R_2+R_3\leq \halflog\left(1+\frac{2P}{N_2}\right)\\
&&R_1+R_3\leq \halflog\left(1+\frac{P}{N_1}\right)+\halflog\left(\frac{2P+N_3}{P+N_3}\right).
\end{eqnarray*}
\begin{IEEEproof}
\begin{eqnarray*}
&&n(R_1+R_2-\epsilon)\\
&&\ \ \leq I(X_1^n;Y_1^n)+I(X_2^n;Y_2^n)\\
&&\ \ \leq I(X_1^n;Y_1^n)+I(X_2^n;Y_2^n|X_3^n)\\
&&\ \ \leq I(X_1^n;Y_1^n)+I(X_2^n;Y_1^n|X_1^n)\\
&&\ \ \leq I(X_1^n,X_2^n;Y_1^n)\\
&&\ \ = h(Y_1^n)-h(Y_1^n|X_1^n,X_2^n)\\
&&\ \ = h(X_1^n+X_2^n+Z_1^n)-h(Z_1^n)\\
&&\ \ \leq \frac{n}{2}\log\left(\frac{2P+N_1}{N_1}\right)
\end{eqnarray*}
where we used the fact that $I(X_2^n;Y_2^n|X_3^n)=I(X_2^n;X_2^n+Z_2^n)\leq I(X_2^n;X_2^n+Z_1^n)=I(X_2^n;Y_1^n|X_1^n)$.
\begin{eqnarray*}
&&n(R_2+R_3-\epsilon)\\
&&\ \ \leq I(X_2^n;Y_2^n)+I(X_3^n;Y_3^n)\\
&&\ \ \leq I(X_2^n;Y_2^n)+I(X_3^n;Y_3^n|X_1^n)\\
&&\ \ \leq I(X_2^n;Y_2^n)+I(X_3^n;Y_2^n|X_2^n)\\
&&\ \ \leq I(X_2^n,X_3^n;Y_2^n)\\
&&\ \ = h(Y_2^n)-h(Y_2^n|X_2^n,X_3^n)\\
&&\ \ = h(X_2^n+X_3^n+Z_2^n)-h(Z_2^n)\\
&&\ \ \leq \frac{n}{2}\log\left(\frac{2P+N_2}{N_2}\right)
\end{eqnarray*}
where we used the fact that $I(X_3^n;Y_3^n|X_1^n)=I(X_3^n;X_3^n+Z_3^n)\leq I(X_3^n;X_3^n+Z_2^n)=I(X_3^n;Y_2^n|X_2^n)$.
\begin{eqnarray*}
&&n(R_1+R_3-\epsilon)\\
&&\ \ \leq I(X_1^n;Y_1^n)+I(X_3^n;Y_3^n)\\
&&\ \ \leq I(X_1^n;Y_1^n|X_2^n)+I(X_3^n;Y_3^n)\\
&&\ \ = h(Y_1^n|X_2^n)-h(Y_1^n|X_1^n,X_2^n)\\
&&\ \ \ \ \ \ \ +h(Y_3^n)-h(Y_3^n|X_3^n)\\
&&\ \ = h(X_1^n+Z_1^n)-h(Z_1^n)\\
&&\ \ \ \ \ \ \ +h(X_1^n+X_3^n+Z_3^n)-h(X_1^n+Z_3^n)\\
&&\ \ \leq \frac{n}{2}\log\left(\frac{P+N_1}{N_1}\right)+\frac{n}{2}\log\left(\frac{2P+N_3}{P+N_3}\right)
\end{eqnarray*}
%\begin{eqnarray}
%&&n(R_2+R_3-\epsilon)\\
%&&\ \ \leq I(X_2^n;Y_2^n)+I(X_3^n;Y_3^n)\\
%&&\ \ \leq I(X_2^n;Y_2^n|X_3^n)+I(X_3^n;Y_3^n)\\
%&&\ \ \leq I(X_2^n;Y_1^n|X_1^n)+I(X_3^n;Y_2^n)\mbox{ check?}\\
%&&\ \ = h(Y_1^n|X_1^n)-h(Y_1^n|X_1^n,X_2^n)\\
%&&\ \ \ \ \ \ \ +h(Y_2^n)-h(Y_2^n|X_3^n)\\
%&&\ \ = h(X_2^n+Z_1^n)-h(Z_1^n)\\
%&&\ \ \ \ \ \ \ +h(X_2^n+X_3^n+Z_2^n)-h(X_2^n+Z_2^n)\\
%&&\ \ \leq \frac{n}{2}\log\left(\frac{P+N_1}{N_1}\right)+\frac{n}{2}\log\left(\frac{2P+N_2}{P+N_2}\right)
%\end{eqnarray}
%\begin{eqnarray}
%&&n(R_1+R_2-\epsilon)\\
%&&\ \ \leq I(X_1^n;Y_1^n)+I(X_2^n;Y_2^n)\\
%&&\ \ \leq I(X_1^n;Y_1^n)+I(X_2^n;Y_2^n|X_3^n)\\
%&&\ \ = h(Y_1^n)-h(Y_1^n|X_1^n)\\
%&&\ \ \ \ \ \ \ +h(Y_2^n|X_3^n)-h(Y_2^n|X_2^n,X_3^n)\\
%&&\ \ = h(X_1^n+X_2^n+Z_1^n)-h(X_2^n+Z_1^n)\\
%&&\ \ \ \ \ \ \ +h(X_2^n+Z_2^n)-h(Z_2^n)\\
%&&\mbox{EPI fail}
%\end{eqnarray}
%\begin{eqnarray}
%&&n(R_2+R_3-\epsilon)\\
%&&\ \ \leq I(X_2^n;Y_2^n)+I(X_3^n;Y_3^n)\\
%&&\ \ \leq I(X_2^n;Y_2^n)+I(X_3^n;Y_3^n|X_1^n)\\
%&&\ \ = h(Y_2^n)-h(Y_2^n|X_2^n)\\
%&&\ \ \ \ \ \ \ +h(Y_3^n|X_1^n)-h(Y_3^n|X_1^n,X_3^n)\\
%&&\ \ = h(X_2^n+X_3^n+Z_2^n)-h(X_3^n+Z_2^n)\\
%&&\ \ \ \ \ \ \ +h(X_3^n+Z_3^n)-h(Z_3^n)\\
%&&\mbox{EPI fail}
%\end{eqnarray}
\end{IEEEproof}
\end{theorem}

\subsection{Relaxed Outer Bounds}
For ease of gap calculation, we also derive relaxed outer bounds. First, we can see that for $N_j\leq N_k$,
\begin{eqnarray*}
\halflog\left(1+\frac{P}{N_j}\right)+\halflog\left(\frac{2P+N_k}{P+N_k}\right)\leq \halflog\left(1+\frac{2P}{N_j}\right).
\end{eqnarray*}
Five outer bound theorems in this section, together with this inequality, give the sum-rate bound expression in Theorem 1.

Next, we can assume that $P\geq 3N_j$ for $j=1,2,3$. Otherwise, showing one-bit gap capacity is trivial as the capacity region is included in the unit hypercube, i.e., $R_j\leq \halflog\left(1+\frac{P}{N_j}\right)< 1$. For $P\geq 3N_j$,
\begin{eqnarray*}
&&\halflog\left(1+\frac{2P}{N_j}\right)=\halflog\left(\frac{P}{N_j}\right)+\halflog\left(\frac{N_j}{P}+2\right)\\
&&\ \ \ \ \ \ \ \ \ \ \ \ \ \ \ \ \ \ \ \ \ \leq \halflog\left(\frac{P}{N_j}\right)+\halflog\left(\frac{7}{3}\right)\\
&&\halflog\left(1+\frac{P}{N_j}\right)\leq \halflog\left(\frac{P}{N_j}\right)+\halflog\left(\frac{4}{3}\right).
\end{eqnarray*}
The resulting relaxed outer bounds $\mathcal{R}_o'$ are summarized in Table \ref{tab:outerBounds}.

\section{Inner Bound: Channel Type 1}
\begin{theorem}
Given $\alpha=(\alpha_0,\alpha_2)\in [0,1]^2$, the rate region $\mathcal{R}_\alpha$ is defined by
\begin{eqnarray*}
&&R_1 \leq \halflog^+\left(\frac{1-\alpha_0}{2-\alpha_0}+\frac{(1-\alpha_0) P}{(\alpha_0+\alpha_2) P+N_2}\right)\\
&&\ \ \ \ \ \ \ \ \ \ \ \ \ \ \ \ \ \ \ \ \ \ \ \ \ \ \ \ \ \ +\halflog\left(1+\frac{{\alpha}_0 P}{N_1}\right)\\
&&R_2 \leq \halflog\left(1+\frac{\alpha_2 P}{\alpha_0 P+N_2}\right)\\
&&R_3 \leq \halflog^+\left(\frac{1}{2-\alpha_0}+\frac{P}{(\alpha_0+\alpha_2) P+N_3}\right)
\end{eqnarray*}
where $\log^+(\cdot)=\max\{0,\log(\cdot)\}$. And, \[\mathcal{R}=\textsc{conv}\left(\bigcup_{\alpha}\mathcal{R}_\alpha\right)\] is achievable where $\textsc{conv}(\cdot)$ is convex hull operator.
\end{theorem}

\subsection{Preliminaries: Lattice Coding}
Lattice $\Lambda$ is a discrete subgroup of $\mathbb{R}^n$, $\Lambda =\{\mathbf{t}=\mathbf{G}\mathbf{u}: \mathbf{u}\in\mathbb{Z}^n\}$ where $\mathbf{G}\in\mathbb{R}^{n\times n}$ is a real generator matrix. Quantization with respect to $\Lambda$ is $Q_{\Lambda}(\mathbf{x})=\arg \min_{\lambda \in \Lambda} \|\mathbf{x}-\lambda\|$. Modulo operation with respect to $\Lambda$ is $M_{\Lambda}(\mathbf{x})=[\mathbf{x}]\modLambda=\mathbf{x}-Q_{\Lambda}(\mathbf{x})$. For convenience, we use both notations $M_{\Lambda}(\cdot)$ and $[\cdot]\modLambda$ interchangeably. Fundamental Voronoi region of $\Lambda$ is $\mathcal{V}(\Lambda)=\{\mathbf{x}:Q_{\Lambda}(\mathbf{x})=\mathbf{0}\}$. Volume of the Voronoi region of $\Lambda$ is $V(\Lambda)=\int_{\mathcal{V}(\Lambda)} d\mathbf{x}$. Normalized second moment of $\Lambda$ is $G(\Lambda)=\frac{\sigma^2(\Lambda)}{V(\Lambda)^{2/n}}$ where $\sigma^2(\Lambda)=\frac{1}{nV(\Lambda)}\int_{\mathcal{V}(\Lambda)} \|\mathbf{x}\|^2 d\mathbf{x}$. Lattices $\Lambda_1$, $\Lambda_2$ and $\Lambda$ are said to be nested if $\Lambda\subseteq\Lambda_2\subseteq\Lambda_1$. For nested lattices $\Lambda_2\subset\Lambda_1$,
$\Lambda_1/\Lambda_2=\Lambda_1\cap\mathcal{V}(\Lambda_2)$.

We briefly review the lattice decoding procedure in \cite{ErezZamir04}. We use nested lattices $\Lambda\subseteq \Lambda_t$ with $\sigma^2(\Lambda)=S$, $G(\Lambda)=\frac{1}{2\pi e}$, and $V(\Lambda)=(2\pi e S)^{\frac{n}{2}}$. The transmitter sends $\mathbf{x}=[\mathbf{t}+\mathbf{d}]\modLambda$ over the point-to-point Gaussian channel $\mathbf{y}=\mathbf{x}+\mathbf{z}$ where the codeword $\mathbf{t}\in \Lambda_t\cap \mathcal{V}(\Lambda)$, the dither signal $\mathbf{d}\sim\textrm{Unif}(\mathcal{V}(\Lambda))$, the transmit power $\frac{1}{n}\|\mathbf{x}\|^2=S$ and the noise $\mathbf{z}\sim\mathcal{N}(0,N\mathbf{I})$. The code rate is given by $R=\frac{1}{n}\log\left(\frac{V(\Lambda)}{V(\Lambda_t)}\right)$.

After linear scaling, dither removal, and mod-$\Lambda$ operation, we get
\begin{eqnarray}
\mathbf{y}'=[\beta\mathbf{y}-\mathbf{d}]\modLambda = \left[\mathbf{t}+\mathbf{z}_e\right]\modLambda
\end{eqnarray}
where the effective noise is $\mathbf{z}_e=(\beta-1)\mathbf{x}+\beta\mathbf{z}_1$
and its variance
$\sigma_e^2=\frac{1}{n}\mathbb{E}[\left\|\mathbf{z}_e\right\|^2]=(\beta-1)^2 S+\beta^2 N$.
With the MMSE scaling factor $\beta=\frac{S}{S+N}$ plugged in, we get $\sigma_e^2=\beta N=\frac{SN}{S+N}$. The capacity of the mod-$\Lambda$ channel \cite{ErezZamir04} between $\mathbf{t}$ and $\mathbf{y}$ is
\begin{eqnarray*}
\frac{1}{n} I\left(\mathbf{t};\mathbf{y}\right)
&=& \frac{1}{n} h\left(\mathbf{y}\right)-\frac{1}{n} h\left(\mathbf{y}|\mathbf{t}\right)\\
&=& \frac{1}{n} h\left(\mathbf{y}\right)-\frac{1}{n} h\left(\mathbf{z}\modLambda\right)\\
&\geq &\frac{1}{n} h\left(\mathbf{y}\right)-\frac{1}{n} h\left(\mathbf{z}\right)\\
&=& \frac{1}{n} \log V(\Lambda)-\frac{1}{n} h\left(\mathbf{z}\right)\\
&=& \frac{1}{2}\log \left(\frac{S}{\beta N}\right)\\
&=& \frac{1}{2}\log \left(1+\frac{S}{N}\right)\\
&=& C
\end{eqnarray*}
where $I(\cdot)$ and $h(\cdot)$ are mutual information and differential entropy, respectively. For reliable decoding of $\mathbf{t}$, we have the code rate constraint $R\leq C$.
With the choice of lattice parameters, $\sigma^2(\Lambda_t)\geq \beta N$, $G(\Lambda_t)=\frac{1}{2\pi e}$ and $V(\Lambda_t)^{\frac{n}{2}}=\frac{\sigma^2(\Lambda_t)}{G(\Lambda_t)}\geq 2\pi e \beta N$,
\begin{eqnarray*}
R &=& \frac{1}{n}\log\left(\frac{V(\Lambda)}{V(\Lambda_t)}\right)\\
&\leq & \frac{1}{n}\log\left(\frac{(2\pi e S)^{\frac{n}{2}}}{(2\pi e \beta N)^{\frac{n}{2}}}\right)\\
&=& \frac{1}{2}\log\left(\frac{S}{\beta N}\right).
\end{eqnarray*}
Thus, the constraint $R\leq C$ can be satisfied. By \emph{lattice decoding} \cite{ErezZamir04}, we can recover $\mathbf{t}$, i.e.,
\begin{equation}
Q_{\Lambda_t}(\mathbf{y}')=\mathbf{t},
\end{equation}
with probability $1-P_e$ where
\begin{equation}
P_e=\textrm{Pr}[Q_{\Lambda_t}\left(\mathbf{y}'\right)\neq \mathbf{t}]
\end{equation}
is the probability of decoding error.
If we choose $\Lambda$ to be Poltyrev-good \cite{Zamir14}, then $P_e\rightarrow 0$ as $n\rightarrow \infty$.

\subsection{Achievable Scheme}
We present an achievable scheme for the proof of \emph{Theorem 8}. The achievable scheme is based on rate-splitting, lattice coding, and interference alignment. Message $M_1\in\{1,2,\ldots,2^{nR_1}\}$ is split into two parts: $M_{11}\in\{1,2,\ldots,2^{nR_{11}}\}$ and $M_{10}\in\{1,2,\ldots,2^{nR_{10}}\}$, so $R_1=R_{11}+R_{10}$. Transmitter 1 sends $\mathbf{x}_1=\mathbf{x}_{11}+\mathbf{x}_{10}$ where $\mathbf{x}_{11}$ and $\mathbf{x}_{10}$ are coded signals of $M_{11}$ and $M_{10}$, respectively. Transmitters 2 and 3 send $\mathbf{x}_2$ and $\mathbf{x}_3$, coded signals of $M_2\in\{1,2,\ldots,2^{nR_2}\}$ and $M_3\in\{1,2,\ldots,2^{nR_3}\}$. In particular, $\mathbf{x}_{11}$ and $\mathbf{x}_3$ are lattice-coded signals.

We use the lattice construction of \cite{NamChungLee10,NamChungLee11} with the lattice partition chain $\Lambda_c/\Lambda_1/\Lambda_3$, so $\Lambda_3\subset \Lambda_1\subset \Lambda_c$ are nested lattices. $\Lambda_c$ is the coding lattice for both $\mathbf{x}_{11}$ and $\mathbf{x}_3$. $\Lambda_1$ and $\Lambda_3$ are shaping lattices for $\mathbf{x}_{11}$ and $\mathbf{x}_3$, respectively.  The lattice signals are formed by
\begin{eqnarray}
&&\mathbf{x}_{11}=[\mathbf{t}_{11}+\mathbf{d}_{11}]\modLambda_1\\
&&\ \mathbf{x}_3=[\mathbf{t}_3+\mathbf{d}_3]\modLambda_3
\end{eqnarray}
where $\mathbf{t}_{11}\in \Lambda_c\cap\mathcal{V}(\Lambda_1)$ and $\mathbf{t}_3\in \Lambda_c\cap\mathcal{V}(\Lambda_3)$ are lattice codewords. The dither signals $\mathbf{d}_{11}$ and $\mathbf{d}_3$ are uniformly distributed over $\mathcal{V}(\Lambda_1)$ and $\mathcal{V}(\Lambda_3)$, respectively.
To satisfy power constraints, we choose $\mathbb{E}[\|\mathbf{x}_{11}\|^2]=n\sigma^2(\Lambda_1)=(1-\alpha_1) nP$, $\mathbb{E}[\|\mathbf{x}_{10}\|^2]={\alpha}_1 nP$,
$\mathbb{E}[\|\mathbf{x}_2\|^2]={\alpha}_2 nP$, $\mathbb{E}[\|\mathbf{x}_3\|^2]=n\sigma^2(\Lambda_3)=nP$.

\begin{figure}[tp]
  \begin{center}
    \mbox{
      \subfigure[Channel type 1]{\includegraphics[width=0.45\textwidth]{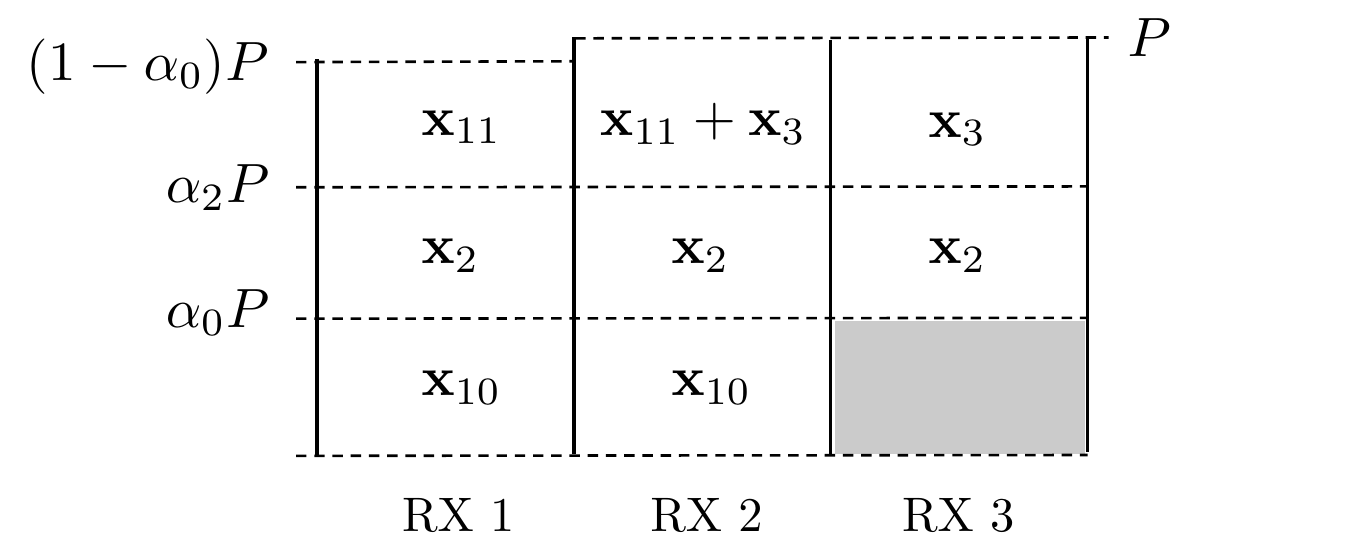}}
      }
    \mbox{
      \subfigure[Channel type 2]{\includegraphics[width=0.45\textwidth]{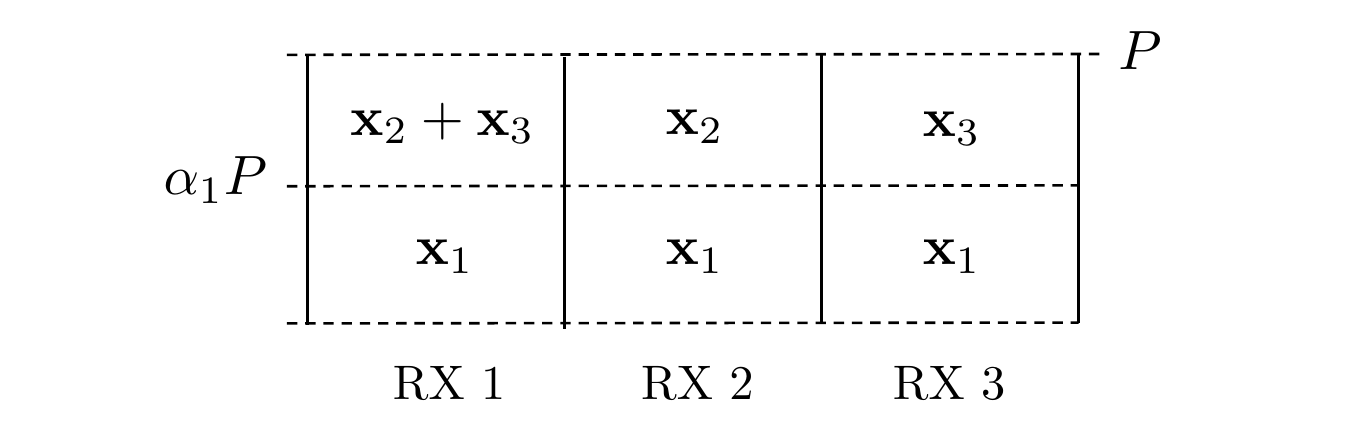}}
      }
    \mbox{
      \subfigure[Channel type 3]{\includegraphics[width=0.45\textwidth]{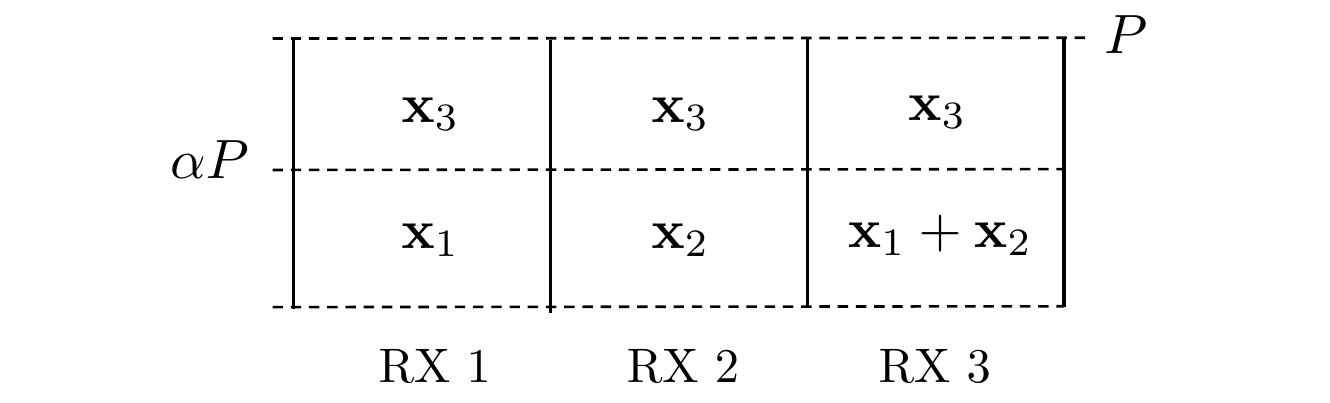}}
      }
\caption{Signal scale diagram.}
\label{fig:signalScale}
  \end{center}
\end{figure}

With the choice of transmit signals, the received signals are given by
\begin{eqnarray*}
&&\mathbf{y}_1=\mathbf{x}_{11}+\mathbf{x}_2+\mathbf{x}_{10}+\mathbf{z}_1\\
&&\mathbf{y}_2=[\mathbf{x}_{11}+\mathbf{x}_3]+\mathbf{x}_2+\mathbf{z}_2'\\
&&\mathbf{y}_3=\mathbf{x}_3+\mathbf{z}_3'.
\end{eqnarray*}
where $\mathbf{x}_f=[\mathbf{x}_{11}+\mathbf{x}_3]$ is the sum of interference, and $\mathbf{z}_2'=\mathbf{x}_{10}+\mathbf{z}_2$ and $\mathbf{z}_3'=\mathbf{x}_2+\mathbf{z}_3$ are the effective Gaussian noise.
The signal scale diagram at each receiver is shown in Fig. \ref{fig:signalScale} (a).

At the receivers, successive decoding is performed in the following order: $\mathbf{x}_{11}\rightarrow \mathbf{x}_2\rightarrow \mathbf{x}_{10}$ at receiver 1, $\mathbf{x}_f\rightarrow \mathbf{x}_2$ at receiver 2, and receiver 3 only decodes $\mathbf{x}_3$.

Note that the aligned lattice codewords $\mathbf{t}_{11}+\mathbf{t}_3\in \Lambda_c$, and $\mathbf{t}_f=[\mathbf{t}_{11}+\mathbf{t}_3]\modLambda_1 \in\Lambda_c\cap\mathcal{V}(\Lambda_1)$. We state the relationship between $\mathbf{x}_f$ and $\mathbf{t}_f$ in the following lemmas.
\begin{lemma} The following holds.
\begin{equation*}
[\mathbf{x}_f-\mathbf{d}_f]\modLambda_1=\mathbf{t}_f
\end{equation*}
where $\mathbf{d}_f=\mathbf{d}_{11}+\mathbf{d}_3$.
\end{lemma}
\begin{IEEEproof}
\begin{eqnarray*}
&&[\mathbf{x}_f-\mathbf{d}_f]\modLambda_1\\
&&=[M_{\Lambda_1}(\mathbf{t}_{11}+\mathbf{d}_{11})+M_{\Lambda_3}(\mathbf{t}_3+\mathbf{d}_3)-\mathbf{d}_f]\modLambda_1\\
&&=[M_{\Lambda_1}(\mathbf{t}_{11}+\mathbf{d}_{11})+M_{\Lambda_1}(\mathbf{t}_3+\mathbf{d}_3)-\mathbf{d}_f]\modLambda_1\\
&&=[\mathbf{t}_{11}+\mathbf{d}_{11}+\mathbf{t}_3+\mathbf{d}_3-\mathbf{d}_f]\modLambda_1\\
&&=[\mathbf{t}_{11}+\mathbf{t}_3]\modLambda_1\\
&&=\mathbf{t}_f
\end{eqnarray*}
The second and third equalities are due to distributive law and the identity in the following lemma.
\end{IEEEproof}

\begin{lemma} For any nested lattices $\Lambda_3\subset\Lambda_1$ and\\ any $\mathbf{x}\in\mathbb{R}^n$, it holds that
\begin{equation*}
[M_{\Lambda_3}(\mathbf{x})]\modLambda_1=[\mathbf{x}]\modLambda_1.
\end{equation*}
\end{lemma}
\begin{IEEEproof}
\begin{eqnarray*}
&&[M_{\Lambda_3}(\mathbf{x})]\modLambda_1\\
&&=[\mathbf{x}-\lambda_3]\modLambda_1\\
&&=[M_{\Lambda_1}(\mathbf{x})-M_{\Lambda_1}(\lambda_3)]\modLambda_1\\
&&=[M_{\Lambda_1}(\mathbf{x})-\lambda_3 +Q_{\Lambda_1}(\lambda_3)]\modLambda_1\\
&&=[M_{\Lambda_1}(\mathbf{x})]\modLambda_1\\
&&=[\mathbf{x}]\modLambda_1
\end{eqnarray*}
where $\lambda_3=Q_{\Lambda_3}(\mathbf{x})\in\Lambda_1$, thus $Q_{\Lambda_1}(\lambda_3)=\lambda_3$.
\end{IEEEproof}

\begin{lemma} The following holds.
\begin{equation*}
[\mathbf{t}_f+\mathbf{d}_f]\modLambda_1=[\mathbf{x}_f]\modLambda_1.
\end{equation*}
\end{lemma}
\begin{IEEEproof}
\begin{eqnarray*}
&&[\mathbf{t}_f+\mathbf{d}_f]\modLambda_1\\
&&=[M_{\Lambda_1}(\mathbf{t}_{11}+\mathbf{t}_3)+\mathbf{d}_f]\modLambda_1\\
&&=[\mathbf{t}_{11}+\mathbf{t}_3+\mathbf{d}_f]\modLambda_1\\
&&=[M_{\Lambda_1}(\mathbf{t}_{11}+\mathbf{d}_{11})+M_{\Lambda_1}(\mathbf{t}_3+\mathbf{d}_3)]\modLambda_1\\
&&=[M_{\Lambda_1}(\mathbf{t}_{11}+\mathbf{d}_{11})+M_{\Lambda_3}(\mathbf{t}_3+\mathbf{d}_3)]\modLambda_1\\
&&=[\mathbf{x}_{11}+\mathbf{x}_3]\modLambda_1\\
&&=[\mathbf{x}_f]\modLambda_1
\end{eqnarray*}
\end{IEEEproof}

Receiver 2 does not need to recover the codewords $\mathbf{t}_{11}$ and $\mathbf{t}_3$ but the real sum $\mathbf{x}_f$ to remove the interference from $\mathbf{y}_2$. Since $\mathbf{x}_f=M_{\Lambda_1}(\mathbf{x}_f)+Q_{\Lambda_1}(\mathbf{x}_f)$, we first recover the modulo part and then the quantized part to cancel out $\mathbf{x}_f$. This idea appeared in \cite{GastparNazer11} as an achievable scheme for the many-to-one interference channel.

The mod-$\Lambda_1$ channel between $\mathbf{t}_f$ and $\mathbf{y}_2'$ is given by
\begin{eqnarray}
&&\mathbf{y}_2'=[\beta_2\mathbf{y}_2-\mathbf{d}_f]\modLambda_1\\
&&\ \ \ \ = [\mathbf{x}_f-\mathbf{d}_f +\mathbf{z}_{e2}]\modLambda_1\\
&&\ \ \ \ = [\mathbf{t}_f +\mathbf{z}_{e2}]\modLambda_1
\end{eqnarray}
where the effective noise $\mathbf{z}_{e2}=(\beta_2-1)\mathbf{x}_f+\beta_2(\mathbf{x}_2+\mathbf{x}_{10}+\mathbf{z}_2)$. Note that $\mathbb{E}[\|\mathbf{x}_f\|^2]=(\bar{\alpha}_0+1)nP$, and the effective noise variance $\sigma_{e2}^2=\frac{1}{n}\mathbb{E}[\|\mathbf{z}_{e2}\|^2]=(\beta_2-1)^2(\bar{\alpha}_0+1)P+\beta_2^2 N_{e2}$ where $N_{e2}=(\alpha_0+\alpha_2) P+N_2$. With the MMSE scaling factor $\beta_2=\frac{(\bar{\alpha}_0+1)P}{(\bar{\alpha}_0+1)P+N_{e2}}$ plugged in, we get $\sigma_{e2}^2=\beta_2 N_{e2}=\frac{(\bar{\alpha}_0+1)P N_{e2}}{(\bar{\alpha}_0+1)P+N_{e2}}$. The capacity of the mod-$\Lambda_1$ channel between $\mathbf{t}_f$ and $\mathbf{y}_2'$ is
\begin{eqnarray*}
&&\frac{1}{n} I\left(\mathbf{t}_f;\mathbf{y}_2'\right)\\
&&\geq \frac{1}{n}\log\left(\frac{V(\Lambda_1)}{2^{h(\mathbf{z}_{e2})}}\right)\\
&&= \frac{1}{2}\log \left(\frac{\bar{\alpha}_0 P}{\beta_2 N_{e2}}\right)\\
&&= \frac{1}{2}\log \left(\frac{\bar{\alpha}_0(\bar{\alpha}_0+1)P+\bar{\alpha}_0 N_{e2}}{(\bar{\alpha}_0+1)N_{e2}}\right)\\
&&= \frac{1}{2}\log \left(\frac{\bar{\alpha}_0}{\bar{\alpha}_0+1}+\frac{\bar{\alpha}_0 P}{N_{e2}}\right)\\
&&= \frac{1}{2}\log \left(\frac{\bar{\alpha}_0}{\bar{\alpha}_0+1}+\frac{\bar{\alpha}_0 P}{(\alpha_0+\alpha_2) P+N_2}\right)\\
&&= C_f
\end{eqnarray*}

For reliable decoding of $\mathbf{t}_f$ at receiver 2, we have the code rate constraint $R_{11}=\frac{1}{n}\log\left(\frac{V(\Lambda_1)}{V(\Lambda_c)}\right)\leq C_f$. This also implies that $R_3=\frac{1}{n}\log\left(\frac{V(\Lambda_2)}{V(\Lambda_c)}\right)\leq C_f+\frac{1}{n}\log\left(\frac{V(\Lambda_2)}{V(\Lambda_1)}\right)=\frac{1}{2}\log\left(\frac{P}{\beta_2 N_{e2}}\right)=\frac{1}{2}\log \left(\frac{1}{\bar{\alpha}_0+1}+\frac{P}{(\alpha_0+\alpha_2) P+N_2}\right)$.
By lattice decoding, we can recover the modulo sum of interference codewords $\mathbf{t}_f$ from $\mathbf{y}_2'$. Then, we can recover the real sum $\mathbf{x}_f$ in the following way.
\begin{itemize}
\item Recover $M_{\Lambda_1}(\mathbf{x}_f)$ by calculating $[\mathbf{t}_f+\mathbf{d}_f]\modLambda_1$ (lemma 3).
\item Subtract it from the received signal,
\begin{equation}
\mathbf{y}_2-M_{\Lambda_1}(\mathbf{x}_f)=Q_{\Lambda_1}(\mathbf{x}_f) +\mathbf{z}_2''
\end{equation}
where $\mathbf{z}_2''=\mathbf{x}_2+\mathbf{x}_{10}+\mathbf{z}_2$.
\item Quantize it to recover $Q_{\Lambda_1}(\mathbf{x}_f)$,
\begin{equation}
Q_{\Lambda_1}\left(Q_{\Lambda_1}(\mathbf{x}_f) +\mathbf{z}_2''\right)=Q_{\Lambda_1}(\mathbf{x}_f)
\end{equation}
with probability $1-P_e$ where
\begin{equation}
P_e=\textrm{Pr}[Q_{\Lambda_1}\left(Q_{\Lambda_1}(\mathbf{x}_f) +\mathbf{z}_2''\right)\neq Q_{\Lambda_1}(\mathbf{x}_f)]
\end{equation}
is the probability of decoding error.
If we choose $\Lambda_1$ to be simultaneously Rogers-good and Poltyrev-good \cite{Zamir14} with $V(\Lambda_1)\geq V(\Lambda_c)$, then $P_e\rightarrow 0$ as $n\rightarrow \infty$.
\item Recover $\mathbf{x}_f$ by adding two vectors,
\begin{equation}
M_{\Lambda_1}(\mathbf{x}_f)+Q_{\Lambda_1}(\mathbf{x}_f)=\mathbf{x}_f.
\end{equation}
\end{itemize}
We now proceed to decoding $\mathbf{x}_2$ from $\mathbf{y}_2-\mathbf{x}_f=\mathbf{x}_2+\mathbf{z}_2'$. Since $\mathbf{x}_2$ is a codeword from an i.i.d. random code for point-to-point channel, we can achieve rate up to
\begin{equation}
R_2\leq \halflog\left(\frac{\alpha_2 P}{\alpha_0 P+N_2}\right).
\end{equation}

At receiver 1, we first decode $\mathbf{x}_{11}$ while treating other signals $\mathbf{x}_2+\mathbf{x}_{10}+\mathbf{z}_1$ as noise. The effective noise in the mod-$\Lambda_1$ channel is $\mathbf{z}_{e1}=(\beta_1-1)^2\mathbf{x}_{11}+\beta_1(\mathbf{x}_2+\mathbf{x}_{10}+\mathbf{z}_1)$ with variance $\sigma_{e1}^2=\frac{1}{n}\mathbb{E}[\|\mathbf{z}_{e1}\|^2]=(\beta_1-1)^2 \bar{\alpha}_0 P+\beta_1^2 N_{e1}$ where $N_{e1}=(\alpha_0+\alpha_2)P+N_1$. For reliable decoding, the rate $R_{11}$ must satisfy
\[R_{11}\leq \halflog\left(\frac{\sigma^2(\Lambda_1)}{\beta_1\sigma_{e1}^2}\right)=\halflog\left(1+\frac{\bar{\alpha}_0 P}{(\alpha_0+\alpha_2) P+N_1}\right)\]
where the MMSE scaling parameter $\beta_1=\frac{\bar{\alpha}_0 P}{\bar{\alpha}_0 P+N_{e1}}$. Similarly, we have the other rate constraints at receiver 1:
\begin{eqnarray}
&& R_2\leq \halflog\left(1+\frac{{\alpha}_2 P}{{\alpha}_0 P +N_1}\right)\\
&& R_{10}\leq \halflog\left(1+\frac{{\alpha}_0 P}{N_1}\right).
\end{eqnarray}

At receiver 3, the signal $\mathbf{x}_3$ is decoded with the effective noise $\mathbf{x}_2+\mathbf{z}_3$. For reliable decoding, $R_3$ must satisfy
\begin{equation}
R_3\leq \halflog\left(1+\frac{P}{\alpha_2 P+N_3}\right).
\end{equation}

In summary,
\begin{itemize}
\item $\mathbf{x}_{11}$ decoded at receivers 1 and 2
\begin{eqnarray*}
&& R_{11}\leq T_{11}'=\halflog\left(1+\frac{(1-\alpha_0) P}{(\alpha_0+\alpha_2) P+N_1}\right)\\
&& R_{11}\leq T_{11}''=\halflog\left(c_{11}+\frac{(1-\alpha_0) P}{(\alpha_0+\alpha_2) P+N_2}\right)
\end{eqnarray*}
where $c_{11}=\frac{(1-\alpha_0)P}{(1-\alpha_0)P+P}=\frac{1-\alpha_0}{2-\alpha_0}$.
\item $\mathbf{x}_{10}$ decoded at receiver 1
\begin{eqnarray}
R_{10}\leq T_{10}=\halflog\left(1+\frac{{\alpha}_0 P}{N_1}\right)
\end{eqnarray}
\item $\mathbf{x}_2$ decoded at receivers 1 and 2
\begin{eqnarray}
&& R_2\leq T_2' =\halflog\left(1+\frac{{\alpha}_2 P}{{\alpha}_0 P +N_1}\right)\\
&& R_2\leq T_2'' =\halflog\left(1+\frac{{\alpha}_2 P}{{\alpha}_0 P +N_2}\right)
\end{eqnarray}
\item $\mathbf{x}_3$ decoded at receivers 2 and 3
\begin{eqnarray}
&& R_3\leq T_3' =\halflog\left(c_3+\frac{P}{(\alpha_0+\alpha_2) P+N_2}\right)\nonumber\\
&& R_3\leq T_3'' =\halflog\left(1+\frac{P}{\alpha_2 P+N_3}\right)
\end{eqnarray}
where $c_3=\frac{P}{(1-\alpha_0)P+P}=\frac{1}{2-\alpha_0}$.
\end{itemize}
Note that $0\leq c_{11}\leq \frac{1}{2}$, $c_{11}+c_3=1$, and $\frac{1}{2}\leq c_3\leq 1$. Putting together, we can see that the following rate region is achievable.
\begin{eqnarray*}
&& R_1 \leq T_1=\min\{T_{11}',T_{11}''\}+T_{10}=T_{11}''+T_{10}\\
&& R_2 \leq T_2=\min\{T_2',T_2''\}=T_2''\\
&& R_3 \leq T_3=\min\{T_3',T_3''\}
\end{eqnarray*}
where
\begin{eqnarray}
&& T_1=\halflog\left(c_{11}+\frac{(1-\alpha_0) P}{(\alpha_0+\alpha_2) P+N_2}\right)\nonumber\\
&&\ \ \ \ \ \ \ \ \ \ \ \ \ \ \ \ \ \ \ \ \ \ \ \ \ \ \ \ \ \ +\halflog\left(1+\frac{{\alpha}_0 P}{N_1}\right)\\
&& T_2=\halflog\left(1+\frac{{\alpha}_2 P}{{\alpha}_0 P +N_2}\right)\\
&& T_3 \geq\halflog\left(c_3+\frac{P}{(\alpha_0+\alpha_2) P+N_3}\right).
\end{eqnarray}
Thus, \emph{Theorem 8} is proved.

\subsection{The Gap}

We choose the parameter $\alpha_0=\frac{N_2}{P}$, which is suboptimal but good enough to achieve a constant gap. This choice of parameter, inspired by \cite{EtkinTseWang08}, ensures making efficient use of signal scale difference between $N_1$ and $N_2$ at receiver 1, while keeping the interference of $\mathbf{x}_{10}$ at the noise level $N_2$ at receiver 2. By substitution, we get
\begin{eqnarray}
&&T_1 = \halflog\left(c_{11}+\frac{P-N_2}{\alpha_2 P+2N_2}\right)\nonumber\\
&&\ \ \ \ \ \ \ \ \ \ \ \ \ \ \ \ \ \ \ \ \ \ \ \ \ \ \ \ \ \ +\halflog\left(1+\frac{N_2}{N_1}\right)\\
&&T_2 = \halflog\left(1+\frac{\alpha_2 P}{2 N_2}\right)\\
&&T_3 \geq \halflog\left(c_3+\frac{P}{\alpha_2 P+N_2+N_3}\right).
\end{eqnarray}
Since $\alpha_0=\frac{N_2}{P}\in\left[0,\frac{1}{3}\right]$, it follows that $c_{11}=\frac{1-N_2/P}{2-N_2/P}\geq \frac{2}{5}$, and $c_3=\frac{1}{2-N_2/P}\geq \frac{1}{2}$.

Starting from $\mathcal{R}_o$ from Table \ref{tab:outerBounds}, we can express the two-dimensional outer bound region at $R_2$ as
\begin{eqnarray*}
&& R_1 \leq \min\left\{\halflog\left(1+\frac{2P}{N_1}\right)-R_2,C_1\right\}\\
&& \ \ \ \ \leq \min\left\{\halflog\left(\frac{P}{N_1}\cdot\frac{7}{3}\right)-R_2,\halflog\left(\frac{P}{N_1}\cdot\frac{4}{3}\right)\right\}\\
&& R_3 \leq \min\left\{\halflog\left(1+\frac{2P}{N_2}\right)-R_2,C_3\right\}\\
&& \ \ \ \ \leq \min\left\{\halflog\left(\frac{P}{N_2}\cdot\frac{7}{3}\right)-R_2,\halflog\left(\frac{P}{N_3}\cdot\frac{4}{3}\right)\right\}.
\end{eqnarray*}
Depending on the bottleneck of $\min\{\cdot,\cdot\}$ expressions, there are three cases:
\begin{itemize}
\item $R_2\leq \halflog\left(\frac{7}{4}\right)$
\item $\halflog\left(\frac{7}{4}\right)\leq R_2\leq \halflog\left(\frac{N_3}{N_2}\cdot\frac{7}{4}\right)$
\item $R_2\geq \halflog\left(\frac{N_3}{N_2}\cdot\frac{7}{4}\right)$.
\end{itemize}
At $R_2=\halflog\left(\frac{\alpha_2 P}{N_2}\cdot\frac{7}{4}\right)$, the outer bound region is
\begin{eqnarray*}
&& R_1 \leq \min\left\{\halflog\left(\frac{P}{\alpha_2 P}\cdot\frac{N_2}{N_1}\cdot\frac{4}{3}\right),\halflog\left(\frac{P}{N_1}\cdot\frac{4}{3}\right)\right\}\\
&& R_3 \leq \min\left\{\halflog\left(\frac{P}{\alpha_2 P}\cdot\frac{4}{3}\right),\halflog\left(\frac{P}{N_3}\cdot\frac{4}{3}\right)\right\}.
\end{eqnarray*}
Depending on the bottleneck of $\min\{\cdot,\cdot\}$ expressions, we consider the following three cases:
\begin{itemize}
\item $\alpha_2 P\geq N_3$
\item $N_2\leq \alpha_2 P\leq N_3$
\item $\alpha_2 P\leq N_2$.
\end{itemize}

\emph{Case i}) $\alpha_2 P\geq N_3$:
The outer bound region at $R_2=\halflog\left(\frac{\alpha_2 P}{N_2}\cdot\frac{7}{4}\right)$ is
\begin{eqnarray}
R_1 \leq \halflog\left(\frac{P}{\alpha_2 P}\cdot\frac{N_2}{N_1}\cdot\frac{4}{3}\right), R_3 \leq \halflog\left(\frac{P}{\alpha_2 P}\cdot\frac{4}{3}\right).
\end{eqnarray}

For comparison, let us take a look at the achievable rate region. The first term of $T_1$ is lower bounded by
\begin{eqnarray}
&&T_{11}'' =\halflog\left(c_{11}+\frac{P-N_2}{\alpha_2 P+2N_2}\right)\\
&&\ \ \ \ \ \geq \halflog\left(\frac{2}{5}+\frac{P-\alpha_2 P}{3\alpha_2 P}\right)\\
&&\ \ \ \ \ > \halflog\left(\frac{P}{3\alpha_2 P}\right).
\end{eqnarray}
We get the lower bounds:
\begin{eqnarray}
&& T_1 = T_{11}''+T_{10}\\
&&\ \ \ \ > \halflog\left(\frac{P}{3\alpha_2 P}\right)+\halflog\left(1+\frac{N_2}{N_1}\right)\\
&&\ \ \ \ > \halflog\left(\frac{P}{3\alpha_2 P}\cdot\frac{N_2}{N_1}\right)\\
&&T_3\geq \halflog\left(\frac{1}{2}+\frac{P}{\alpha_2 P+N_2+N_3}\right)\\
&&\ \ \ \ > \halflog\left(\frac{P}{3\alpha_2 P}\right).
\end{eqnarray}
For fixed $\alpha_2$ and $R_2=\halflog\left(\frac{\alpha_2 P}{2N_2}\right)$, the two-dimensional achievable rate region is given by
\begin{eqnarray}
R_1 \leq \halflog\left(\frac{P}{3\alpha_2 P}\cdot\frac{N_2}{N_1}\right),\ R_3 \leq \halflog\left(\frac{P}{3\alpha_2 P}\right).
\end{eqnarray}

\emph{Case ii}) $N_2\leq \alpha_2 P\leq N_3$:
The outer bound region at $R_2=\halflog\left(\frac{\alpha_2 P}{N_2}\cdot\frac{7}{4}\right)$ is
\begin{eqnarray}
R_1 \leq \halflog\left(\frac{P}{\alpha_2 P}\cdot\frac{N_2}{N_1}\cdot\frac{4}{3}\right),\ R_3 \leq \halflog\left(\frac{P}{N_3}\cdot\frac{4}{3}\right).
\end{eqnarray}

Now, let us take a look at the achievable rate region. We have the lower bounds:
\begin{eqnarray}
&& T_1 > \halflog\left(\frac{P}{3\alpha_2 P}\cdot\frac{N_2}{N_1}\right)\\
&& T_3 \geq \halflog\left(\frac{1}{2}+\frac{P}{\alpha_2 P+N_2+N_3}\right)\\
&&\ \ \ \ > \halflog\left(\frac{P}{3N_3}\right).
\end{eqnarray}
For fixed $\alpha_2$ and $R_2=\halflog\left(\frac{\alpha_2 P}{2N_2}\right)$, the two-dimensional achievable rate region is given by
\begin{eqnarray}
R_1\leq \halflog\left(\frac{P}{3\alpha_2 P}\cdot\frac{N_2}{N_1}\right),\ R_3\leq \halflog\left(\frac{P}{3N_3}\right).
\end{eqnarray}

\emph{Case iii}) $\alpha_2 P\leq N_2$:
The outer bound region at $R_2=\halflog\left(\frac{\alpha_2 P}{N_2}\cdot\frac{7}{4}\right)$ is
\begin{eqnarray}
R_1 \leq \halflog\left(\frac{P}{N_1}\cdot\frac{4}{3}\right),\ R_3 \leq \halflog\left(\frac{P}{N_3}\cdot\frac{4}{3}\right).
\end{eqnarray}
For this range of $\alpha_2$, the rate $R_2$ is small, i.e., $R_2 = \halflog\left(\frac{\alpha_2 P}{N_2}\cdot\frac{7}{4}\right)\leq \halflog\left(\frac{7}{4}\right)<\frac{1}{2}$, and $R_1$ and $R_3$ are close to single user capacities $C_1$ and $C_3$, respectively.

Let us take a look at the achievable rate region.
The first term of $T_1$ is lower bounded by
\begin{eqnarray}
&&T_{11}'' =\halflog\left(c_{11}+\frac{P-N_2}{\alpha_2 P+2N_2}\right)\\
&&\ \ \ \ \ \geq \halflog\left(\frac{2}{5}+\frac{P-N_2}{3N_2}\right)\\
&&\ \ \ \ \ > \halflog\left(\frac{P}{3N_2}\right).
\end{eqnarray}
We get the lower bounds:
\begin{eqnarray}
&& T_1 = T_{11}''+T_{10}\\
&&\ \ \ \ > \halflog\left(\frac{P}{3N_2}\right)+\halflog\left(1+\frac{N_2}{N_1}\right)\\
&&\ \ \ \ > \halflog\left(\frac{P}{3N_1}\right)\\
&& T_3 \geq \halflog\left(\frac{1}{2}+\frac{P}{\alpha_2 P+N_2+N_3}\right)\\
&&\ \ \ \ > \halflog\left(\frac{P}{3N_3}\right).
\end{eqnarray}
For fixed $\alpha_2$ and $R_2=\halflog\left(\frac{\alpha_2 P}{2N_2}\right)$, the following two-dimensional rate region is achievable.
\begin{eqnarray}
R_1 \leq \halflog\left(\frac{P}{3N_1}\right),\ R_3 \leq \halflog\left(\frac{P}{3N_3}\right).
\end{eqnarray}

In all three cases above, by comparing the inner and outer bound regions, we can see that $\delta_1\leq \halflog\left(3\cdot\frac{4}{3}\right)=1$, $\delta_2\leq \halflog\left(2\cdot\frac{7}{4}\right) = 0.91$ and $\delta_3\leq \halflog\left(3\cdot\frac{4}{3}\right)=1$. Therefore, we can conclude that the gap is to within one bit per message.

\section{Inner Bound: Channel Type 2}
\begin{theorem}
Given $\alpha_1\in [0,1]$, the region $\mathcal{R}_\alpha$ is defined by
\begin{eqnarray*}
&&R_1 \leq \halflog\left(1+\frac{\alpha_1 P}{N_1}\right)\\
&&R_2 \leq \halflog^+\left(\frac{1}{2}+\frac{P}{\alpha_1 P+N_2}\right)\\
&&R_3 \leq \halflog^+\left(\frac{1}{2}+\frac{P}{\alpha_1 P+N_3}\right),
\end{eqnarray*}
and $\mathcal{R}=\textsc{conv}\left(\bigcup_{\alpha_1}\mathcal{R}_\alpha\right)$ is achievable.
\end{theorem}

\subsection{Achievable Scheme}
For this channel type, rate splitting is not necessary. Transmit signal $\mathbf{x}_k$ is a coded signal of $M_k\in\{1,2,\ldots,2^{nR_k}\},k=1,2,3$. In particular, $\mathbf{x}_2$ and $\mathbf{x}_3$ are lattice-coded signals using the same pair of coding and shaping lattices. As a result, the sum $\mathbf{x}_2+\mathbf{x}_3$ is a dithered lattice codeword. The power allocation satisfies
$\mathbb{E}[\|\mathbf{x}_1\|^2]=\alpha_1 nP$, $\mathbb{E}[\|\mathbf{x}_2\|^2]=nP$, and $\mathbb{E}[\|\mathbf{x}_3\|^2]=nP$.
The received signals are
\begin{eqnarray*}
&&\mathbf{y}_1=[\mathbf{x}_2+\mathbf{x}_3]+\mathbf{x}_1+\mathbf{z}_1\\
&&\mathbf{y}_2=\mathbf{x}_2+\mathbf{x}_1+\mathbf{z}_2\\
&&\mathbf{y}_3=\mathbf{x}_3+\mathbf{x}_1+\mathbf{z}_3.
\end{eqnarray*}
The signal scale diagram at each receiver is shown in Fig. \ref{fig:signalScale} (b).
Decoding is performed in the following way.
\begin{itemize}
\item At receiver 1, $[\mathbf{x}_2+\mathbf{x}_3]$ is first decoded while treating $\mathbf{x}_1+\mathbf{z}_1$ as noise. Next, $\mathbf{x}_1$ is decoded from $\mathbf{y}_1-[\mathbf{x}_2+\mathbf{x}_3]=\mathbf{x}_1+\mathbf{z}_1$. For reliable decoding, the code rates should satisfy
\begin{eqnarray}
&& R_2 \leq T_2' =\halflog\left(\frac{1}{2}+\frac{P}{\alpha_1 P+N_1}\right)\\
&& R_3 \leq T_3' =\halflog\left(\frac{1}{2}+\frac{P}{\alpha_1 P+N_1}\right)\\
&& R_1 \leq T_1 =\halflog\left(1+\frac{\alpha_1 P}{N_1}\right).
\end{eqnarray}
\item At receiver 2, $\mathbf{x}_2$ is decoded while treating $\mathbf{x}_1+\mathbf{z}_2$ as noise. Similarly at receiver 3, $\mathbf{x}_3$ is decoded while treating $\mathbf{x}_1+\mathbf{z}_3$ as noise. For reliable decoding, the code rates should satisfy
\begin{eqnarray}
&&R_2\leq T_2'' =\halflog\left(1+\frac{P}{\alpha_1 P+N_2}\right)\\
&&R_3\leq T_3'' =\halflog\left(1+\frac{P}{\alpha_1 P+N_3}\right).
\end{eqnarray}
\end{itemize}
Putting together, we get
\begin{eqnarray*}
&& R_1 \leq T_1\\
&& R_2 \leq T_2 =\min\{T_2',T_2''\}\\
&& R_3 \leq T_3 =\min\{T_3',T_3''\}
\end{eqnarray*}
where
\begin{eqnarray}
&& T_1 =\halflog\left(1+\frac{\alpha_1 P}{N_1}\right)\\
&& T_2 \geq \halflog\left(\frac{1}{2}+\frac{P}{\alpha_1 P+N_2}\right)\\
&& \ \ \ \ \geq \halflog\left(\frac{1}{2}+\frac{P}{2\cdot\max\{\alpha_1 P,N_2\} }\right)\\
&& T_3 \geq \halflog\left(\frac{1}{2}+\frac{P}{\alpha_1 P+N_3}\right)\\
&& \ \ \ \ \geq \halflog\left(\frac{1}{2}+\frac{P}{2\cdot\max\{\alpha_1 P,N_3\} }\right).
\end{eqnarray}

\subsection{The Gap}
Starting from $\mathcal{R}_o$ from Table \ref{tab:outerBounds}, we can express the two-dimensional outer bound region at $R_1$ as
\begin{eqnarray*}
&& R_2 \leq \min\left\{\halflog\left(1+\frac{2P}{N_1}\right)-R_1,C_2\right\}\\
&& \ \ \ \ \leq \min\left\{\halflog\left(\frac{P}{N_1}\cdot\frac{7}{3}\right)-R_1,\halflog\left(\frac{P}{N_2}\cdot\frac{4}{3}\right)\right\}\\
&& R_3 \leq \min\left\{\halflog\left(1+\frac{2P}{N_1}\right)-R_1,C_3\right\}\\
&& \ \ \ \ \leq \min\left\{\halflog\left(\frac{P}{N_1}\cdot\frac{7}{3}\right)-R_1,\halflog\left(\frac{P}{N_3}\cdot\frac{4}{3}\right)\right\}.
\end{eqnarray*}
Depending on the bottleneck of $\min\{\cdot,\cdot\}$ expressions, there are three cases:
\begin{itemize}
\item $R_1\leq \halflog\left(\frac{N_2}{N_1}\cdot\frac{7}{4}\right)$
\item $\halflog\left(\frac{N_2}{N_1}\cdot\frac{7}{4}\right)\leq R_1\leq \halflog\left(\frac{N_3}{N_1}\cdot\frac{7}{4}\right)$
\item $R_1\geq \halflog\left(\frac{N_3}{N_1}\cdot\frac{7}{4}\right)$.
\end{itemize}
At $R_1=\halflog\left(\frac{\alpha_1 P}{N_1}\cdot\frac{7}{4}\right)$, the region can be expressed as
\begin{eqnarray*}
&& R_2 \leq \min\left\{\halflog\left(\frac{P}{\alpha_1 P}\cdot\frac{4}{3}\right),\halflog\left(\frac{P}{N_2}\cdot\frac{4}{3}\right)\right\}\\
&& R_3 \leq \min\left\{\halflog\left(\frac{P}{\alpha_1 P}\cdot\frac{4}{3}\right),\halflog\left(\frac{P}{N_3}\cdot\frac{4}{3}\right)\right\}.
\end{eqnarray*}
Depending on the bottleneck of $\min\{\cdot,\cdot\}$ expressions, we consider the following three cases.

\emph{Case i}) $\alpha_1 P\geq N_3$: The two-dimensional outer bound region at $R_1=\halflog\left(\frac{\alpha_1 P}{N_1}\cdot\frac{7}{4}\right)$ is
\begin{eqnarray}
R_2 \leq \halflog\left(\frac{P}{\alpha_1 P}\cdot\frac{4}{3}\right),\ R_3 \leq \halflog\left(\frac{P}{\alpha_1 P}\cdot\frac{4}{3}\right).
\end{eqnarray}
%For $\alpha_1\geq N_3$, the following rate region is achievable.
%\begin{eqnarray}
%&& R_1 \leq \halflog\left(1+\frac{\alpha_1 P}{N_1}\right)\\
%&& R_2 \leq \halflog\left(1+\frac{P}{\alpha_1 P}\right)-\frac{1}{2}\\
%&& R_3 \leq \halflog\left(1+\frac{P}{\alpha_1 P}\right)-\frac{1}{2}.
%\end{eqnarray}
For fixed $\alpha_1$ and $R_1=\halflog\left(\frac{\alpha_1 P}{N_1}\right)$, the following two-dimensional region is achievable.
\begin{eqnarray}
R_2 \leq \halflog\left(\frac{P}{2\alpha_1 P}\right),\ R_3 \leq \halflog\left(\frac{P}{2\alpha_1 P}\right).
\end{eqnarray}

\emph{Case ii}) $N_2\leq \alpha_1 P\leq N_3$: The two-dimensional outer bound region at $R_1=\halflog\left(\frac{\alpha_1 P}{N_1}\cdot\frac{7}{4}\right)$ is
\begin{eqnarray}
R_2 \leq \halflog\left(\frac{P}{\alpha_1 P}\cdot\frac{4}{3}\right),\ R_3 \leq \halflog\left(\frac{P}{N_3}\cdot\frac{4}{3}\right).
\end{eqnarray}
%For $N_2\leq \alpha_1 P\leq N_3$, the following rate region is achievable.
%\begin{eqnarray}
%&& R_1 \leq \halflog\left(1+\frac{\alpha_1 P}{N_1}\right)\\
%&& R_2 \leq \halflog\left(1+\frac{P}{\alpha_1 P}\right)-\frac{1}{2}\\
%&& R_3 \leq \halflog\left(1+\frac{P}{N_3}\right)-\frac{1}{2}.
%\end{eqnarray}
For fixed $\alpha_1$ and $R_1=\halflog\left(\frac{\alpha_1 P}{N_1}\right)$, the following two-dimensional region is achievable.
\begin{eqnarray}
R_2 \leq \halflog\left(\frac{P}{2\alpha_1 P}\right),\ R_3 \leq \halflog\left(\frac{P}{2N_3}\right).
\end{eqnarray}

\emph{Case iii}) $\alpha_1 P\leq N_2$: The two-dimensional outer bound region at $R_1=\halflog\left(\frac{\alpha_1 P}{N_1}\cdot\frac{7}{4}\right)$ is
\begin{eqnarray}
R_2 \leq \halflog\left(\frac{P}{N_2}\cdot\frac{4}{3}\right),\ R_3 \leq \halflog\left(\frac{P}{N_3}\cdot\frac{4}{3}\right).
\end{eqnarray}
%For $\alpha_1 P\leq N_2$, the following rate region is achievable.
%\begin{eqnarray}
%&& R_1 \leq \halflog\left(1+\frac{\alpha_1 P}{N_1}\right)\\
%&& R_2 \leq \halflog\left(1+\frac{P}{N_2}\right)-\frac{1}{2}\\
%&& R_3 \leq \halflog\left(1+\frac{P}{N_3}\right)-\frac{1}{2}.
%\end{eqnarray}
For fixed $\alpha_1$ and $R_1=\halflog\left(\frac{\alpha_1 P}{N_1}\right)$, the following two-dimensional region is achievable.
\begin{eqnarray}
R_2 \leq \halflog\left(\frac{P}{2N_2}\right),\ R_3 \leq \halflog\left(\frac{P}{2N_3}\right).
\end{eqnarray}

In all three cases above, by comparing the inner and outer bounds, we can see that $\delta_1\leq \halflog\left(\frac{7}{4}\right)<0.41$, $\delta_2 \leq \halflog\left(2\cdot\frac{4}{3}\right)<0.71$, and $\delta_3 \leq\halflog\left(2\cdot\frac{4}{3}\right)<0.71$. We can conclude that the inner and outer bounds are to within one bit.

\section{Inner Bound: Channel Type 3}
\begin{theorem}
Given $\alpha \in [0,1]$, the region $\mathcal{R}_\alpha$ is defined by
\begin{eqnarray*}
&&R_1 \leq \halflog\left(1+\frac{\alpha P}{N_1}\right)\\
&&R_2 \leq \halflog\left(1+\frac{\alpha P}{N_2}\right)\\
&&R_3 \leq \halflog\left(1+\frac{P}{2\alpha P+N_3}\right),
\end{eqnarray*}
and $\mathcal{R}=\textsc{conv}\left(\bigcup_{\alpha}\mathcal{R}_\alpha\right)$ is achievable.
\end{theorem}
\subsection{Achievable Scheme}
For this channel type, neither rate splitting nor aligned interference decoding is necessary.
Transmit signal $\mathbf{x}_k$ is a coded signal of $M_k\in\{1,2,\ldots,2^{nR_k}\},k=1,2,3$. The power allocation satisfies $\mathbb{E}[\|\mathbf{x}_1\|^2]=\alpha nP$, $\mathbb{E}[\|\mathbf{x}_2\|^2]=\alpha nP$, and $\mathbb{E}[\|\mathbf{x}_3\|^2]=nP$.
The received signals are
\begin{eqnarray*}
&&\mathbf{y}_1=\mathbf{x}_3+\mathbf{x}_1+\mathbf{z}_1\\
&&\mathbf{y}_2=\mathbf{x}_3+\mathbf{x}_2+\mathbf{z}_2\\
&&\mathbf{y}_3=\mathbf{x}_3+\mathbf{x}_1+\mathbf{x}_2+\mathbf{z}_3.
\end{eqnarray*}
The signal scale diagram at each receiver is shown in Fig. \ref{fig:signalScale} (c). Decoding is performed in the following way.
\begin{itemize}
\item At receiver 1, $\mathbf{x}_3$ is first decoded while treating $\mathbf{x}_1+\mathbf{z}_1$ as noise. Next, $\mathbf{x}_1$ is decoded from $\mathbf{y}_1-\mathbf{x}_3=\mathbf{x}_1+\mathbf{z}_1$. For reliable decoding, the code rates should satisfy
\begin{eqnarray}
&& R_3 \leq T_3' =\halflog\left(1+\frac{P}{\alpha P+N_1}\right)\\
&& R_1 \leq T_1 =\halflog\left(1+\frac{\alpha P}{N_1}\right).
\end{eqnarray}
\item At receiver 2, $\mathbf{x}_3$ is first decoded while treating $\mathbf{x}_2+\mathbf{z}_2$ as noise. Next, $\mathbf{x}_2$ is decoded from $\mathbf{y}_2-\mathbf{x}_3=\mathbf{x}_2+\mathbf{z}_2$. For reliable decoding, the code rates should satisfy
\begin{eqnarray}
&& R_3 \leq T_3'' =\halflog\left(1+\frac{P}{\alpha P+N_2}\right)\\
&& R_2 \leq T_2 =\halflog\left(1+\frac{\alpha P}{N_2}\right).
\end{eqnarray}
\item At receiver 3, $\mathbf{x}_3$ is decoded while treating $\mathbf{x}_1+\mathbf{x}_2+\mathbf{z}_3$ as noise. For reliable decoding, the code rates should satisfy
\begin{eqnarray}
&& R_3\leq T_3''' =\halflog\left(1+\frac{P}{2\alpha P+N_3}\right).
\end{eqnarray}
\end{itemize}
Putting together, we get
\begin{eqnarray*}
&& R_1 \leq T_1\\
&& R_2 \leq T_2\\
&& R_3 \leq T_3 =\min\{T_3',T_3'',T_3'''\}
\end{eqnarray*}
where
\begin{eqnarray}
&& T_1 =\halflog\left(1+\frac{\alpha P}{N_1}\right)\\
&& T_2 =\halflog\left(1+\frac{\alpha P}{N_2}\right)\\
&& T_3 =\halflog\left(1+\frac{P}{2\alpha P+N_3}\right)\\
&& \ \ \ \ \geq\halflog\left(1+\frac{P}{3\cdot\max\{\alpha P,N_3\} }\right).
\end{eqnarray}

\subsection{The Gap}
Starting from $\mathcal{R}_o$ from Table \ref{tab:outerBounds}, we can express the two-dimensional outer bound region at $R_3$ as
\begin{eqnarray*}
&& R_1 \leq \min\left\{\halflog\left(1+\frac{2P}{N_1}\right)-R_3,C_1\right\}\\
&& \ \ \ \ \leq \min\left\{\halflog\left(\frac{P}{N_1}\cdot\frac{7}{3}\right)-R_3,\halflog\left(\frac{P}{N_1}\cdot\frac{4}{3}\right)\right\}\\
&& R_2 \leq \min\left\{\halflog\left(1+\frac{2P}{N_2}\right)-R_3,C_2\right\}\\
&& \ \ \ \ \leq \min\left\{\halflog\left(\frac{P}{N_2}\cdot\frac{7}{3}\right)-R_3,\halflog\left(\frac{P}{N_2}\cdot\frac{4}{3}\right)\right\}.
\end{eqnarray*}
Depending on the bottleneck of $\min\{\cdot,\cdot\}$ expressions, there are two cases:
$R_3\leq \halflog\left(\frac{7}{4}\right)$ and $R_3\geq \halflog\left(\frac{7}{4}\right)$.
We assume that $R_3\geq \halflog\left(\frac{7}{4}\right)$, equivalently $\alpha\leq\frac{4}{7}$. We also assume that $R_3\leq\halflog\left(\frac{P}{N_3}\right)$, equivalently $\alpha P\geq N_3$. The other cases are trivial.

The two-dimensional outer bound region at $R_3=\halflog\left(\frac{P}{\alpha P}\right)$ is
\begin{eqnarray*}
&& R_1 \leq \min\left\{\halflog\left(\frac{\alpha P}{N_1}\cdot\frac{7}{3}\right),\halflog\left(\frac{P}{N_1}\cdot\frac{4}{3}\right)\right\}\\
&& R_2 \leq \min\left\{\halflog\left(\frac{\alpha P}{N_2}\cdot\frac{7}{3}\right),\halflog\left(\frac{P}{N_2}\cdot\frac{4}{3}\right)\right\}.
\end{eqnarray*}
For $\alpha\leq\frac{4}{7}$, the two-dimensional outer bound region is
\begin{eqnarray}
R_1 \leq \halflog\left(\frac{\alpha P}{N_1}\cdot\frac{7}{3}\right),\ R_2 \leq \halflog\left(\frac{\alpha P}{N_2}\cdot\frac{7}{3}\right).
\end{eqnarray}

%For $\alpha P\geq N_3$, the following rate region is achievable.
%\begin{eqnarray}
%&& R_1 \leq \halflog\left(1+\frac{\alpha P}{N_1}\right)\\
%&& R_2 \leq \halflog\left(1+\frac{\alpha P}{N_2}\right)\\
%&& R_3 \leq \halflog\left(3+\frac{P}{\alpha P}\right)-\halflog\left(3\right)
%\end{eqnarray}

For $\alpha P\geq N_3$, the two-dimensional achievable rate region at $R_3=\halflog\left(\frac{P}{3\alpha P}\right)$ is
\begin{eqnarray}
R_1 \leq \halflog\left(\frac{\alpha P}{N_1}\right),\ R_2 \leq \halflog\left(\frac{\alpha P}{N_2}\right).
\end{eqnarray}
By comparing the inner and outer bounds, we can see that $\delta_1\leq \halflog\left(\frac{7}{3}\right)<0.62$, $\delta_2 \leq \halflog\left(\frac{7}{3}\right)<0.62$, and $\delta_3 \leq\halflog\left(3\right)<0.8$. We can conclude that the inner and outer bounds are to within one bit.

\begin{figure}[t]
  \begin{center}
    \mbox{
      \subfigure[Large $R_1$]{\includegraphics[width=0.2\textwidth]{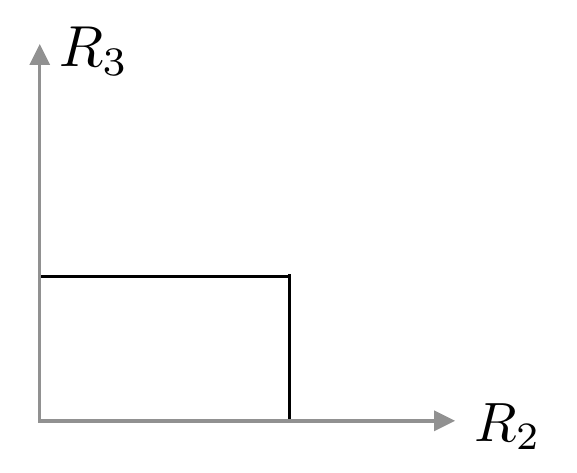}}
      }
    \mbox{
      \subfigure[Small $R_1$]{\includegraphics[width=0.2\textwidth]{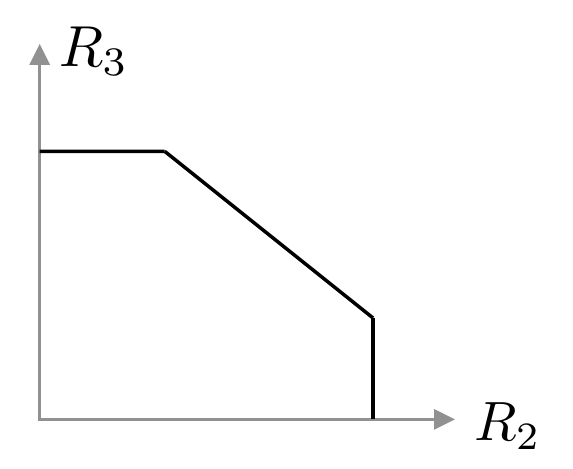}}
      }
\caption{The cross-section of the type 4 outer bound region at a relatively small or large $R_1$.}
\label{fig:outerBoundCrossSection4}
  \end{center}
\end{figure}

\section{Inner Bound: Channel Type 4}
The relaxed outer bound region $\mathcal{R}_o'$ given by
\begin{eqnarray*}
&& \ \ \ \ \ \ \ R_k \leq \halflog\left(\frac{P}{N_k}\right)+\halflog\left(\frac{4}{3}\right),\ k=1,2,3\\
&& R_1+R_2 \leq \halflog\left(\frac{P}{N_1}\right)+\halflog\left(\frac{7}{3}\right)\\
&& R_1+R_3 \leq \halflog\left(\frac{P}{N_1}\right)+\halflog\left(\frac{7}{3}\right)\\
&& R_2+R_3 \leq \halflog\left(\frac{P}{N_2}\right)+\halflog\left(\frac{7}{3}\right).
\end{eqnarray*}
The cross-sectional region at a given $R_1$ is described by
\begin{eqnarray*}
&& R_2\leq \min\left\{\halflog\left(\frac{P}{N_1}\cdot\frac{7}{3}\right)-R_1,\halflog\left(\frac{P}{N_2}\cdot\frac{4}{3}\right)\right\}\\
&& R_3\leq \min\left\{\halflog\left(\frac{P}{N_1}\cdot\frac{7}{3}\right)-R_1,\halflog\left(\frac{P}{N_3}\cdot\frac{4}{3}\right)\right\}\\
&& R_2+R_3 \leq \halflog\left(\frac{P}{N_2}\cdot\frac{7}{3}\right).
\end{eqnarray*}
Depending on the bottleneck of $\min\{\cdot,\cdot\}$ expressions, there are three cases:
\begin{itemize}
\item $R_1\leq \halflog\left(\frac{N_2}{N_1}\cdot\frac{7}{4}\right)$
\item $\halflog\left(\frac{N_2}{N_1}\cdot\frac{7}{4}\right)\leq R_1\leq \halflog\left(\frac{N_3}{N_1}\cdot\frac{7}{4}\right)$
\item $R_1\geq \halflog\left(\frac{N_3}{N_1}\cdot\frac{7}{4}\right)$.
\end{itemize}
In this section, we focus on the third case. The other cases can be proved similarly.
If the sum of the righthand sides of $R_2$ and $R_3$ bounds is smaller than the righthand side of $R_2+R_3$ bound, i.e.,
\begin{eqnarray}
\log\left(\frac{P}{N_1}\cdot\frac{7}{3}\right)-2R_1\leq \halflog\left(\frac{P}{N_2}\cdot\frac{7}{3}\right),
\end{eqnarray}
then the $R_2+R_3$ bound is not active at the $R_1$.
This condition can be expressed as a threshold on $R_1$ given by
\begin{eqnarray}
&& R_1 > R_{1,th}= \frac{1}{2}\log\left(\frac{P}{N_1}\cdot\frac{7}{3}\right)-\frac{1}{4}\log\left(\frac{P}{N_2}\cdot\frac{7}{3}\right)\nonumber\\
&& \ \ \ \ \ \ \ \ \ \ \ \ \ \ = \frac{1}{4}\log\left(\frac{P}{N_1}\cdot\frac{7}{3}\right)+\frac{1}{4}\log\left(\frac{N_2}{N_1}\right).
\end{eqnarray}
For this relatively large $R_1$, the cross-sectional region is a rectangle as described in Fig. \ref{fig:outerBoundCrossSection4} (a). In contrast, for a relatively small $R_1$, when the threshold condition does not hold, the cross-sectional region is a MAC-like region as described in Fig. \ref{fig:outerBoundCrossSection4} (b). In the rest of the section, we present achievable schemes for each case.

\subsection{Achievable Scheme for Relatively Large $R_1$}
\begin{figure}[tp]
  \begin{center}
    \mbox{
      \subfigure[Channel type 4: relatively large $R_1$]{\includegraphics[width=0.45\textwidth]{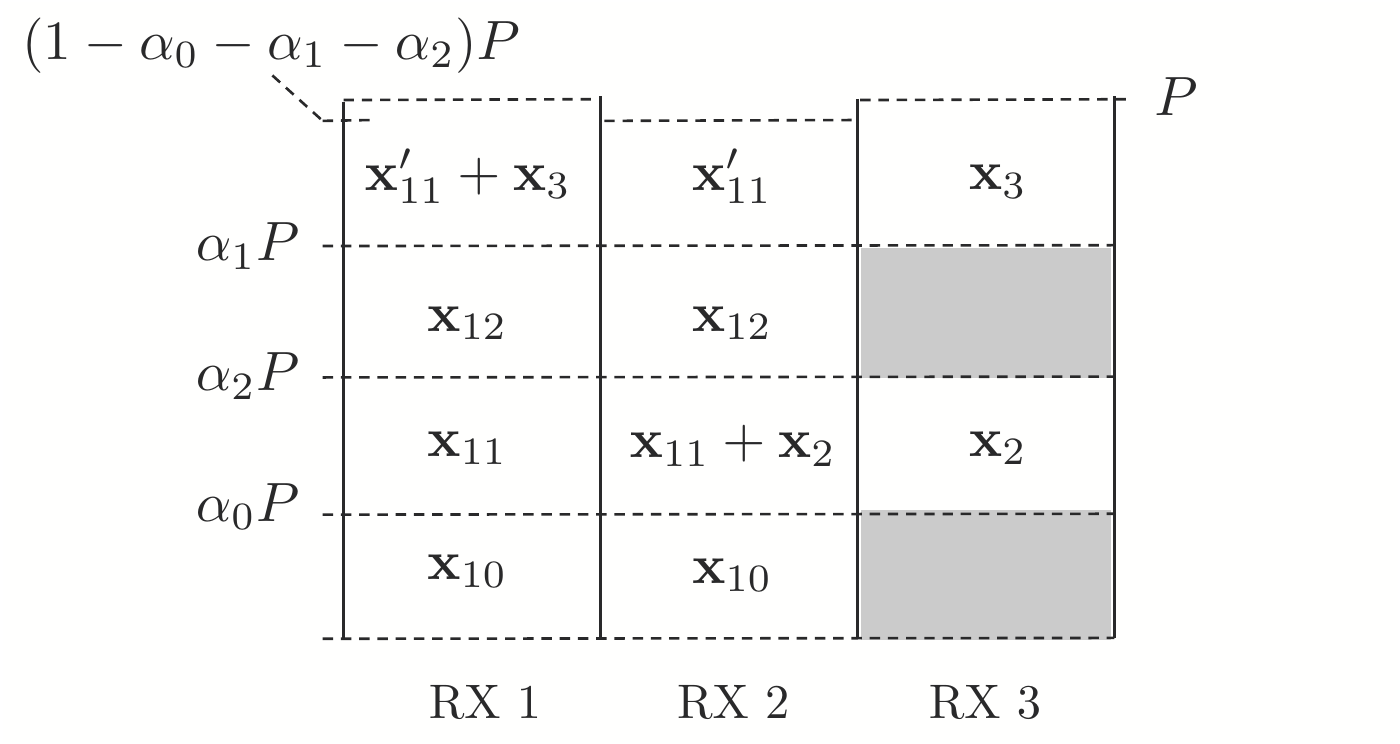}}
      }
    \mbox{
      \subfigure[Channel type 4: relatively small $R_1$]{\includegraphics[width=0.45\textwidth]{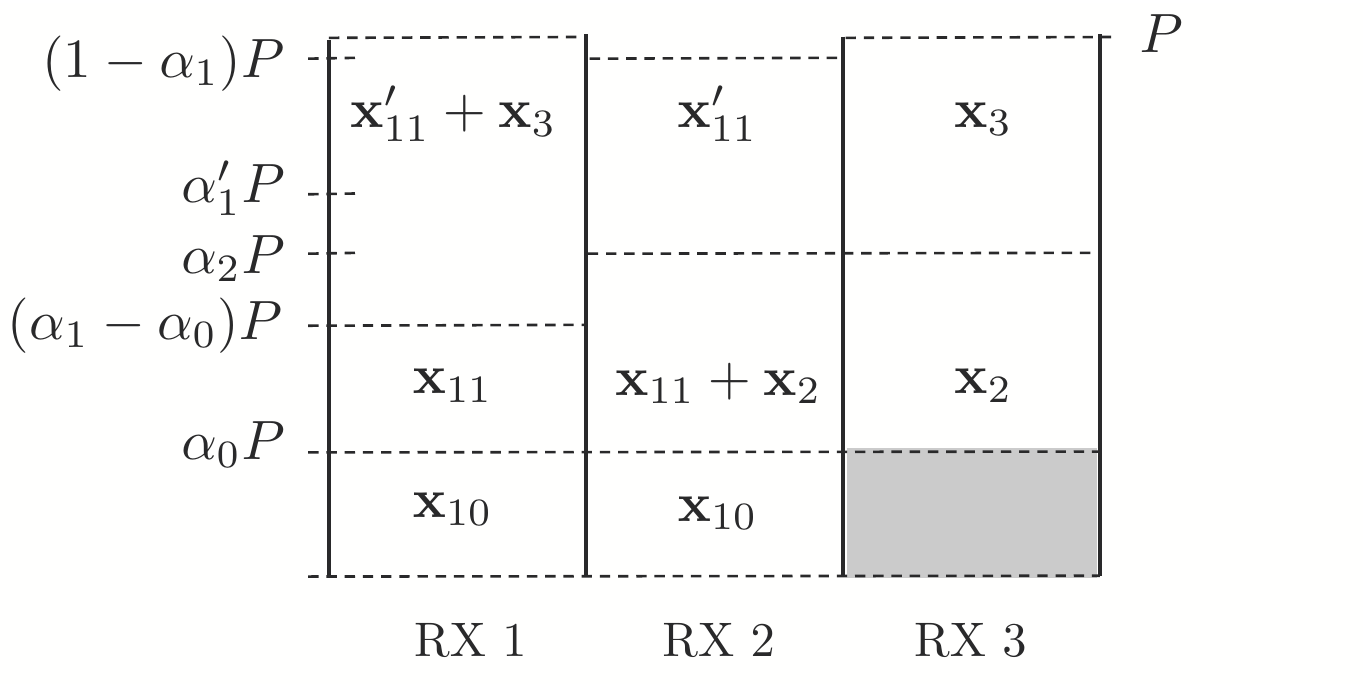}}
      }
\caption{Signal scale diagram.}
\label{fig:signalScale4}
  \end{center}
\end{figure}

\begin{theorem}
Given $\alpha=(\alpha_0,\alpha_1,\alpha_2) \in [0,1]^3$, the region $\mathcal{R}_\alpha$ is defined by
\begin{eqnarray*}
&& R_1 \leq \min\left\{\halflog^+\left(c_{11}+\frac{(1-\alpha_0-\alpha_1-\alpha_2)P}{(\alpha_0+\alpha_1+2\alpha_2)P+N_2}\right),\right.\\
&&\ \ \ \ \ \ \ \ \ \ \ \ \ \ \ \ \ \ \ \ \ \ \ \ \ \ \ \ \ \ \ \left.\halflog\left(1+\frac{\alpha_2 P}{\alpha_0 P+N_1}\right)\right\}\\
&&\ \ \ \ \ \ \ \ +\halflog\left(1+\frac{\alpha_1 P}{(\alpha_0+\alpha_2) P+N_2}\right)\\
&&\ \ \ \ \ \ \ \ +\halflog\left(1+\frac{\alpha_0 P}{N_1}\right)\\
&& R_2 \leq \halflog\left(1+\frac{\alpha_2 P}{\alpha_0 P+N_2}\right)\\
&& R_3 \leq \halflog^+\left(c_3+\frac{P}{(\alpha_0+\alpha_1+\alpha_2) P+N_3}\right)
\end{eqnarray*}
where $c_{11}=\frac{1-\alpha_0-\alpha_1-\alpha_2}{2-\alpha_0-\alpha_1-\alpha_2}$ and $c_3=\frac{1}{2-\alpha_0-\alpha_1-\alpha_2}$,
and $\mathcal{R}=\textsc{conv}\left(\bigcup_{\alpha}\mathcal{R}_\alpha\right)$ is achievable.
\end{theorem}

We present an achievable scheme for the case of $R_1 > R_{1,th}$. Message $M_1\in\{1,2,\ldots,2^{nR_1}\}$ is split into three parts: $M_{10}\in\{1,2,\ldots,2^{nR_{10}}\}$, $M_{11}\in\{1,2,\ldots,2^{nR_{11}}\}$ and $M_{12}\in\{1,2,\ldots,2^{nR_{12}}\}$, so $R_1=R_{10}+R_{11}+R_{12}$. We generate the signals in the following way: $\mathbf{x}_{11}$ and $\mathbf{x}_{11}'$ are differently coded signals of $M_{11}$, and $\mathbf{x}_{10}$ and $\mathbf{x}_{12}$ are coded signal of $M_{10}$ and $M_{12}$, respectively. The transmit signal is the sum
\[\mathbf{x}_1=\mathbf{x}_{10}+\mathbf{x}_{11}+\mathbf{x}_{12}+\mathbf{x}_{11}'.\] The power allocation satisfies $\mathbb{E}[\|\mathbf{x}_{10}\|^2]=\alpha_0 nP$, $\mathbb{E}[\|\mathbf{x}_{11}\|^2]=\alpha_2 nP$, $\mathbb{E}[\|\mathbf{x}_{12}\|^2]=\alpha_1 nP$, and $\mathbb{E}[\|\mathbf{x}_{11}'\|^2]=(1-\alpha_0-\alpha_1-\alpha_2) nP$.

The transmit signals $\mathbf{x}_2$ and $\mathbf{x}_3$ are coded signals of the messages $M_2\in\{1,2,\ldots,2^{nR_2}\}$ and $M_3\in\{1,2,\ldots,2^{nR_3}\}$, satisfying $\mathbb{E}[\|\mathbf{x}_2\|^2]=\alpha_2 nP$ and $\mathbb{E}[\|\mathbf{x}_3\|^2]=nP$.

The signals $\mathbf{x}_{11}'$ and $\mathbf{x}_3$ are lattice-coded signals using the same coding lattice but different shaping lattices. As a result, the sum $\mathbf{x}_{11}'+\mathbf{x}_3$ is a dithered lattice codeword.

The received signals are
\begin{eqnarray*}
&&\mathbf{y}_1=[\mathbf{x}_{11}'+\mathbf{x}_3]+\mathbf{x}_{12}+\mathbf{x}_{11}+\mathbf{x}_{10}+\mathbf{z}_1\\
&&\mathbf{y}_2=\mathbf{x}_{11}'+\mathbf{x}_{12}+\mathbf{x}_{11}+\mathbf{x}_{2}+\mathbf{x}_{10}+\mathbf{z}_2\\
&&\mathbf{y}_3=\mathbf{x}_3+\mathbf{x}_2+\mathbf{z}_3.
\end{eqnarray*}
The signal scale diagram at each receiver is shown in Fig. \ref{fig:signalScale4} (a).
Decoding is performed in the following way.
\begin{itemize}
\item At receiver 1, $[\mathbf{x}_{11}'+\mathbf{x}_3]$ is first decoded while treating other signals as noise and removed from $\mathbf{y}_1$. Next, $\mathbf{x}_{12}$, $\mathbf{x}_{11}$, and $\mathbf{x}_{10}$ are decoded successively. For reliable decoding, the code rates should satisfy
\begin{eqnarray*}
&&R_{11}\leq T_{11}' =\halflog\left(c_{11}+\frac{(1-\alpha_0-\alpha_1-\alpha_2)P}{(\alpha_0+\alpha_1+\alpha_2)P+N_1}\right)\\
&&R_3\ \leq T_3'\ =\halflog\left(c_3+\frac{P}{(\alpha_0+\alpha_1+\alpha_2)P+N_1}\right)\\
&&R_{12}\leq T_{12}' =\halflog\left(1+\frac{\alpha_1 P}{(\alpha_0+\alpha_2) P+N_1}\right)\\
&&R_{11}\leq T_{11}'' =\halflog\left(1+\frac{\alpha_2 P}{\alpha_0 P+N_1}\right)\\
&&R_{10}\leq T_{10} =\halflog\left(1+\frac{\alpha_0 P}{N_1}\right)
\end{eqnarray*}
where $c_{11}=\frac{(1-\alpha_0-\alpha_1-\alpha_2)P}{(1-\alpha_0-\alpha_1-\alpha_2)P+P}=\frac{1-\alpha_0-\alpha_1-\alpha_2}{2-\alpha_0-\alpha_1-\alpha_2}$ and $c_3=\frac{P}{(1-\alpha_0-\alpha_1-\alpha_2)P+P}=\frac{1}{2-\alpha_0-\alpha_1-\alpha_2}$. Note that $0\leq c_{11}\leq \frac{1}{2}$, $c_{11}+c_3=1$, and $\frac{1}{2}\leq c_3\leq 1$.
\item At receiver 2, $\mathbf{x}_{11}'$ is first decoded while treating other signals as noise. Having successfully recovered $M_{11}$, receiver 2 can generate $\mathbf{x}_{11}$ and $\mathbf{x}_{11}'$, and cancel them from $\mathbf{y}_2$. Next, $\mathbf{x}_{12}$ is decoded from $\mathbf{x}_{12}+\mathbf{x}_{2}+\mathbf{x}_{10}+\mathbf{z}_2$. Finally, $\mathbf{x}_2$ is decoded from $\mathbf{x}_2+\mathbf{x}_{10}+\mathbf{z}_2$. For reliable decoding, the code rates should satisfy
\begin{eqnarray*}
&& R_{11}\leq T_{11}''' =\halflog\left(1+\frac{(1-\alpha_0-\alpha_1-\alpha_2)P}{(\alpha_0+\alpha_1+2\alpha_2)P+N_2}\right)\\
&& R_{12}\leq T_{12}'' =\halflog\left(1+\frac{\alpha_1 P}{(\alpha_0+\alpha_2)P+N_2}\right)\\
&& R_2\ \leq T_2\ =\halflog\left(1+\frac{\alpha_2 P}{\alpha_0 P+N_2}\right).
\end{eqnarray*}
\item At receiver 3, $\mathbf{x}_3$ is decoded while treating $\mathbf{x}_2+\mathbf{z}_3$ as noise. Reliable decoding is possible if
\begin{eqnarray}
&&R_3\ \leq T_3'' =\halflog\left(1+\frac{P}{\alpha_2 P+N_3}\right).
\end{eqnarray}
\end{itemize}
Putting together, we can see that given $\alpha_0,\alpha_1,\alpha_2\in[0,1]$, the following rate region is achievable.
\begin{eqnarray*}
&& R_1 \leq T_1=\min\{T_{11}',T_{11}'',T_{11}'''\}+\min\{T_{12}',T_{12}''\}+T_{10}\\
&& R_2 \leq T_2\\
&& R_3 \leq T_3=\min\{T_3',T_3''\}
\end{eqnarray*}
where
\begin{eqnarray*}
&& T_1 = \min\{T_{11}',T_{11}'',T_{11}'''\}+\min\{T_{12}',T_{12}''\}+T_{10}\\
&&\ \ \ \ = \min\{\min\{T_{11}',T_{11}'''\},T_{11}''\}+T_{12}''+T_{10}\\
&&\ \ \ \ \geq \min\left\{\halflog\left(c_{11}+\frac{(1-\alpha_0-\alpha_1-\alpha_2)P}{(\alpha_0+\alpha_1+2\alpha_2)P+N_2}\right),\right.\\
&&\ \ \ \ \ \ \ \ \ \ \ \ \ \ \ \ \ \ \ \ \ \ \ \ \ \ \ \ \ \ \ \left.\halflog\left(1+\frac{\alpha_2 P}{\alpha_0 P+N_1}\right)\right\}\\
&&\ \ \ \ \ \ \ \ +\halflog\left(1+\frac{\alpha_1 P}{(\alpha_0+\alpha_2) P+N_2}\right)\\
&&\ \ \ \ \ \ \ \ +\halflog\left(1+\frac{\alpha_0 P}{N_1}\right)\\
&& T_2 =\halflog\left(1+\frac{\alpha_2 P}{\alpha_0 P+N_2}\right)\\
&& T_3 \geq \halflog\left(c_3+\frac{P}{(\alpha_0+\alpha_1+\alpha_2) P+N_3}\right).
\end{eqnarray*}

\subsection{The Gap for Relatively Large $R_1$}
We choose $\alpha_0$, $\alpha_1$ and $\alpha_2$ such that $\alpha_1\leq \frac{3}{8}$, that  $\alpha_1\geq 3(\alpha_0+\alpha_2)$, that $\alpha_2 P\geq 3N_3$, and that $\alpha_0 P=N_2$. It follows that $\alpha_0+\alpha_1+\alpha_2\leq \frac{4}{3}\alpha_1\leq  \frac{1}{2}$, that $c_{11}\geq \frac{1}{3}$, and that $(\alpha_0+\alpha_1+2\alpha_2)P+N_2 = 2(\alpha_0+\alpha_2)P+\alpha_1 P\leq \frac{5}{3}\alpha_1 P$. We get the lower bounds for each term of $T_1$ expression above.
\begin{eqnarray*}
&& \min\{T_{11}',T_{11}'''\}\\
&&\geq\halflog\left(c_{11}+\frac{(1-\alpha_0-\alpha_1-\alpha_2)P}{(\alpha_0+\alpha_1+2\alpha_2)P+N_2}\right)\\
&&\geq\halflog\left(\frac{1}{3}+\frac{(1-(4/3)\alpha_1)P}{(5/3)\alpha_1 P}\right)\\
&&= \halflog\left(\frac{P}{(5/3)\alpha_1 P}-\frac{7}{15}\right)\\
&&= \halflog\left(\frac{P}{(5/3)\alpha_1 P}\right)+\halflog\left(1-\frac{7}{15}\cdot\frac{5}{3}\alpha_1\right)\\
&&\geq \halflog\left(\frac{P}{(5/3)\alpha_1 P}\right)+\halflog\left(\frac{17}{24}\right)\\
&&\geq \halflog\left(\frac{P}{\alpha_1 P}\cdot\frac{17}{40}\right)
\end{eqnarray*}
and
\begin{eqnarray}
&& T_{11}''=\halflog\left(1+\frac{\alpha_2 P}{\alpha_0 P+N_1}\right)\\
&&\ \ \ \ \ = \halflog\left(\frac{(\alpha_0+\alpha_2) P+N_1}{\alpha_0 P+N_1}\right)\\
&&\ \ \ \ \ \geq \halflog\left(\frac{(\alpha_0+\alpha_2) P}{\alpha_0 P+N_2}\right)\\
&&\ \ \ \ \ =\halflog\left(\frac{(\alpha_0+\alpha_2) P}{2 N_2}\right).
\end{eqnarray}
Since $(\alpha_0+\alpha_2)P\geq N_2+3N_3\geq 4 N_2$,
\begin{eqnarray}
&&T_{12}'' = \halflog\left(1+\frac{\alpha_1 P}{(\alpha_0+\alpha_2)P+N_2}\right)\\
&& \ \ \ \ \ \geq \halflog\left(\frac{\alpha_1 P}{(5/4)(\alpha_0+\alpha_2)P}\right).
\end{eqnarray}
Putting together,
\begin{eqnarray*}
&& T_1 \geq \min\left\{\halflog\left(\frac{P}{\alpha_1 P}\cdot\frac{17}{40}\right),\halflog\left(\frac{(\alpha_0+\alpha_2) P}{2 N_2}\right)\right\}\\
&&\ \ \ \ \ \ \ +\halflog\left(\frac{\alpha_1 P}{(5/4)(\alpha_0+\alpha_2) P}\right)+\halflog\left(\frac{N_2}{N_1}\right)\\
&&\ \ \ \ = \min\left\{\halflog\left(\frac{P}{(\alpha_0+\alpha_2) P}\cdot\frac{N_2}{N_1}\cdot\frac{17}{40}\cdot\frac{4}{5}\right),\right.\\
&&\ \ \ \ \ \ \ \ \ \ \ \ \ \ \ \ \ \ \ \ \ \ \ \ \ \ \ \ \ \ \ \ \ \ \ \ \left.\halflog\left(\frac{\alpha_1 P}{ N_1}\cdot\frac{1}{2}\cdot\frac{4}{5}\right)\right\}\\
&&\ \ \ \ = \min\left\{\halflog\left(\frac{P}{(\alpha_0+\alpha_2) P}\cdot\frac{N_2}{N_1}\cdot\frac{17}{50}\right),\right.\\
&&\ \ \ \ \ \ \ \ \ \ \ \ \ \ \ \ \ \ \ \ \ \ \ \ \ \ \ \ \ \ \ \ \ \ \ \ \left.\halflog\left(\frac{\alpha_1 P}{ N_1}\cdot\frac{2}{5}\right)\right\}.
\end{eqnarray*}
Given $\alpha_1$, we choose $\alpha_2$ that satisfies $\halflog\left(\frac{P}{\alpha_1 P}\cdot\frac{17}{40}\right)=\halflog\left(\frac{(\alpha_0+\alpha_2) P}{2N_2}\right)$. As a result, we can write $T_1\geq \halflog\left(\frac{\alpha_1 P}{ N_1}\cdot\frac{2}{5}\right)$, and also
\begin{eqnarray}
&& T_2 =\halflog\left(1+\frac{\alpha_2 P}{\alpha_0 P+N_2}\right)\\
&& \ \ \ \ \geq\halflog\left(\frac{(\alpha_0+\alpha_2) P}{2 N_2}\right)\\
&& \ \ \ \ =\halflog\left(\frac{P}{\alpha_1 P}\cdot\frac{17}{40}\right).
\end{eqnarray}
Since $N_3\leq \frac{1}{3}\alpha_2 P\leq\frac{1}{3}(\alpha_0+\alpha_2)P \leq \frac{1}{9}\alpha_1 P$,
\begin{eqnarray*}
&& T_3 \geq \halflog\left(c_3+\frac{P}{(\alpha_0+\alpha_1+\alpha_2) P+N_3}\right)\\
&& \ \ \ \ \geq \halflog\left(\frac{1}{2}+\frac{P}{(4/3)\alpha_1 P+(1/9)\alpha_1 P}\right)\\
&& \ \ \ \ \geq \halflog\left(\frac{P}{(13/9)\alpha_1 P}\right).
\end{eqnarray*}
The following rate region is achievable.
\begin{eqnarray}
&& R_1\leq \halflog\left(\frac{\alpha_1 P}{N_1}\cdot\frac{2}{5}\right)\\
&& R_2\leq \halflog\left(\frac{P}{\alpha_1 P}\cdot\frac{17}{40}\right)\\
&& R_3\leq \halflog\left(\frac{P}{\alpha_1 P}\cdot\frac{9}{13}\right).
\end{eqnarray}
For fixed $\alpha_1$ and $R_1=\halflog\left(\frac{\alpha_1 P}{N_1}\cdot\frac{2}{5}\right)$, the two-dimensional rate region, given by
\begin{eqnarray*}
&& R_2\leq \halflog\left(\frac{P}{\alpha_1 P}\cdot\frac{17}{40}\right),\ R_3\leq \halflog\left(\frac{P}{\alpha_1 P}\cdot\frac{9}{13}\right)
\end{eqnarray*}
is achievable.

In comparison, the two-dimensional outer bound region at $R_1=\halflog\left(\frac{\alpha_1 P}{N_1}\cdot\frac{2}{5}\right)+1$, given by
\begin{eqnarray*}
&&R_2 \leq \halflog\left(\frac{P}{N_1}\cdot\frac{7}{3}\right)-\halflog\left(\frac{\alpha_1 P}{N_1}\cdot\frac{2}{5}\right)-1\\
&&\ \ \ \ = \halflog\left(\frac{P}{\alpha_1 P}\right)+\halflog\left(\frac{7}{3}\cdot\frac{5}{2}\cdot\frac{1}{4}\right)\\
&&R_3 \leq \halflog\left(\frac{P}{N_1}\cdot\frac{7}{3}\right)-\halflog\left(\frac{\alpha_1 P}{N_1}\cdot\frac{2}{5}\right)-1\\
&&\ \ \ \ = \halflog\left(\frac{P}{\alpha_1 P}\right)+\halflog\left(\frac{7}{3}\cdot\frac{5}{2}\cdot\frac{1}{4}\right).
\end{eqnarray*}
As discussed above, the sum-rate bound on $R_2+R_3$ is loose for $R_1$ larger than the threshold, so the rate region is a rectangle.
By comparing the inner and outer bound rate regions, we can see that $\delta_2< \halflog\left(\frac{40}{17}\cdot\frac{7}{3}\cdot\frac{5}{2}\cdot\frac{1}{4}\right) < 0.89$ and $\delta_3< \halflog\left(\frac{13}{9}\cdot\frac{7}{3}\cdot\frac{5}{2}\cdot\frac{1}{4}\right) < 0.54$. Therefore, we can conclude that the gap is to within one bit per message.

\subsection{Achievable Scheme for Relatively Small $R_1$}
\begin{theorem}
Given $\alpha=(\alpha_0,\alpha_1,\alpha_2) \in [0,1]^3$, the region $\mathcal{R}_\alpha$ is defined by
\begin{eqnarray*}
&& R_1 \leq \min\left\{\halflog^+\left(c_{11}+\frac{(1-\alpha_1)P}{(\alpha_1+\alpha_2) P+N_2}\right),\right.\\
&&\ \ \ \ \ \ \ \left.\halflog\left(1+\frac{(\alpha_1-\alpha_0) P}{\alpha_0 P+N_1}\right)\right\}+\halflog\left(1+\frac{\alpha_0 P}{N_1}\right)\\
&& R_2 \leq\halflog\left(1+\frac{\alpha_2 P}{\alpha_0 P+N_2}\right)\\
&& R_3 \leq \halflog^+\left(c_3+\frac{P}{\max\{\alpha_1,\alpha_2\} P+N_3}\right)
\end{eqnarray*}
where $c_{11}=\frac{1-\alpha_1}{2-\alpha_1}$ and $c_3=\frac{1}{2-\alpha_1}$,
and $\mathcal{R}=\textsc{conv}\left(\bigcup_{\alpha}\mathcal{R}_\alpha\right)$ is achievable.
\end{theorem}

For the case of $R_1 < R_{1,th}$, we present the following achievable scheme. At transmitter 1, we split $M_1$ into $M_{10}$ and $M_{11}$, so $R_{1}=R_{10}+R_{11}$. The transmit signal is the sum
\[\mathbf{x}_1=\mathbf{x}_{10}+\mathbf{x}_{11}+\mathbf{x}_{11}'.\]

The power allocation satisfies $\mathbb{E}[\|\mathbf{x}_{10}\|^2]=\alpha_0 nP$, $\mathbb{E}[\|\mathbf{x}_{11}\|^2]=(\alpha_1-\alpha_0) nP$, and $\mathbb{E}[\|\mathbf{x}_{11}'\|^2]=(1-\alpha_1)nP$ at receiver 1, $\mathbb{E}[\|\mathbf{x}_2\|^2]=\alpha_2 nP$ at receiver 2, and $\mathbb{E}[\|\mathbf{x}_3\|^2]=nP$ at receiver 3.

The signals $\mathbf{x}_{11}'$ and $\mathbf{x}_3$ are lattice codewords using the same coding lattice but different shaping lattices. As a result, the sum $\mathbf{x}_{11}'+\mathbf{x}_3$ is a lattice codeword.

The received signals are
\begin{eqnarray*}
&&\mathbf{y}_1 = [\mathbf{x}_{11}'+\mathbf{x}_3]+\mathbf{x}_{11}+\mathbf{x}_{10}+\mathbf{z}_1\\
&&\mathbf{y}_2 = \mathbf{x}_{11}'+\mathbf{x}_{11}+\mathbf{x}_2+\mathbf{x}_{10}+\mathbf{z}_2\\
&&\mathbf{y}_3 = \mathbf{x}_3+\mathbf{x}_2+\mathbf{z}_3.
\end{eqnarray*}
The signal scale diagram at each receiver is shown in Fig. \ref{fig:signalScale4} (b).
Decoding is performed in the following way.
\begin{itemize}
\item At receiver 1, $[\mathbf{x}_{11}'+\mathbf{x}_3]$ is first decoded while treating other signals as noise and removed from $\mathbf{y}_1$. Next, $\mathbf{x}_{11}$ and then $\mathbf{x}_{10}$ is decoded successively. For reliable decoding, the code rates should satisfy
\begin{eqnarray*}
&&R_{11}\leq T_{11}' =\halflog\left(c_{11}+\frac{(1-\alpha_1)P}{\alpha_1 P+N_1}\right)\\
&&R_3\ \leq T_3'\ =\halflog\left(c_3+\frac{P}{\alpha_1 P+N_1}\right)\\
&&R_{11}\leq T_{11}'' =\halflog\left(1+\frac{(\alpha_1-\alpha_0) P}{\alpha_0 P+N_1}\right)\\
&&R_{10}\leq T_{10} =\halflog\left(1+\frac{\alpha_0 P}{N_1}\right)
\end{eqnarray*}
where $c_{11}=\frac{(1-\alpha_1)P}{(1-\alpha_1)P+P}=\frac{1-\alpha_1}{2-\alpha_1}$ and $c_3=\frac{P}{(1-\alpha_1)P+P}=\frac{1}{2-\alpha_1}$. Note that $0\leq c_{11}\leq \frac{1}{2}$, $c_{11}+c_3=1$, and $\frac{1}{2}\leq c_3\leq 1$.
\item At receiver 2, $\mathbf{x}_{11}'$
is first decoded while treating other signals as noise. Having successfully recovered $M_{11}$, receiver 1 can generate $\mathbf{x}_{11}$ and $\mathbf{x}_{11}'$, and cancel them from $\mathbf{y}_2$. Next, $\mathbf{x}_2$ is decoded from $\mathbf{x}_2+\mathbf{x}_{10}+\mathbf{z}_2$. At receiver 2, $\mathbf{x}_{10}$ is not decoded. For reliable decoding, the code rates should satisfy
\begin{eqnarray*}
&& R_{11}\leq T_{11}''' =\halflog\left(1+\frac{(1-\alpha_1)P}{(\alpha_1+\alpha_2) P+N_2}\right)\\
&& R_2\ \leq T_2\ =\halflog\left(1+\frac{\alpha_2 P}{\alpha_0 P+N_2}\right).
\end{eqnarray*}
\item At receiver 3, $\mathbf{x}_3$ is decoded while treating $\mathbf{x}_2+\mathbf{z}_3$ as noise. Reliable decoding is possible if
\begin{eqnarray}
&&R_3\ \leq T_3'' =\halflog\left(1+\frac{P}{\alpha_2 P+N_3}\right).
\end{eqnarray}
\end{itemize}

Putting together, we can see that given $\alpha_0,\alpha_1\alpha_2\in[0,1]$, the following rate region is achievable.
\begin{eqnarray}
&& R_1 \leq T_1 =\min\{T_{11}',T_{11}'',T_{11}'''\}+T_{10}\\
&& R_2 \leq T_2 \\
&& R_3 \leq T_3 = \min\{T_3',T_3''\}
\end{eqnarray}
where
\begin{eqnarray*}
&& T_1 = \min\{T_{11}',T_{11}'',T_{11}'''\}+T_{10}\\
&&\ \ \ \ = \min\{\min\{T_{11}',T_{11}'''\},T_{11}''\}+T_{10}\\
&&\ \ \ \ \geq \min\left\{\halflog\left(c_{11}+\frac{(1-\alpha_1)P}{(\alpha_1+\alpha_2) P+N_2}\right),\right.\\
&&\ \ \ \ \ \ \ \left.\halflog\left(1+\frac{(\alpha_1-\alpha_0) P}{\alpha_0 P+N_1}\right)\right\}+\halflog\left(1+\frac{\alpha_0 P}{N_1}\right)\\
&& T_2 =\halflog\left(1+\frac{\alpha_2 P}{\alpha_0 P+N_2}\right)\\
&& T_3 \geq \halflog\left(c_3+\frac{P}{\max\{\alpha_1,\alpha_2\} P+N_3}\right).
\end{eqnarray*}

\begin{figure}[tp]
  \begin{center}
    \mbox{
      \subfigure[Channel type 4: small $R_1$]{\includegraphics[width=0.2\textwidth]{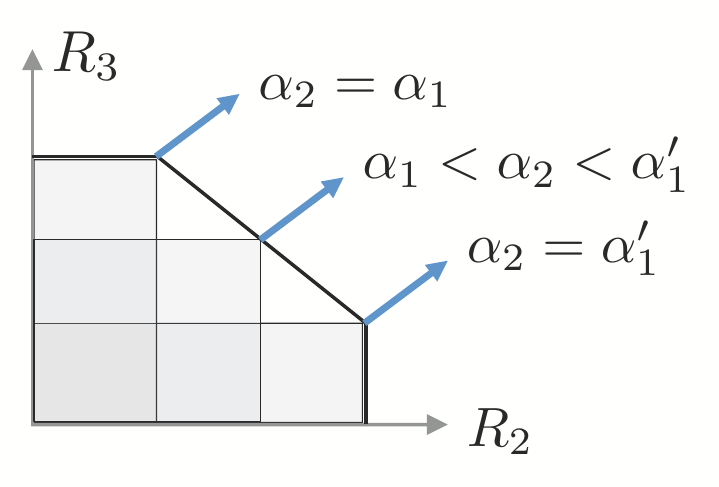}}
      }
    \mbox{
      \subfigure[Channel type 5: small $R_2$]{\includegraphics[width=0.2\textwidth]{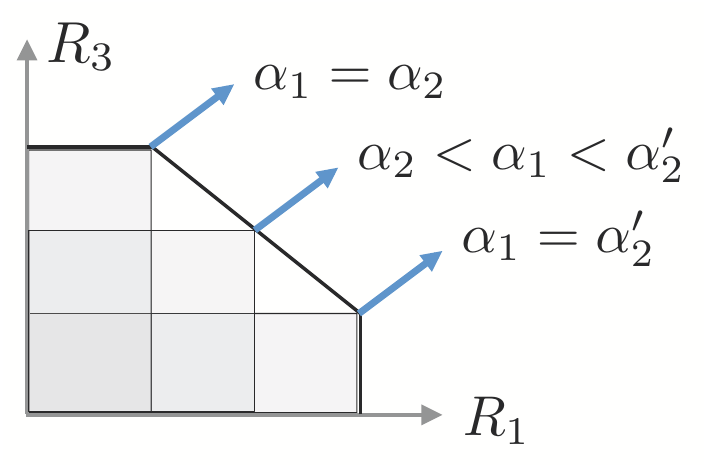}}
      }
\caption{MAC-like region.}
\label{fig:MAClike}
  \end{center}
\end{figure}

\subsection{The Gap for Relatively Small $R_1$}
We choose $\alpha_0$, $\alpha_1$, and $\alpha_2$ such that $\alpha_1 \leq \alpha_2\leq \frac{1}{2}$, that $\alpha_1 P\geq 3N_2$, that $\alpha_2 P\geq 3N_3$, and that $\alpha_0 P=\frac{4}{5} N_2$. It follows that $c_{11}\geq \frac{1}{3}$ and that $(\alpha_1+\alpha_2)P+N_2\leq \frac{4}{3}\alpha_1 P+\alpha_2P\leq \frac{7}{3}\alpha_2 P$.
\begin{eqnarray*}
&&\min\{T_{11}',T_{11}'''\}\\
&&=\halflog\left(c_{11}+\frac{(1-\alpha_1)P}{(\alpha_1+\alpha_2)P+N_2}\right)\\
&&\geq \halflog\left(\frac{1}{3}+\frac{(1-\alpha_2)P}{(7/3)\alpha_2 P}\right)\\
&&=\halflog\left(\frac{P}{(7/3)\alpha_2 P}-\frac{2}{21}\right)\\
&&=\halflog\left(\frac{P}{(7/3)\alpha_2 P}\right)+\halflog\left(1-\frac{2}{21}\cdot\frac{7}{3}\alpha_2\right)\\
&&\geq \halflog\left(\frac{P}{(7/3)\alpha_2 P}\right)+\halflog\left(\frac{8}{9}\right)\\
&&\geq \halflog\left(\frac{P}{\alpha_2 P}\cdot\frac{8}{21}\right)
\end{eqnarray*}
and
\begin{eqnarray}
&& T_{11}''=\halflog\left(1+\frac{(\alpha_1-\alpha_0) P}{\alpha_0 P+N_1}\right)\\
&& \ \ \ \ \ = \halflog\left(\frac{\alpha_1 P+N_1}{\alpha_0 P+N_1}\right)\\
&& \ \ \ \ \ \geq \halflog\left(\frac{\alpha_1 P}{\alpha_0 P+N_2}\right)\\
&& \ \ \ \ \ =\halflog\left(\frac{\alpha_1 P}{(9/5)N_2}\right).
\end{eqnarray}
Putting together,
\begin{eqnarray*}
&& T_1 \geq \min\left\{\halflog\left(\frac{P}{\alpha_2 P}\cdot\frac{8}{21}\right),\halflog\left(\frac{\alpha_1 P}{(9/5) N_2}\right) \right\}\\
&&\ \ \ \ \ \ \ +\halflog\left(\frac{N_2}{N_1}\cdot\frac{4}{5}\right).
\end{eqnarray*}
Let us define $\alpha_1'$ by the equality $\halflog\left(\frac{P}{\alpha_1' P}\cdot\frac{8}{21}\right)=\halflog\left(\frac{\alpha_1 P}{(9/5) N_2}\right)$. If we choose $\alpha_2\leq \alpha_1'$, then $\halflog\left(\frac{P}{\alpha_2 P}\cdot\frac{8}{21}\right)\geq \halflog\left(\frac{\alpha_1 P}{(9/5) N_2}\right)$, and
\begin{eqnarray*}
&& T_1 \geq \halflog\left(\frac{\alpha_1 P}{(9/5)N_2}\cdot\frac{N_2}{N_1}\cdot\frac{4}{5}\right)=\halflog\left(\frac{\alpha_1 P}{N_1}\cdot\frac{4}{9}\right).
\end{eqnarray*}
We can see that the following rate region is achievable.
\begin{eqnarray}
&& R_1\leq \halflog\left(\frac{\alpha_1 P}{N_1}\cdot\frac{4}{9}\right)\\
&& R_2\leq \halflog\left(\frac{\alpha_2 P}{(9/5)N_2}\right)\\
&& R_3\leq \halflog\left(\frac{P}{(4/3)\alpha_2 P}\right).
\end{eqnarray}
For fixed $\alpha_2\in [\alpha_1,\alpha_1']$ and $R_1=\halflog\left(\frac{\alpha_1 P}{N_1}\cdot\frac{4}{9}\right)$, the two-dimensional rate region $\mathcal{R}_\alpha$, given by
\begin{eqnarray}
&& R_2\leq \halflog\left(\frac{\alpha_2 P}{(9/5)N_2}\right)\\
&& R_3\leq \halflog\left(\frac{P}{(4/3)\alpha_2 P}\right)
\end{eqnarray}
is achievable. The union $\bigcup_{\alpha_2\in [\alpha_1,\alpha_1']}\mathcal{R}_\alpha$ is a MAC-like region, given by
\begin{eqnarray}
&& \ \ \ \ \ \ \ R_2\leq \halflog\left(\frac{\alpha_1' P}{(9/5)N_2}\right)\\
&& \ \ \ \ \ \ \ \ \ \ \ \leq \halflog\left(\frac{P}{\alpha_1 P}\cdot\frac{8}{21}\right)\\
&& \ \ \ \ \ \ \ R_3\leq \halflog\left(\frac{P}{\alpha_1 P}\cdot\frac{3}{4}\right)\\
&& R_2+R_3 \leq \halflog\left(\frac{\alpha_2 P}{(9/5)N_2}\cdot\frac{P}{(4/3)\alpha_2 P}\right)\\
&& \ \ \ \ \ \ \ \ \ \ \ \leq \halflog\left(\frac{P}{N_2}\cdot\frac{15}{36}\right).
\end{eqnarray}
This region is described in Fig. \ref{fig:MAClike} (a).

In comparison, the two-dimensional outer bound region at $R_1=\halflog\left(\frac{\alpha_1 P}{N_1}\cdot\frac{4}{9}\right)+1$, given by
\begin{eqnarray*}
&&\ \ \ \ \ \ \ R_2 \leq \halflog\left(\frac{P}{N_1}\cdot\frac{7}{3}\right)-\halflog\left(\frac{\alpha_1 P}{N_1}\cdot\frac{4}{9}\right)-1\\
&&\ \ \ \ \ \ \ \ \ \ \ = \halflog\left(\frac{P}{\alpha_1 P}\right)+\halflog\left(\frac{7}{3}\cdot\
\frac{9}{4}\cdot\frac{1}{4}\right)\\
&&\ \ \ \ \ \ \ R_3 \leq \halflog\left(\frac{P}{N_1}\cdot\frac{7}{3}\right)-\halflog\left(\frac{\alpha_1 P}{N_1}\cdot\frac{4}{9}\right)-1\\
&&\ \ \ \ \ \ \ \ \ \ \ = \halflog\left(\frac{P}{\alpha_1 P}\right)+\halflog\left(\frac{7}{3}\cdot \frac{9}{4}\cdot\frac{1}{4}\right)\\
&&R_2+R_3 \leq \halflog\left(\frac{P}{N_2}\right)+\halflog\left(\frac{7}{3}\right).
\end{eqnarray*}
Since $\delta_2 < \halflog\left(\frac{21}{8}\cdot\frac{7}{3}\cdot \frac{9}{4}\cdot\frac{1}{4}\right) < 0.90$, $\delta_3 < \halflog\left(\frac{4}{3}\cdot \frac{7}{3}\cdot \frac{9}{4}\cdot\frac{1}{4}\right) < 0.41$ and $\delta_{23} < \halflog\left(\frac{36}{15}\cdot\frac{7}{3}\right) < 1.25 < \sqrt{2}$, we can conclude that the gap is to within one bit per message.

\section{Inner Bound: Channel Type 5}

\begin{figure}[t]
  \begin{center}
    \mbox{
      \subfigure[Large $R_2$]{\includegraphics[width=0.2\textwidth]{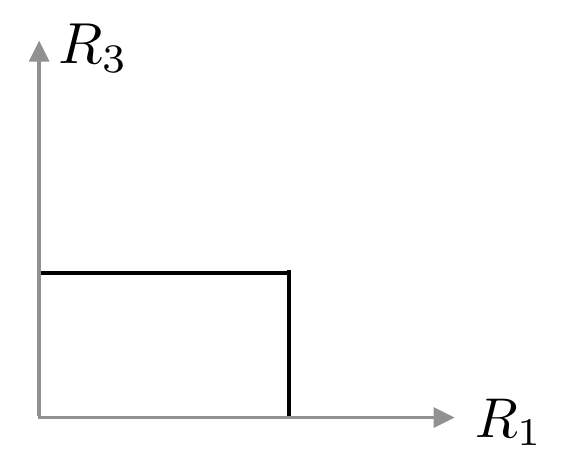}}
      }
    \mbox{
      \subfigure[Small $R_2$]{\includegraphics[width=0.2\textwidth]{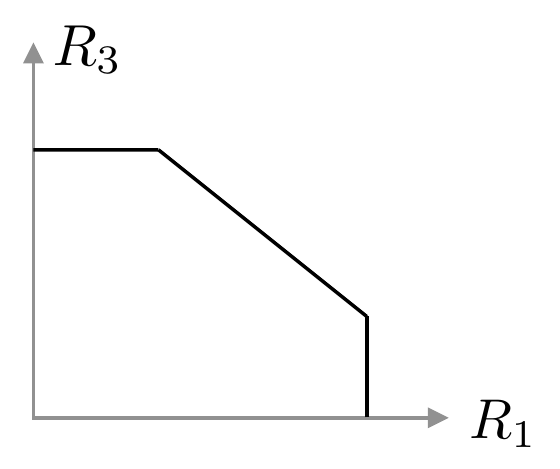}}
      }
\caption{The cross-section of the type 5 outer bound region at a relatively small or large $R_2$.}
\label{fig:outerBoundCrossSection5}
  \end{center}
\end{figure}

Let us consider the relaxed outer bound region $\mathcal{R}_o'$ given by
\begin{eqnarray*}
&&\ \ \ \ \ \ \ R_k \leq \halflog\left(\frac{P}{N_k}\right)+\halflog\left(\frac{4}{3}\right),\ k=1,2,3\\
&&R_1+R_2 \leq \halflog\left(\frac{P}{N_1}\right)+\halflog\left(\frac{7}{3}\right)\\
&&R_2+R_3 \leq \halflog\left(\frac{P}{N_2}\right)+\halflog\left(\frac{7}{3}\right)\\
&&R_1+R_3 \leq \halflog\left(\frac{P}{N_1}\right)+\halflog\left(\frac{7}{3}\right).
\end{eqnarray*}
The cross-sectional region at a given $R_2$ is described by
\begin{eqnarray*}
&& R_1\leq \min\left\{\halflog\left(\frac{P}{N_1}\cdot\frac{7}{3}\right)-R_2,\halflog\left(\frac{P}{N_1}\cdot\frac{4}{3}\right)\right\}\\
&& R_3\leq \min\left\{\halflog\left(\frac{P}{N_2}\cdot\frac{7}{3}\right)-R_2,\halflog\left(\frac{P}{N_3}\cdot\frac{4}{3}\right)\right\}\\
&& R_1+R_3 \leq \halflog\left(\frac{P}{N_1}\cdot\frac{7}{3}\right).
\end{eqnarray*}
Depending on the bottleneck of $\min\{\cdot,\cdot\}$ expressions, there are three cases:
\begin{itemize}
\item $R_2\leq \halflog\left(\frac{7}{4}\right)$
\item $\halflog\left(\frac{7}{4}\right)\leq R_2\leq \halflog\left(\frac{N_3}{N_2}\cdot\frac{7}{4}\right)$
\item $R_2\geq \halflog\left(\frac{N_3}{N_2}\cdot\frac{7}{4}\right)$.
\end{itemize}
In this section, we focus on the third case. The other cases can be proved similarly.
If the sum of the righthand sides of $R_1$ and $R_3$ bounds is smaller than the righthand side of $R_1+R_3$ bound, i.e.,
\begin{eqnarray*}
 \halflog\left(\frac{P}{N_1}\cdot\frac{7}{3}\right)+\halflog\left(\frac{P}{N_2}\cdot\frac{7}{3}\right)-2R_2 \leq \halflog\left(\frac{P}{N_1}\cdot\frac{7}{3}\right),
\end{eqnarray*}
then the $R_1+R_3$ bound is not active at the $R_2$.
By rearranging, the threshold condition is given by
\begin{eqnarray}
R_2 > R_{2,th}=\frac{1}{4}\log\left(\frac{P}{N_2}\cdot\frac{7}{3}\right).
\end{eqnarray}
Note that $R_{2,th}$ is roughly half of $C_2$.
For this relatively large $R_2$, the cross-sectional region is a rectangle as described in Fig. \ref{fig:outerBoundCrossSection5} (a). In contrast, for a relatively small $R_1$, when the threshold condition does not hold, the cross-sectional region is a MAC-like region as described in Fig. \ref{fig:outerBoundCrossSection5} (b). In the following subsections, we present achievable schemes for each case.

\subsection{Achievable Scheme for Relatively Large $R_2$}
\begin{theorem}
Given $\alpha=(\alpha_1,\alpha_2,\alpha_2') \in [0,1]^3$, the region $\mathcal{R}_\alpha$ is defined by
\begin{eqnarray*}
&& R_1 \leq \halflog\left(1+\frac{\alpha_1 P}{N_1}\right)\\
&& R_2 \leq \min\left\{\halflog^+\left(c_{21}+\frac{(1-\alpha_2-\alpha_2')P}{(\alpha_1+\alpha_2+\alpha_2')P+N_2}\right),\right.\\
&&\ \ \ \ \ \ \ \left.\halflog\left(1+\frac{\alpha_2' P}{N_2}\right)\right\}+\halflog\left(1+\frac{\alpha_2 P}{\alpha_2' P+N_2}\right) \nonumber\\
&& R_3 \leq \halflog^+\left(c_3+\frac{P}{\max\{\alpha_1,\alpha_2+\alpha_2'\} P+N_3}\right)
\end{eqnarray*}
where $c_{21}=\frac{1-\alpha_2-\alpha_2'}{2-\alpha_2-\alpha_2'}$ and $c_3=\frac{1}{2-\alpha_2-\alpha_2'}$,
and $\mathcal{R}=\textsc{conv}\left(\bigcup_{\alpha}\mathcal{R}_\alpha\right)$ is achievable.
\end{theorem}

We present an achievable scheme for the case of $R_2 > R_{2,th}$. Message $M_2\in\{1,2,\ldots,2^{nR_2}\}$ for receiver 2 is split into two parts: $M_{21}\in\{1,2,\ldots,2^{nR_{21}}\}$ and $M_{22}\in\{1,2,\ldots,2^{nR_{22}}\}$, so $R_2=R_{21}+R_{22}$. We generate the signals in the following way: $\mathbf{x}_{21}$ and $\mathbf{x}_{21}'$ are differently coded signals of $M_{21}$, and $\mathbf{x}_{22}$ is a coded signal of $M_{22}$. The transmit signal is the sum
\[\mathbf{x}_2=\mathbf{x}_{21}+\mathbf{x}_{22}+\mathbf{x}_{21}'.\]
The power allocation satisfies $\mathbb{E}[\|\mathbf{x}_1\|^2]=\alpha_1 nP$, at receiver 1, $\mathbb{E}[\|\mathbf{x}_{21}\|^2]=\alpha_2' nP$, $\mathbb{E}[\|\mathbf{x}_{22}\|^2]=\alpha_2 nP$, and $\mathbb{E}[\|\mathbf{x}_{21}'\|^2]=(1-\alpha_2-\alpha_2')P$ at receiver 2, and $\mathbb{E}[\|\mathbf{x}_3\|^2]=nP$ at receiver 3.

The signals $\mathbf{x}_{21}'$ and $\mathbf{x}_3$ are lattice codewords using the same coding lattice but different shaping lattices. As a result, the sum $\mathbf{x}_{21}'+\mathbf{x}_3$ is a lattice codeword.

\begin{figure}[tp]
  \begin{center}
    \mbox{
      \subfigure[Channel type 5: relatively large $R_2$]{\includegraphics[width=0.45\textwidth]{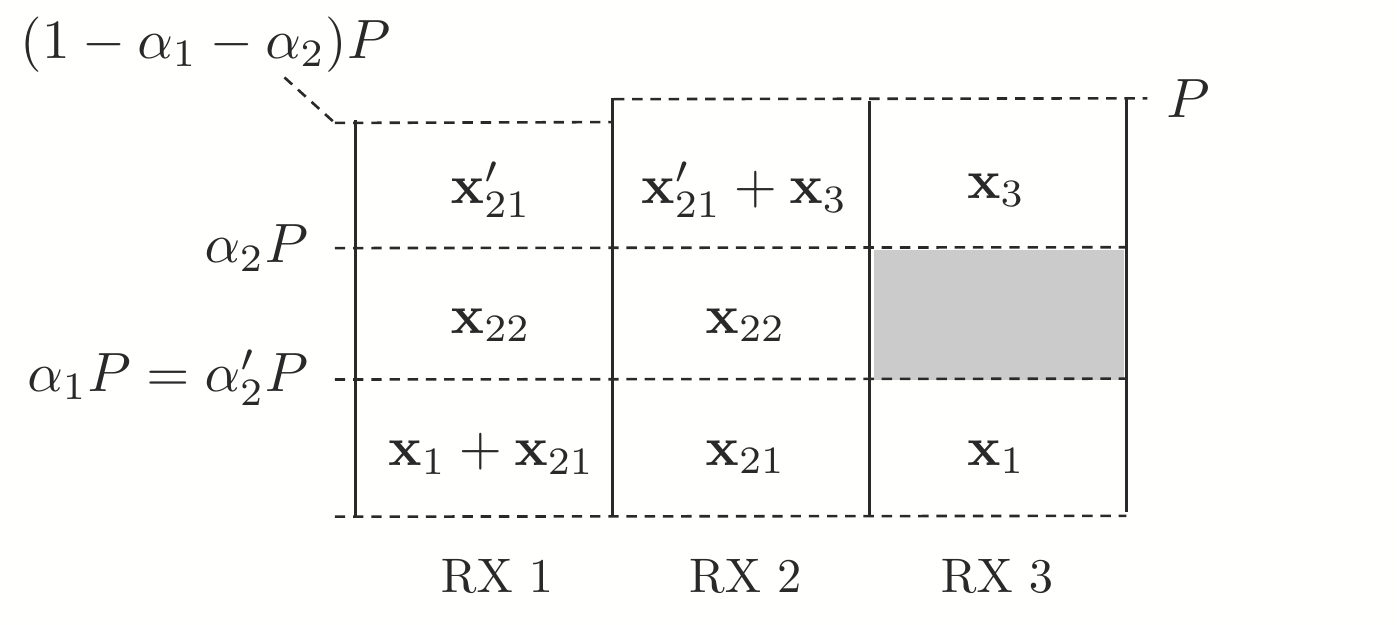}}
      }
    \mbox{
      \subfigure[Channel type 5: relatively small $R_2$]{\includegraphics[width=0.45\textwidth]{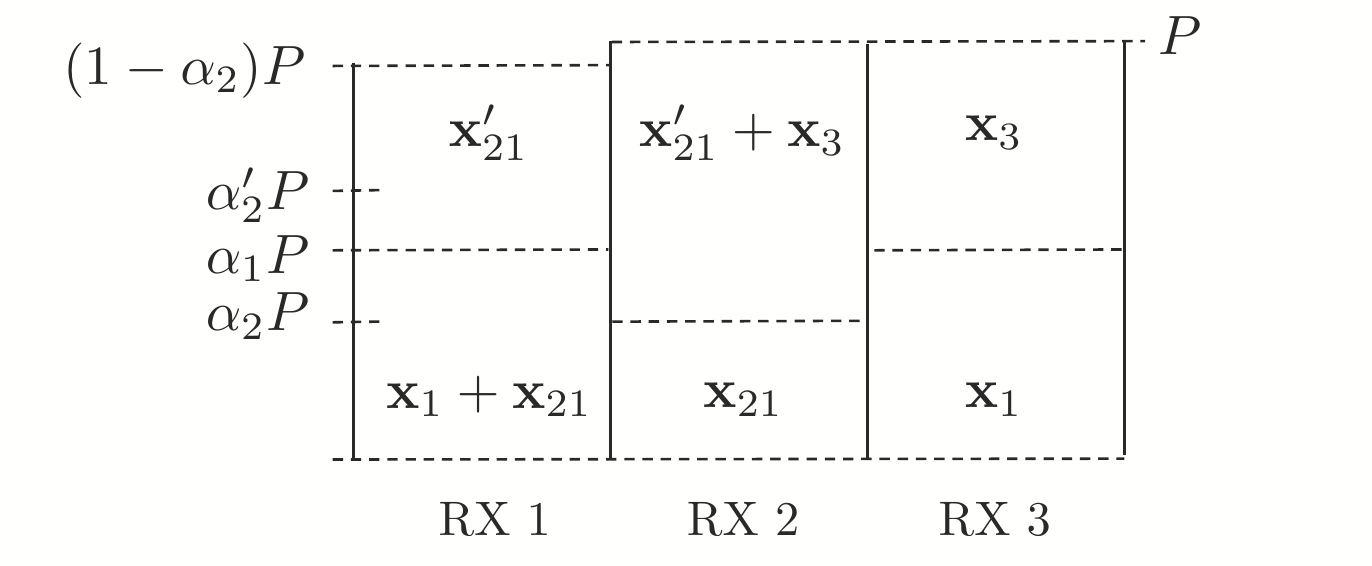}}
      }
\caption{Signal scale diagram.}
\label{fig:signalScale5}
  \end{center}
\end{figure}

The received signals are
\begin{eqnarray*}
&&\mathbf{y}_1=\mathbf{x}_{21}'+\mathbf{x}_{22}+\mathbf{x}_{21}+\mathbf{x}_1+\mathbf{z}_1\\
&&\mathbf{y}_2=[\mathbf{x}_{21}'+\mathbf{x}_3]+\mathbf{x}_{22}+\mathbf{x}_{21}+\mathbf{z}_2\\
&&\mathbf{y}_3=\mathbf{x}_3+\mathbf{x}_1+\mathbf{z}_3.
\end{eqnarray*}
The signal scale diagram at each receiver is shown in Fig. \ref{fig:signalScale5} (a). Decoding is performed in the following way.
\begin{itemize}
\item At receiver 1, $\mathbf{x}_{21}'$ is first decoded while treating other signals as noise. Having successfully recovered $M_{21}$, receiver 1 can generate $\mathbf{x}_{21}$ and $\mathbf{x}_{21}'$, and cancel them from $\mathbf{y}_1$. Next, $\mathbf{x}_{22}$ is decoded from $\mathbf{x}_{22}+\mathbf{x}_1+\mathbf{z}_1$. Finally, $\mathbf{x}_1$ is decoded from $\mathbf{x}_1+\mathbf{z}_1$. For reliable decoding, the code rates should satisfy
\begin{eqnarray*}
&& R_{21}\leq T_{21}' =\halflog\left(1+\frac{(1-\alpha_2-\alpha_2')P}{(\alpha_1+\alpha_2+\alpha_2')P+N_1}\right)\\
&& R_{22}\leq T_{22}' =\halflog\left(1+\frac{\alpha_2 P}{\alpha_1 P+N_1}\right)\\
&& R_1\ \leq T_1\ =\halflog\left(1+\frac{\alpha_1 P}{N_1}\right).
\end{eqnarray*}
\item At receiver 2, $[\mathbf{x}_{21}'+\mathbf{x}_3]$ first decoded while treating other signals as noise and removed from $\mathbf{y}_2$. Next, $\mathbf{x}_{22}$ and $\mathbf{x}_{21}$ are decoded successively. For reliable decoding, the code rates should satisfy
\begin{eqnarray*}
&&R_{21}\leq T_{21}'' =\halflog\left(c_{21}+\frac{(1-\alpha_2-\alpha_2')P}{(\alpha_2+\alpha_2')P+N_2}\right)\\
&&R_3\ \leq T_3'\ =\halflog\left(c_3+\frac{P}{(\alpha_2+\alpha_2')P+N_2}\right)\\
&&R_{22}\leq T_{22}'' =\halflog\left(1+\frac{\alpha_2 P}{\alpha_2' P+N_2}\right)\\
&&R_{21}\leq T_{21}''' =\halflog\left(1+\frac{\alpha_2' P}{N_2}\right)
\end{eqnarray*}
where $c_{21}=\frac{(1-\alpha_2-\alpha_2')P}{(1-\alpha_2-\alpha_2')P+P}=\frac{1-\alpha_2-\alpha_2'}{2-\alpha_2-\alpha_2'}$ and $c_3=\frac{P}{(1-\alpha_2-\alpha_2')P+P}=\frac{1}{2-\alpha_2-\alpha_2'}$. Note that $0\leq c_{21}\leq \frac{1}{2}$, $c_{21}+c_3=1$, and $\frac{1}{2}\leq c_3\leq 1$.
\item At receiver 3, $\mathbf{x}_3$ is decoded while treating $\mathbf{x}_1+\mathbf{z}_3$ as noise. Reliable decoding is possible if
\begin{eqnarray}
&&R_3\ \leq T_3'' =\halflog\left(1+\frac{P}{\alpha_1 P+N_3}\right).
\end{eqnarray}
\end{itemize}
Putting together, we can see that given $\alpha_1,\alpha_2,\alpha_2'\in[0,1]$, the following rate region is achievable.
\begin{eqnarray*}
&& R_1 \leq T_1\\
&& R_2 \leq T_2 =\min\{T_{21}',T_{21}'',T_{21}'''\}+\min\{T_{22}',T_{22}''\}\\
&& R_3 \leq T_3=\min\{T_3',T_3''\}
\end{eqnarray*}
where
\begin{eqnarray*}
&& T_1 =\halflog\left(1+\frac{\alpha_1 P}{N_1}\right)\\
&& T_2 = \min\{T_{21}',T_{21}'',T_{21}'''\}+T_{22}''\\
&&\ \ \ \ = \min\{\min\{T_{21}',T_{21}''\},T_{21}'''\}+T_{22}''\\
&&\ \ \ \ \geq \min\left\{\halflog\left(c_{21}+\frac{(1-\alpha_2-\alpha_2')P}{(\alpha_1+\alpha_2+\alpha_2')P+N_2}\right),\right.\\
&&\ \ \ \ \ \ \ \left.\halflog\left(1+\frac{\alpha_2' P}{N_2}\right)\right\}+\halflog\left(1+\frac{\alpha_2 P}{\alpha_2' P+N_2}\right) \nonumber\\
%&&\ \ \ \ \geq \min\left\{\halflog\left(c_{21}+\frac{(1-\alpha_2-\alpha_2')P}{(\alpha_1+\alpha_2+\alpha_2') P+N_2}\right)\right.\\
%&&\ \ \ \ \ \ \ \ \ \ \ \ \ \ \ +\halflog\left(1+\frac{\alpha_2 P}{\alpha_2' P+N_2}\right),\\
%&&\ \ \ \ \ \ \ \ \ \ \ \ \ \ \ \ \ \ \ \ \ \ \ \ \ \ \ \ \ \ \left.\halflog\left(1+\frac{(\alpha_2+\alpha_2' )P}{N_2}\right)\right\}\\
&& T_3 \geq \halflog\left(c_3+\frac{P}{\max\{\alpha_1,\alpha_2+\alpha_2'\} P+N_3}\right).
\end{eqnarray*}

\subsection{The Gap for Relatively Large $R_2$}

We choose $\alpha_1$ and $\alpha_2$ such that $\alpha_1 P\geq N_2$, that $\alpha_2 P\geq N_3$, that $\alpha_1=\alpha_2'\leq \alpha_2$, and that $\alpha_1+\alpha_2\leq \frac{1}{2}$. It follows that $c_{21}\geq \frac{1}{3}$. We get the lower bounds for each term of $T_2$ expression above.
\begin{eqnarray}
&&\min\{T_{21}',T_{21}''\}\\
&&\geq \halflog\left(c_{21}+\frac{(1-\alpha_1-\alpha_2)P}{(2\alpha_1+\alpha_2)P+N_2}\right)\\
&&\geq \halflog\left(\frac{1}{3}+\frac{(1-\alpha_1-\alpha_2)P}{(3\alpha_1+\alpha_2)P}\right)\\
&&\geq \halflog\left(\frac{P}{(3\alpha_1+\alpha_2)P}\right).
\end{eqnarray}
The first entry of $\min\{\cdot,\cdot\}$ in
\[T_2 = \min\{\min\{T_{21}',T_{21}''\}+T_{22}'',T_{21}'''+T_{22}''\}\] is lower bounded as follows.
\begin{eqnarray*}
&&\min\{T_{21}',T_{21}''\}+T_{22}''\\
&&\geq \halflog\left(\frac{P}{(3\alpha_1+\alpha_2)P}\right)+\halflog\left(\frac{(\alpha_1+\alpha_2) P+N_2}{\alpha_1 P+N_2}\right)\\
&&= \halflog\left(\frac{P}{\alpha_1 P+N_2}\cdot\frac{(\alpha_1+\alpha_2) P+N_2}{(3\alpha_1+\alpha_2)P}\right)\\
&&\geq \halflog\left(\frac{P}{3(\alpha_1 P+N_2)}\right)\\
&&\geq \halflog\left(\frac{P}{6\alpha_1 P}\right).
\end{eqnarray*}
The second entry of $T_2=\min\{\cdot,\cdot\}$ is lower bounded as follows.
\begin{eqnarray*}
&&T_{21}'''+T_{22}''\\
&&=\halflog\left(1+\frac{\alpha_1 P}{N_2}\right)+\halflog\left(1+\frac{\alpha_2 P}{\alpha_1 P+N_2}\right)\\
&&=\halflog\left(1+\frac{(\alpha_1+\alpha_2) P}{N_2}\right)\\
&&\geq\halflog\left(\frac{\alpha_2 P}{N_2}\right).
\end{eqnarray*}
Putting together, we get the lower bound
\begin{eqnarray*}
&& T_2 \geq \min\left\{\halflog\left(\frac{P}{6\alpha_1 P}\right) ,\halflog\left(\frac{\alpha_2 P}{N_2}\right)\right\}.
\end{eqnarray*}
Given $\alpha_2$, we choose $\alpha_1$ that satisfies $\halflog\left(\frac{P}{6\alpha_1 P}\right)=\halflog\left(\frac{\alpha_2 P}{N_2}\right)$. As a result, we can write $T_2\geq \halflog\left(\frac{\alpha_2 P}{N_2}\right)$.
We also have
\begin{eqnarray*}
&& T_3 \geq \halflog\left(\frac{P}{(\alpha_1+\alpha_2) P+N_3}\right) \geq \halflog\left(\frac{P}{3\alpha_2 P}\right).
\end{eqnarray*}
Putting together, we can see that the following rate region is achievable.
\begin{eqnarray}
&& R_1\leq \halflog\left(\frac{\alpha_1 P}{N_1}\right)\\
&& R_2\leq \halflog\left(\frac{\alpha_2 P}{N_2}\right)\\
&& R_3\leq \halflog\left(\frac{P}{3\alpha_2 P}\right).
\end{eqnarray}
For fixed $\alpha_2$ and $R_2=\halflog\left(\frac{\alpha_2 P}{N_2}\right)$, the two-dimensional rate region, given by
\begin{eqnarray}
&& R_1\leq \halflog\left(\frac{\alpha_1 P}{N_1}\right)\\
&& \ \ \ \ = \halflog\left(\frac{P}{6\alpha_2 P}\cdot\frac{N_2}{N_1}\right)\\
&& R_3\leq \halflog\left(\frac{P}{3\alpha_2 P}\right)
\end{eqnarray}
is achievable.

In comparison, the two-dimensional outer bound region at $R_2=\halflog\left(\frac{\alpha_2 P}{N_2}\right)+1$ is given by
\begin{eqnarray*}
&&R_1 \leq \halflog\left(\frac{P}{N_1}\cdot\frac{7}{3}\right)-\halflog\left(\frac{\alpha_2 P}{N_2}\right)-1\\
&& \ \ \ \ = \halflog\left(\frac{P}{\alpha_2 P}\cdot\frac{N_2}{N_1}\right)+\halflog\left(\frac{7}{3}\cdot\frac{1}{4}\right)\\
&&R_3 \leq \halflog\left(\frac{P}{N_2}\cdot\frac{7}{3}\right)-\halflog\left(\frac{\alpha_2 P}{N_2}\right)-1\\
&&\ \ \ \ = \halflog\left(\frac{P}{\alpha_2 P}\right)+\halflog\left(\frac{7}{3}\cdot\frac{1}{4}\right).
\end{eqnarray*}
As discussed above, the sum-rate bound on $R_1+R_3$ is loose for $R_2$ larger than the threshold, so the rate region is a rectangle.

By comparing the inner and outer bound rate regions, we can see that $\delta_1< \halflog\left(6\cdot\frac{7}{3}\cdot\frac{1}{4}\right) < 0.91$ and $\delta_3< \halflog\left(3\cdot\frac{7}{3}\cdot\frac{1}{4}\right) < 0.41$. Therefore, we can conclude that the gap is to within one bit per message.

\subsection{Achievable Scheme for Relatively Small $R_2$}
\begin{theorem}
Given $\alpha=(\alpha_1,\alpha_2) \in [0,1]^2$, the region $\mathcal{R}_\alpha$ is defined by
\begin{eqnarray*}
&& R_1 \leq\halflog\left(1+\frac{\alpha_1 P}{N_1}\right)\\
&& R_2 \leq \min\left\{\halflog^+\left(c_{21}+\frac{(1-\alpha_2)P}{(\alpha_1+\alpha_2) P+N_2}\right),\right.\\
&&\ \ \ \ \ \ \ \ \ \ \ \ \ \ \ \ \ \ \ \ \ \ \ \ \ \ \ \ \ \ \ \ \ \left.\halflog\left(1+\frac{\alpha_2 P}{N_2}\right)\right\}\\
&& R_3 \leq \halflog^+\left(c_3+\frac{P}{\max\{\alpha_1,\alpha_2\} P+N_3}\right)
\end{eqnarray*}
where $c_{21}=\frac{1-\alpha_2}{2-\alpha_2}$ and $c_3=\frac{1}{2-\alpha_2}$,
and $\mathcal{R}=\textsc{conv}\left(\bigcup_{\alpha}\mathcal{R}_\alpha\right)$ is achievable.
\end{theorem}

For the case of $R_2 < R_{2,th}$, we present the following scheme. At transmitter 2, rate splitting is not necessary. The transmit signal is the sum
\[\mathbf{x}_2=\mathbf{x}_{21}+\mathbf{x}_{21}'\]
where $\mathbf{x}_{21}$ and $\mathbf{x}_{21}'$ are differently coded versions of the same message $M_2\in\{1,2,\ldots,2^{nR_2}\}$.

The power allocation: $\mathbb{E}[\|\mathbf{x}_1\|^2]=\alpha_1 nP$ at receiver 1, $\mathbb{E}[\|\mathbf{x}_{21}\|^2]=\alpha_2 nP$, and $\mathbb{E}[\|\mathbf{x}_{21}'\|^2]=(1-\alpha_2)nP$ at receiver 2, and $\mathbb{E}[\|\mathbf{x}_3\|^2]=nP$ at receiver 3.

The signals $\mathbf{x}_{21}'$ and $\mathbf{x}_3$ are lattice codewords using the same coding lattice but different shaping lattices. As a result, the sum $\mathbf{x}_{21}'+\mathbf{x}_3$ is a lattice codeword.

The received signals are
\begin{eqnarray*}
&&\mathbf{y}_1=\mathbf{x}_{21}'+\mathbf{x}_{21}+\mathbf{x}_1+\mathbf{z}_1\\
&&\mathbf{y}_2=[\mathbf{x}_{21}'+\mathbf{x}_3]+\mathbf{x}_{21}+\mathbf{z}_2\\
&&\mathbf{y}_3=\mathbf{x}_3+\mathbf{x}_1+\mathbf{z}_3.
\end{eqnarray*}
The signal scale diagram at each receiver is shown in Fig. \ref{fig:signalScale5} (b).
Decoding is performed in the following way.
\begin{itemize}
\item At receiver 1, $\mathbf{x}_{21}'$
is first decoded while treating other signals as noise. Having successfully recovered $M_{21}$, receiver 1 can generate $\mathbf{x}_{21}$ and $\mathbf{x}_{21}'$, and cancel them from $\mathbf{y}_1$. Next, $\mathbf{x}_1$ is decoded from $\mathbf{x}_1+\mathbf{z}_1$. For reliable decoding, the code rates should satisfy
\begin{eqnarray*}
&& R_{21}\leq T_{21}' =\halflog\left(1+\frac{(1-\alpha_2)P}{(\alpha_1+\alpha_2) P+N_1}\right)\\
&& R_1\ \leq T_1\ =\halflog\left(1+\frac{\alpha_1 P}{N_1}\right).
\end{eqnarray*}
\item At receiver 2, $[\mathbf{x}_{21}'+\mathbf{x}_3]$ first decoded while treating other signals as noise and removed from $\mathbf{y}_2$. Next, $\mathbf{x}_{21}$ is decoded from $\mathbf{x}_{21}+\mathbf{z}_2$. For reliable decoding, the code rates should satisfy
\begin{eqnarray*}
&&R_{21}\leq T_{21}'' =\halflog\left(c_{21}+\frac{(1-\alpha_2)P}{\alpha_2 P+N_2}\right)\\
&&R_3\ \leq T_3'\ =\halflog\left(c_3+\frac{P}{\alpha_2 P+N_2}\right)\\
&&R_{21}\leq T_{21}''' =\halflog\left(1+\frac{\alpha_2 P}{N_2}\right)
\end{eqnarray*}
where $c_{21}=\frac{(1-\alpha_2)P}{(1-\alpha_2)P+P}=\frac{1-\alpha_2}{2-\alpha_2}$ and $c_3=\frac{P}{(1-\alpha_2)P+P}=\frac{1}{2-\alpha_2}$. Note that $0\leq c_{21}\leq \frac{1}{2}$, $c_{21}+c_3=1$, and $\frac{1}{2}\leq c_3\leq 1$.
\item At receiver 3, $\mathbf{x}_3$ is decoded while treating $\mathbf{x}_1+\mathbf{z}_3$ as noise. Reliable decoding is possible if
\begin{eqnarray}
&&R_3\ \leq T_3'' =\halflog\left(1+\frac{P}{\alpha_1 P+N_3}\right).
\end{eqnarray}
\end{itemize}
Putting together, we get
\begin{eqnarray}
&& R_1 \leq T_1\\
&& R_2 \leq T_2 =\min\{T_{21}',T_{21}'',T_{21}'''\}\\
&& R_3 \leq T_3=\min\{T_3',T_3''\}
\end{eqnarray}
where
\begin{eqnarray*}
&& T_1 =\halflog\left(1+\frac{\alpha_1 P}{N_1}\right)\\
&& T_2 = \min\{T_{21}',T_{21}'',T_{21}'''\}\\
&&\ \ \ \ = \min\{\min\{T_{21}',T_{21}''\}, T_{21}'''\}\\
&&\ \ \ \ \geq \min\left\{\halflog\left(c_{21}+\frac{(1-\alpha_2)P}{(\alpha_1+\alpha_2) P+N_2}\right),\right.\\
&&\ \ \ \ \ \ \ \ \ \ \ \ \ \ \ \ \ \ \ \ \ \ \ \ \ \ \ \ \ \ \ \ \ \left.\halflog\left(1+\frac{\alpha_2 P}{N_2}\right)\right\}\\
&& T_3 \geq \halflog\left(c_3+\frac{P}{\max\{\alpha_1,\alpha_2\} P+N_3}\right).
\end{eqnarray*}

\subsection{The Gap for Relatively Small $R_2$}
We choose $\alpha_1$ and $\alpha_2$ such that $\alpha_1 P\geq N_2$, that $\alpha_2 P\geq N_3$, that $\alpha_1+\alpha_2\leq \frac{1}{2}$, and that $\alpha_1\geq \alpha_2$. It follows that $c_{21}\geq \frac{1}{3}$. We get the lower bound
\begin{eqnarray}
&&\min\{T_{21}',T_{21}''\}\\
&&= \halflog\left(c_{21}+\frac{(1-\alpha_2)P}{(\alpha_1+\alpha_2)P+N_2}\right)\\
&&\geq \halflog\left(\frac{1}{3}+\frac{(1-\alpha_1)P}{3\alpha_1 P}\right)\\
&&= \halflog\left(\frac{P}{3\alpha_1 P}\right)
\end{eqnarray}
and
\begin{eqnarray*}
&& T_2 \geq \min\left\{\halflog\left(\frac{P}{3\alpha_1 P}\right), \halflog\left(\frac{\alpha_2 P}{N_2}\right)\right\}.
\end{eqnarray*}
Let us define $\alpha_2'$ by the equality $\halflog\left(\frac{P}{3\alpha_2' P}\right)=\halflog\left(\frac{\alpha_2 P}{N_2}\right)$. If we choose $\alpha_1\leq \alpha_2'$, then $T_2 \geq \halflog\left(\frac{\alpha_2 P}{N_2}\right)$.
We can see that the following rate region is achievable.
\begin{eqnarray}
&& R_1\leq \halflog\left(\frac{\alpha_1 P}{N_1}\right)\\
&& R_2\leq \halflog\left(\frac{\alpha_2 P}{N_2}\right)\\
&& R_3\leq \halflog\left(\frac{P}{2\alpha_1 P}\right).
\end{eqnarray}

For fixed $\alpha_1\in [\alpha_2,\alpha_2']$ and $R_2=\halflog\left(\frac{\alpha_2 P}{N_2}\right)$, the two-dimensional rate region $\mathcal{R}_\alpha$, given by
\begin{eqnarray}
&& R_1\leq \halflog\left(\frac{\alpha_1 P}{N_1}\right)\\
&& R_3\leq \halflog\left(\frac{P}{2\alpha_1 P}\right)
\end{eqnarray}
is achievable. The union $\bigcup_{\alpha_1\in [\alpha_2,\alpha_2']}\mathcal{R}_\alpha$ is a MAC-like region, given by
\begin{eqnarray}
&& \ \ \ \ \ \ \ R_1\leq \halflog\left(\frac{\alpha_2' P}{N_1}\right)\\
&& \ \ \ \ \ \ \ \ \ \ \ = \halflog\left(\frac{P}{3\alpha_2 P}\cdot\frac{N_2}{N_1}\right)\\
&& \ \ \ \ \ \ \ R_3\leq \halflog\left(\frac{P}{2\alpha_2 P}\right)\\
&& R_1+R_3=\halflog\left(\frac{P}{2N_1}\right).
\end{eqnarray}

In comparison, the two-dimensional outer bound region at $R_2=\halflog\left(\frac{\alpha_2 P}{N_2}\right)+1$ is given by
\begin{eqnarray*}
&&\ \ \ \ \ \ \ R_1 \leq \halflog\left(\frac{P}{N_1}\cdot\frac{7}{3}\right)-\halflog\left(\frac{\alpha_2 P}{N_2}\right)-1\\
&&\ \ \ \ \ \ \ \ \ \ \ = \halflog\left(\frac{P}{\alpha_2 P}\cdot\frac{N_2}{N_1}\right)+\halflog\left(\frac{7}{3}\cdot\frac{1}{4}\right)\\
&&\ \ \ \ \ \ \ R_3 \leq \halflog\left(\frac{P}{N_2}\cdot\frac{7}{3}\right)-\halflog\left(\frac{\alpha_2 P}{N_2}\right)-1\\
&&\ \ \ \ \ \ \ \ \ \ \ = \halflog\left(\frac{P}{\alpha_2 P}\right)+\halflog\left(\frac{7}{3}\cdot\frac{1}{4}\right)\\
&&R_1+R_3 \leq \halflog\left(\frac{P}{N_1}\cdot\frac{8}{3}\right).
\end{eqnarray*}
Since $\delta_1 < \halflog\left(3\cdot\frac{7}{3}\cdot\frac{1}{4}\right) < 0.41$, $\delta_3 < \halflog\left(2\cdot\frac{7}{3}\cdot\frac{1}{4}\right) < 0.12$ and $\delta_{13} < \halflog\left(2\cdot\frac{7}{3}\right) < 1.12 < \sqrt{2}$, we can conclude that the gap is to within one bit per message.

\section{Conclusion}
We presented approximate capacity region of five important cases of partially connected interference channels. The outer bounds based on $Z$-channel type argument are derived. Achievable schemes are developed and shown to approximately achieve the capacity to within a constant bit. 

For future work, the channels with fully general coefficients may be considered. In this paper, we presented different schemes for each channel type although they share some principle. A universal scheme is to be developed for unified capacity characterization of all possible topologies. The connection between interference channel and index coding problems is much to explore. In particular, the results on the capacity region for index coding in \cite{ArbabjolfaeiBandemerKimSasogluWang13} seem to have an interesting connection to our work.

\appendices

\section{Random Coding Achievability: Channel Type 4}
At transmitter 1, message $M_1$ is split into three parts $(M_{12},M_{11},M_{10})$, and the transmit signal is $\mathbf{x}_1=\mathbf{x}_{12}+\mathbf{x}_{11}+\mathbf{x}_{10}$. The signals satisfy $\mathbb{E}[\|\mathbf{x}_{12}\|^2]=n(P-N_2-N_3)$, $\mathbb{E}[\|\mathbf{x}_{11}\|^2]=nN_3$, and $\mathbb{E}[\|\mathbf{x}_{10}\|^2]=nN_2$.

At transmitter 2, message $M_2$ is split into three parts $(M_{21},M_{20})$, and the transmit signal is $\mathbf{x}_2=\mathbf{x}_{21}+\mathbf{x}_{20}$. The signals satisfy $\mathbb{E}[\|\mathbf{x}_{21}\|^2]=n(P-N_3)$ and $\mathbb{E}[\|\mathbf{x}_{20}\|^2]=nN_3$. Rate-splitting is not performed at transmitter 3, and $\mathbb{E}[\|\mathbf{x}_3\|^2]=nP$.

The top layer codewords $(\mathbf{x}_{12},\mathbf{x}_{21},\mathbf{x}_{3})$ are from a joint random codebook for $(M_{12},M_{21},M_3)$. The mid-layer codewords $(\mathbf{x}_{11},\mathbf{x}_{20})$ are from a joint random codebook for $(M_{11},M_{20})$. The bottom layer codeword $\mathbf{x}_{10}$ is from a single-user random codebook for $M_{10}$.

The received signals are
\begin{eqnarray*}
&&\mathbf{y}_1=(\mathbf{x}_{12}+\mathbf{x}_3)+\mathbf{x}_{11}+\mathbf{x}_{10}+\mathbf{z}_1\\
&&\mathbf{y}_2=(\mathbf{x}_{12}+\mathbf{x}_{21})+(\mathbf{x}_{11}+\mathbf{x}_{20})+\mathbf{x}_{10}+\mathbf{z}_2\\
&&\mathbf{y}_3=(\mathbf{x}_{21}+\mathbf{x}_3)+\mathbf{x}_{20}+\mathbf{z}_3
\end{eqnarray*}
Decoding is performed from the top layer to the bottom layer. At receiver 1, simultaneous decoding of $(\mathbf{x}_{12},\mathbf{x}_{3})$ is performed while treating other signals as noise. And then, $\mathbf{x}_{11}$ and $\mathbf{x}_{10}$ are decoded successively. At receiver 2, simultaneous decoding of $(\mathbf{x}_{12},\mathbf{x}_{21})$ is performed while treating other signals as noise. And then, simultaneous decoding of $(\mathbf{x}_{11},\mathbf{x}_{20})$ is performed. At receiver 3, simultaneous decoding of $(\mathbf{x}_{21},\mathbf{x}_3)$ is performed while treating other signals as noise. For reliable decoding, code rates should satisfy
\begin{eqnarray*}
&& \ \ \ \ \ \ \ \ R_{12}\leq I_1=\halflog\left(1+\frac{P-N_2-N_3}{N_1+N_2+N_3}\right)\\
&& \ \ \ \ \ \ \ \ R_{3}\ \leq I_2=\halflog\left(1+\frac{P}{N_1+N_2+N_3}\right)\\
&& R_{12}+R_3\ \leq I_3=\halflog\left(1+\frac{2P-N_2-N_3}{N_1+N_2+N_3}\right)\\
&& \ \ \ \ \ \ \ \ R_{11}\leq I_4=\halflog\left(1+\frac{N_3}{N_1+N_2}\right)\\
&& \ \ \ \ \ \ \ \ R_{10}\leq I_5=\halflog\left(1+\frac{N_2}{N_1}\right)
\end{eqnarray*}
at receiver 1,
\begin{eqnarray*}
&& \ \ \ \ \ \ \ \ R_{12}\leq I_6=\halflog\left(1+\frac{P-N_2-N_3}{2N_2+2N_3}\right)\\
&& \ \ \ \ \ \ \ \ R_{21}\leq I_7=\halflog\left(1+\frac{P-N_3}{2N_2+2N_3}\right)\\
&& R_{12}+R_{21}\leq I_8=\halflog\left(1+\frac{2P-N_2-2N_3}{2N_2+2N_3}\right)\\
&& \ \ \ \ \ \ \ \ R_{11}\leq I_9=\halflog\left(1+\frac{N_3}{2N_2}\right)\\
&& \ \ \ \ \ \ \ \ R_{20}\leq I_{10}=\halflog\left(1+\frac{N_3}{2N_2}\right)\\
&& R_{11}+R_{20}\leq I_{11}=\halflog\left(1+\frac{2N_3}{2N_2}\right)
\end{eqnarray*}
at receiver 2,
\begin{eqnarray*}
&& \ \ \ \ \ \ \ \ R_{21}\leq I_{12}=\halflog\left(1+\frac{P-N_3}{2N_3}\right)\\
&& \ \ \ \ \ \ \ \ R_{3}\ \leq I_{13}=\halflog\left(1+\frac{P}{2N_3}\right)\\
&& R_{21}+R_{3}\ \leq I_{14}=\halflog\left(1+\frac{2P-N_3}{2N_3}\right)
\end{eqnarray*}
at receiver 3. Putting together,
\begin{eqnarray*}
&& \ \ \ \ \ \ \ \ R_{12}\leq T_1=\min\{I_1,I_6\}=I_6\\
&& \ \ \ \ \ \ \ \ R_{21}\leq T_2=\min\{I_7,I_{12}\}=I_7\\
&& \ \ \ \ \ \ \ \ R_{3}\ \leq T_3=\min\{I_2,I_{13}\}\\
&& R_{12}+R_{21}\leq T_4=I_8\\
&& R_{12}+R_{3}\ \leq T_5=I_3\\
&& R_{21}+R_{3}\ \leq T_6=I_{14}
\end{eqnarray*}
at the top layer,
\begin{eqnarray*}
&& \ \ \ \ \ \ \ \ R_{11}\leq T_7=\min\{I_4,I_9\}=I_9\\
&& \ \ \ \ \ \ \ \ R_{20}\leq T_8=I_{10}\\
&& R_{11}+R_{20}\leq T_9=I_{11}
\end{eqnarray*}
at the mid-layer,
\begin{eqnarray*}
R_{10}\leq T_{10}=I_5
\end{eqnarray*}
at the bottom layer. Note that the rate variables are not coupled between layers.
We get the achievable rate region
\begin{eqnarray*}
&& \ \ \ \ \ \ \ R_{1}=R_{12}+R_{11}+R_{10}\leq T_1+T_7+T_{10}\\
&& \ \ \ \ \ \ \ R_{2}=R_{21}+R_{20}\leq T_2+T_8\\
&& \ \ \ \ \ \ \ R_{3}\leq T_3\\
&& R_{1}+R_{2}\leq T_4+T_9+T_{10}\\
&& R_{1}+R_{3}\leq T_5+T_7+T_{10}\\
&& R_{2}+R_{3}\leq T_6+T_8.
\end{eqnarray*}
This region includes the following region.
\begin{eqnarray*}
&& \ \ \ \ \ \ \ R_{1}\leq \halflog\left(2+\frac{P}{N_1}\right)-1\\
&& \ \ \ \ \ \ \ R_{2}\leq \halflog\left(3+\frac{P}{N_2}\right)-1\\
&& \ \ \ \ \ \ \ R_{3}\leq \halflog\left(3+\frac{P}{N_3}\right)-\halflog(3)\\
&& R_{1}+R_{2}\leq \halflog\left(1+\frac{2P}{N_1}\right)-\frac{1}{2}\\
&& R_{1}+R_{3}\leq \halflog\left(1+\frac{2P}{N_1}\right)-1\\
&& R_{2}+R_{3}\leq \halflog\left(1+\frac{2P}{N_2}\right)-1.
\end{eqnarray*}
Therefore, we can conclude the capacity region to within one bit.

\section{Random Coding Achievability: Channel Type 5}
Transmit signal construction is the same as the one for channel type 4.
The received signals are
\begin{eqnarray*}
&&\mathbf{y}_1=(\mathbf{x}_{12}+\mathbf{x}_{21})+(\mathbf{x}_{11}+\mathbf{x}_{20})+\mathbf{x}_{10}+\mathbf{z}_1\\
&&\mathbf{y}_2=(\mathbf{x}_{21}+\mathbf{x}_{3})+\mathbf{x}_{20}+\mathbf{z}_2\\
&&\mathbf{y}_3=(\mathbf{x}_{12}+\mathbf{x}_3)+\mathbf{x}_{11}+\mathbf{x}_{10}+\mathbf{z}_3
\end{eqnarray*}
Decoding is performed from the top layer to the bottom layer. At receiver 1, simultaneous decoding of $(\mathbf{x}_{12},\mathbf{x}_{21})$ is performed while treating other signals as noise. And then, simultaneous decoding of $\mathbf{x}_{11}$ and $\mathbf{x}_{20}$ is performed. Lastly, $\mathbf{x}_{10}$ is decoded. At receiver 2, simultaneous decoding of $(\mathbf{x}_{21},\mathbf{x}_{3})$ is performed while treating other signals as noise. And then, $\mathbf{x}_{20}$ is decoded. At receiver 3, simultaneous decoding of $(\mathbf{x}_{12},\mathbf{x}_3)$ is performed while treating other signals as noise. And then, $\mathbf{x}_{11}$ and $\mathbf{x}_{10}$ are decoded successively. For reliable decoding, code rates should satisfy
\begin{eqnarray*}
&& \ \ \ \ \ \ \ \ R_{12}\leq I_1=\halflog\left(1+\frac{P-N_2-N_3}{N_1+N_2+2N_3}\right)\\
&& \ \ \ \ \ \ \ \ R_{21}\leq I_2=\halflog\left(1+\frac{P-N_3}{N_1+N_2+2N_3}\right)\\
&& R_{12}+R_{21}\leq I_3=\halflog\left(1+\frac{2P-N_2-2N_3}{N_1+N_2+2N_3}\right)\\
&& \ \ \ \ \ \ \ \ R_{11}\leq I_4=\halflog\left(1+\frac{N_3}{N_1+N_2}\right)\\
&& \ \ \ \ \ \ \ \ R_{20}\leq I_5=\halflog\left(1+\frac{N_3}{N_1+N_2}\right)\\
&& R_{11}+R_{20}\leq I_6=\halflog\left(1+\frac{2N_3}{N_1+N_2}\right)\\
&& \ \ \ \ \ \ \ \ R_{10}\leq I_7=\halflog\left(1+\frac{N_2}{N_1}\right)
\end{eqnarray*}
at receiver 1,
\begin{eqnarray*}
&& \ \ \ \ \ \ \ \ R_{21}\leq I_8=\halflog\left(1+\frac{P-N_3}{N_2+N_3}\right)\\
&& \ \ \ \ \ \ \ \ R_{3}\ \leq I_9=\halflog\left(1+\frac{P}{N_2+N_3}\right)\\
&& R_{21}+R_{3}\ \leq I_{10}=\halflog\left(1+\frac{2P-N_3}{N_2+N_3}\right)\\
&& \ \ \ \ \ \ \ \ R_{20}\leq I_{11}=\halflog\left(1+\frac{N_3}{N_2}\right)
\end{eqnarray*}
at receiver 2,
\begin{eqnarray*}
&& \ \ \ \ \ \ \ \ R_{12}\leq I_{12}=\halflog\left(1+\frac{P-N_2-N_3}{N_2+2N_3}\right)\\
&& \ \ \ \ \ \ \ \ R_{3}\ \leq I_{13}=\halflog\left(1+\frac{P}{N_2+2N_3}\right)\\
&& R_{12}+R_{3}\ \leq I_{14}=\halflog\left(1+\frac{2P-N_2-N_3}{N_2+2N_3}\right)
\end{eqnarray*}
at receiver 3. Putting together,
\begin{eqnarray*}
&& \ \ \ \ \ \ \ \ R_{12}\leq T_1=\min\{I_1,I_{12}\}=I_1\\
&& \ \ \ \ \ \ \ \ R_{21}\leq T_2=\min\{I_2,I_8\}=I_2\\
&& \ \ \ \ \ \ \ \ R_{3}\ \leq T_3=\min\{I_9,I_{13}\}=I_{13}\\
&& R_{12}+R_{21}\leq T_4=I_3\\
&& R_{12}+R_{3}\ \leq T_5=I_{14}\\
&& R_{21}+R_{3}\ \leq T_6=I_{10}
\end{eqnarray*}
at the top layer,
\begin{eqnarray*}
&& \ \ \ \ \ \ \ \ R_{11}\leq T_7=I_4\\
&& \ \ \ \ \ \ \ \ R_{20}\leq T_8=\min\{I_5,I_{11}\}=I_5\\
&& R_{11}+R_{20}\leq T_9=I_6
\end{eqnarray*}
at the mid-layer,
\begin{eqnarray*}
R_{10}\leq T_{10}=I_7
\end{eqnarray*}
at the bottom layer. Note that the rate variables are not coupled between layers.
We get the achievable rate region
\begin{eqnarray*}
&& \ \ \ \ \ \ \ R_{1}=R_{12}+R_{11}+R_{10}\leq T_1+T_7+T_{10}\\
&& \ \ \ \ \ \ \ R_{2}=R_{21}+R_{20}\leq T_2+T_8\\
&& \ \ \ \ \ \ \ R_{3}\leq T_3\\
&& R_{1}+R_{2}\leq T_4+T_9+T_{10}\\
&& R_{1}+R_{3}\leq T_5+T_7+T_{10}\\
&& R_{2}+R_{3}\leq T_6+T_8.
\end{eqnarray*}
This region includes the following region.
\begin{eqnarray*}
&& \ \ \ \ \ \ \ R_{1}\leq \halflog\left(2+\frac{P}{N_1}\right)-\frac{1}{2}\\
&& \ \ \ \ \ \ \ R_{2}\leq \halflog\left(2+\frac{P}{N_2}\right)-1\\
&& \ \ \ \ \ \ \ R_{3}\leq \halflog\left(3+\frac{P}{N_3}\right)-\halflog(3)\\
&& R_{1}+R_{2}\leq \halflog\left(1+\frac{2P}{N_1}\right)\\
&& R_{1}+R_{3}\leq \halflog\left(1+\frac{2P}{N_1}\right)-\frac{1}{2}\\
&& R_{2}+R_{3}\leq \halflog\left(1+\frac{2P}{N_2}\right)-\frac{1}{2}.
\end{eqnarray*}
Therefore, we can conclude the capacity region to within one bit.


\begin{thebibliography}{1}

\bibitem{EtkinTseWang08}
R.~Etkin, D.~Tse and H.~Wang, ``Gaussian interference channel
capacity to within one bit,'' \emph{IEEE Trans. Inf. Theory,} vol. 54, no. 12, pp. 5534--5562, Dec. 2008.

\bibitem{SridharanVishwanathJafar08}
S.~Sridharan, S.~Vishwanath, and S.~A.~Jafar, ``Capacity of the
symmetric $K$-user Gaussian very strong interference channel,''
\emph{Proc. IEEE Global Telecommun. Conf.,} vol. 56, Dec. 2008.

\bibitem{SridharanJafarianVishwanathJafarShamai08}
S.~Sridharan, A.~Jafarian, S.~Vishwanath, S.~Jafar, and
S.~Shamai, ``A layered lattice coding scheme for a class of
three user Gaussian interference channel,'' \emph{46th Annual Allerton Conference on Communication, Control, and Computing,} pp. 531--538, 2008.

\bibitem{BreslerParekhTse10}
G.~Bresler, A.~Parekh, and D.~N.~C.~Tse, ``The approximate capacity
of the many-to-one and one-to-many Gaussian interference channels,''
\emph{IEEE Trans. Inf. Theory,} vol. 56, no. 9, pp. 4566--4592, Sep. 2010.

\bibitem{JafarVishwanath10}
S.~A.~Jafar and S.~Vishwanath, ``Generalized degrees of freedom of the symmetric Gaussian $K$-user interference channel,'' \emph{IEEE Trans. Inf. Theory,} vol. 56, no. 7, pp. 3297--3303, Jul. 2010.

\bibitem{ZhouYu13}
L.~Zhou and W.~Yu, ``On the capacity of the $K$-user cyclic Gaussian
interference channel,'' \emph{IEEE Trans. Inf. Theory,} vol. 59, no. 1,
pp. 154--165, Jan. 2013.

\bibitem{OrdentlichErezNazer14}
O.~Ordentlich, U.~Erez, and B.~Nazer, ``The approximate sum capacity of the symmetric Gaussian $K$-user interference channel,'' \emph{IEEE Trans. Inf. Theory,} vol. 60, no. 6, pp. 3450--3482, Jun. 2014.

\bibitem{Jafar2014}
S.~A.~Jafar, ``Topological interference management through index coding,'' \emph{IEEE Trans. Inf. Theory,} vol. 60, no. 1, pp.
529--568, Jan. 2014.

\bibitem{HuangCadambeJafar12}
C.~Huang, V.~R.~Cadambe, and S.~A.~Jafar, ``Interference alignment and
the generalized degrees of freedom of the X channel,'' \emph{IEEE Trans. Inf. Theory,} vol. 58, no. 8, pp. 5130--5150, Aug. 2012.

\bibitem{NiesenMaddahAli13}
U.~Niesen and M.~A.~Maddah-Ali, ``Interference alignment: From degrees-of-freedom to constant-gap capacity approximations,'' \emph{IEEE Trans. Inf. Theory,} vol. 59, no. 8, pp. 4855--4888, Apr. 2013.

\bibitem{MotahariGharanMaddahAliKhandani14}
A.~S.~Motahari, S.~O.-Gharan, M.-A.~Maddah-Ali, and A.~K.~Khandani, ``Real interference alignment: Exploiting the potential of single antenna systems,'' \emph{IEEE Trans. Inf. Theory,} vol. 60, no. 8, pp. 4799--4810, Aug. 2014.


\bibitem{ErezZamir04}
U.~Erez and R.~Zamir, ``Achieving $\frac{1}{2}\log(1+\snr)$ on the AWGN channel
with lattice encoding and decoding,'' \emph{IEEE Trans. Inf.
Theory,} vol. 50, no. 10, pp. 2293--2314, Oct. 2004.

\bibitem{WilsonNarayananPfisterSprintson10}
M.~P.~Wilson, K.~Narayanan, H.~Pfister, and A.~Sprintson, ``Joint physical
layer coding and network coding for bidirectional relaying,'' \emph{IEEE Trans.
Inf. Theory,} vol. 56, no. 11, pp. 5641--5654, Nov. 2010.

\bibitem{NamChungLee10}
W.~Nam, S.-Y.~Chung, and Y.~H.~Lee, ``Capacity of the Gaussian
two-way relay channel to within $\frac{1}{2}$ bit,'' \emph{IEEE Trans. Inf. Theory,}
vol. 56, no. 11, pp. 5488--5494, Nov. 2010.

\bibitem{NamChungLee11}
W.~Nam, S.-Y.~Chung, and Y.~H.~Lee, ``Nested lattice codes for Gaussian relay networks with interference,'' \emph{IEEE Trans. Inf. Theory,} vol. 57, no. 12, pp. 7733--7745, Dec. 2011.

\bibitem{NazerGastpar11}
B.~Nazer and M.~Gastpar, ``Compute-and-forward: Harnessing interference through structured codes,'' \emph{IEEE Trans. Inf. Theory,} vol. 57, no. 10, pp. 6463--6486, Oct. 2011.

\bibitem{GastparNazer11}
M.~Gastpar and B.~Nazer, ``Algebraic structure in network information theory,'' \emph{IEEE ISIT} Tutorial, 2011.


\bibitem{BirkKol98}
Y.~Birk and T.~Kol, ``Informed-source coding-on-demand (ISCOD) over
broadcast channels,'' in \emph{Proc. IEEE INFOCOM,} vol. 13, pp. 1257--1264, 1998.

\bibitem{Bar-YossefBirkJayramKol06}
Z.~Bar-Yossef, Y.~Birk, T.~S.~Jayram, and T.~Kol, ``Index coding with
side information,'' in \emph{Proc. 47th IEEE Ann. Symp. Found. Comput. Sci.
(FOCS),} 2006, pp. 197–-206.

\bibitem{Bar-YossefBirkJayramKol11}
Z.~Bar-Yossef, Y.~Birk, T.~S.~Jayram, and T.~Kol, ``Index coding with
side information,'' \emph{IEEE Trans. Inf. Theory,} vol. 57, no. 3, pp.
1479--1494, Mar. 2011.

\bibitem{AlonLubetzkyStavWeinsteinHassidim08}
N.~Alon, E.~Lubetzky, U.~Stav, A.~Weinstein, and A.~Hassidim, ``Broadcasting
with side information,'' in \emph{Proc. 49th IEEE Ann. Symp. Found. Comput. Sci.
(FOCS),} 2008, pp. 823--832.

\bibitem{MalekiCadambeJafar2012}
H.~Maleki, V.~Cadambe, and S.~A.~Jafar, ``Index coding--An interference
alignment perspective,'' in \emph{Proc. IEEE Int. Symp. Inf. Theory,} 2012, pp.
2236--2240.

\bibitem{ArbabjolfaeiBandemerKimSasogluWang13}
F.~Arbabjolfaei, B.~Bandemer, Y.~-H.~Kim, E.~Sasoglu, and
L.~Wang, ``On the capacity region for index coding,'' in \emph{Proc.
IEEE Int. Symp. Inf. Theory,} Istanbul, Turkey, Jul. 2013, pp. 962--966.

\bibitem{Ong14}
L.~Ong, ``Linear codes are optimal for index-coding
instances with five or fewer receivers,'' in \emph{Proc. IEEE Int. Symp. Inf. Theory,} 2014, pp.
491--495.

\bibitem{EffrosElRouayhebLangberg15}
M.~Effros, S.~El Rouayheb, M.~Langberg, ``An equivalence between network coding and index coding,'' 
\emph{IEEE Trans. Inf. Theory,} vol. 61, no. 5, pp.
2478--2487, May 2015.

\bibitem{ElGamalKim11}
A.~El~Gamal and Y.-H.~Kim, \emph{Network Information Theory.} Cambridge Univ. Press, 2011.

\bibitem{Zamir14}
R.~Zamir, \emph{Lattice Coding for Signals and Networks.} Cambridge Univ.
Press, 2014.

%\bibitem{SahraeiGastpar14}
%S.~Sahraei and M.~Gastpar, ``Compute-and-forward: Finding the best equation,'' arXiv:1410.3656v1.

%\bibitem{OrdentlichErez13}
%O.~Ordentlich and U.~Erez, ``On the robustness of lattice interference alignment,'' \emph{IEEE Trans. Inf. Theory,} vol. 59, no. 5, pp. 2435--2759, May 2013.

%\bibitem{OrdentlichErezNazer12b}
%O.~Ordentlich, U.~Erez, and B.~Nazer, ``The compute-and-forward transform,''  in \emph{Proc. IEEE
%ISIT,} Jul. 2012, pp. 3008--3012.

%\bibitem{OrdentlichErezNazer13}
%O.~Ordentlich, U.~Erez, and B.~Nazer, ``Successive integer-forcing and its sum-rate optimality,'' in \emph{51st Annual Allerton Conference on Communication, Control, and Computing,} Monticello, IL, Oct. 2013.

%\bibitem{Nazer12}
%B.~Nazer, ``Successive compute-and-forward,'' in \emph{Proceedings of the International Zurich Seminar on Communications (IZS 2012),} Zurich, Switzerland, Mar. 2012.

%\bibitem{NiesenWhiting12}
%U.~Niesen and P.~Whiting, ``The degrees of freedom of compute-and-forward,'' \emph{IEEE Trans. Inf. Theory,} vol. 58, no. 8, pp. 5214--5232, Aug. 2012.

%\bibitem{AvestimehrDiggaviTse11}
%A.~S.~Avestimehr, S.~N.~Diggavi, and D.~N.~C.~Tse, ``Wireless network information flow: A deterministic approach,'' \emph{IEEE Trans. Inf. Theory,} vol. 57, no. 4, pp. 1872--1905, Apr. 2011.
%
%\bibitem{CoverThomas91}
%T.~Cover and J.~Thomas, \emph{Elements of information theory,} New
%York: Wiley, 1991.

%\bibitem{ForneyTrottChung00}
%G.~D.~Forney, Jr., M.~D.~Trott, and S.-Y.~Chung, ``Sphere-bound-achieving coset codes and multilevel coset codes,'' \emph{IEEE Trans. Inf. Theory,} vol. 46, no. 3, pp.
%820-850, Oct. 2000.
%
%\bibitem{Forney}
%G.~D.~Forney Jr., Principles of Digital Communication II, MIT course notes.

%\bibitem{Ben-TalNemirovski2012}
%A.~Ben-Tal and A.~Nemirovski, \emph{Lectures on modern convex optimization: Analysis, algorithms, and engineering applications,} MPS-SIAM series on optimization, 2001.

\end{thebibliography}
\end{document}